\documentclass[vecphys]{svmult}

\usepackage{makeidx}         
\usepackage{graphicx}        
\usepackage{multicol}        

\makeindex             

\begin{document}

\title*{Field Theory Approaches to Nonequilibrium Dynamics}

\author{Uwe Claus T\"auber}

\institute{Department of Physics,
        Center for Stochastic Processes in Science and Engineering \\ 
        Virginia Polytechnic Institute and State University \\
        Blacksburg, Virginia 24061-0435, USA \\
	  email: \texttt{tauber@vt.edu}}

\begin{abstract}
	It is explained how field-theoretic methods and the dynamic 
	renormalisation group (RG) can be applied to study the universal 
	scaling properties of systems that either undergo a continuous phase 
	transition or display generic scale invariance, both near and far 
	from thermal equilibrium.
	Part 1 introduces the response functional field theory representation
	of (nonlinear) Langevin equations. 
	The RG is employed to compute the scaling exponents for several 
	universality classes governing the critical dynamics near second-order
	phase transitions in equilibrium.
	The effects of reversible mode-coupling terms, quenching from random
	initial conditions to the critical point, and violating the detailed
	balance constraints are briefly discussed.
	It is shown how the same formalism can be applied to nonequilibrium
	systems such as driven diffusive lattice gases. 
	Part 2 describes how the master equation for stochastic particle 
	reaction processes can be mapped onto a field theory action.
	The RG is then used to analyse simple diffusion-limited annihilation 
	reactions as well as generic continuous transitions from active to
	inactive, absorbing states, which are characterised by the power laws
	of (critical) directed percolation.
	Certain other important universality classes are mentioned, and some 
	open issues are listed.
\end{abstract}

\maketitle

\section{Critical Dynamics}
\label{sec:I}

Field-theoretic tools and the {\em renormalisation group} (RG) method have had
a tremendous impact in our understanding of the {\em universal power laws} 
that emerge near equilibrium critical points (see, e.g., 
Refs.~\cite{Ramond,Amit,ItzDro,Bellac,Zinn,Cardy}), 
including the associated {\em dynamic critical phenomena} 
\cite{HohHalp,Janssen}.
Our goal here is to similarly describe the scaling properties of systems 
driven far from thermal equilibrium, which either undergo a {\em continuous 
nonequilibrium phase transition} or display {\em generic scale invariance}.
We are then confronted with capturing the (stochastic) dynamics of the 
long-wavelength modes of the `slow' degrees of freedom, namely the order 
parameter for the transition, any conserved quantities, and perhaps additional
relevant variables. 
In these lecture notes, I aim to briefly describe how a representation in 
terms of a {\em field theory action} can be obtained for (1) general nonlinear
Langevin stochastic differential equations \cite{Janssen,BaJaWa}; and (2) for
master equations governing classical particle reaction--diffusion systems 
\cite{JCardy,MatGlas,TauHowLee}.
I will then demonstrate how the dynamic (perturbative) RG can be employed to
derive the asymptotic {\em scaling laws} in stochastic dynamical systems; to 
infer the {\em upper critical dimension} $d_c$ (for dimensions $d \leq d_c$, 
fluctuations strongly affect the universal scaling properties); and to 
systematically compute the {\em critical exponents} as well as to determine 
further universal properties in various intriguing dynamical model systems 
both near and far from equilibrium.
(For considerably more details, especially on the more technical aspects, the
reader is referred to Ref.~\cite{UCT}.)

\subsection{Continuous phase transitions and critical slowing down} 
\label{ssec:I1}
 
The vicinity of a {\em critical point} is characterised by {\em strong 
correlations} and {\em large fluctuations}.
The system under investigation is then behaving in a highly cooperative 
manner, and as a consequence, the standard approximative methods of
statistical mechanics, namely perturbation or cluster expansions that assume 
either weak interactions or short-range correlations, fail.
Upon approaching an equilibrium continuous (second-order) phase transition,
i.e., for $|\tau| \ll 1$, where $\tau = (T - T_c) / T_c$ measures the
deviation from the critical temperature $T_c$, the thermal fluctuations of the
{\em order parameter} $S(x)$ (which characterises the different thermodynamic 
phases, usually chosen such that the thermal average $\langle S \rangle = 0$ 
vanishes in the high-temperature `disordered' phase) are, in the 
thermodynamic limit, governed by a diverging length scale 
\begin{equation}
  \xi(\tau) \sim |\tau|^{- \nu} \ .
\label{I1nuex}
\end{equation} 
Here, we have defined the {\em correlation length} via the typically 
exponential decay of the static {\em cumulant} or {\em connected two-point 
correlation function} $C(\vec{x}) = \langle S(\vec{x}) \, S(0) \rangle - 
\langle S \rangle^2 \sim \E^{- |\vec{x}| / \xi}$, and $\nu$ denotes the 
correlation length {\em critical exponent}.
As $T \to T_c$, $\xi \to \infty$, which entails the absence of any 
characteristic length scale for the order parameter fluctuations at 
criticality.
Hence we expect the {\em critical correlations} to follow a power law 
$C(x) \sim |x|^{- (d - 2 + \eta)}$ in $d$ dimensions, which defines the 
{\em Fisher exponent} $\eta$.
The following {\em scaling ansatz} generalises this power law to 
$T \not= T_c$, but still in the vicinity of the critical point,
\begin{equation}
  C(\tau,\vec{x}) = |\vec{x}|^{- (d - 2 + \eta)} \, 
  {\widetilde C}_\pm(\vec{x} / \xi) \ ,
\label{I1cumx}
\end{equation}
with two distinct regular {\em scaling functions} ${\widetilde C}_+(\vec{y})$
for $T > T_c$ and ${\widetilde C}_-(\vec{y})$ for $T < T_c$, respectively. 
For its Fourier transform $C(\tau,\vec{q}) = \int \! \D^dx \, 
\E^{- \I \vec{q} \cdot \vec{x}} \, C(\tau,\vec{x})$, one obtains the 
corresponding scaling form
\begin{equation}
  C(\tau,\vec{q}) = |\vec{q}|^{- 2 + \eta} \, {\hat C}_\pm(\vec{q} \, \xi)
  \ ,
\label{I1cumq}
\end{equation}
with new scaling functions ${\hat C}_\pm(\vec{p}) = |\vec{p}|^{2 - \eta} \int 
\! \D^dy \, \E^{- \I \vec{p} \cdot \vec{y}} \, |\vec{y}|^{- (d - 2 + \eta)} \,
{\widetilde C}_\pm(\vec{y})$.

As we will see in Section~\ref{ssec:I5}, there are only {\em two} independent 
static critical exponents.
Consequently, it must be possible to use the static scaling hypothesis 
(\ref{I1cumx}) or (\ref{I1cumq}), along with the definition (\ref{I1nuex}), to
express the exponents describing the thermodynamic singularities near a
second-order phase transition in terms of $\nu$ and $\eta$ through 
{\em scaling laws}.  
For example, the order parameter in the low-temperature phase ($\tau < 0$) is 
expected to grow as $\langle S \rangle \sim (-\tau)^\beta$.
Let us consider Eq.~(\ref{I1cumx}) in the limit $|\vec{x}| \to \infty$.
In order for the $|\vec{x}|$ dependence to cancel, 
${\widetilde C}_\pm(\vec{y}) \propto |\vec{y}|^{d - 2 + \eta}$ for large 
$|\vec{y}|$, and therefore $C(\tau,|\vec{x}| \to \infty) \sim 
\xi^{- (d - 2 + \eta)} \sim |\tau|^{\nu (d - 2 + \eta)}$.
On the other hand, $C(\tau,|\vec{x}| \to \infty) \to 
- \langle S \rangle^2 \sim - (-\tau)^{2 \beta}$ for $T < T_c$; thus we 
identify the order parameter critical exponent through the {\em hyperscaling
relation}
\begin{equation}
  \beta = \frac{\nu}{2} \, (d - 2 + \eta) \ .
\label{I1opex}
\end{equation}
Let us next consider the isothermal static {\em susceptibility} $\chi_\tau$, 
which according to the equilibrium fluctuation--response theorem is given in 
terms of the spatial integral of the correlation function $C(\tau,\vec{x})$:
$\chi_\tau(\tau) = (k_{\rm B} T)^{-1} \lim_{\vec{q} \to 0} C(\tau,\vec{q})$.
But ${\hat C}_\pm(\vec{p}) \sim |\vec{p}|^{2 - \eta}$ as $\vec{p} \to 0$ to 
ensure nonsingular behaviour, whence 
$\chi_\tau(\tau) \sim \xi^{2 - \eta} \sim |\tau|^{- \nu (2 - \eta)}$, and
upon defining the associated thermodynamic critical exponent $\gamma$ via 
$\chi_\tau(\tau) \sim |\tau|^{- \gamma}$, we obtain the scaling relation
\begin{equation} 
  \gamma = \nu \, (2 - \eta) \ . 
\label{I1scex}
\end{equation}
  
The scaling laws (\ref{I1cumx}), (\ref{I1cumq}) as well as scaling relations
such as (\ref{I1opex}) and (\ref{I1scex}) can be put on solid foundations by
means of the RG procedure, based on an {\em effective} long-wavelength 
Hamiltonian ${\cal H}[S]$, a functional of $S(\vec{x})$, that captures the 
essential physics of the problem, namely the relevant symmetries in order 
parameter and real space, and the existence of a continuous phase transition.
The probability of finding a configuration $S(\vec{x})$ at given temperature
$T$ is then given by the canonical distribution
\begin{equation} 
  {\cal P}_{\rm eq}[S] \propto \exp\left( -{\cal H}[S] / k_{\rm B}T \right) 
  \ .
\label{I1eqpd}
\end{equation} 
For example, the mathematical description of the critical phenomena for an 
$O(n)$-symmetric order parameter field $S^\alpha(\vec{x})$, with vector index
$\alpha = 1,\ldots,n$, is based on the {\em Landau--Ginzburg--Wilson
functional} \cite{Ramond,Amit,ItzDro,Bellac,Zinn,Cardy}
\begin{eqnarray} 
  {\cal H}[S] &=& \int \! \D^dx \sum_\alpha \biggl[ \frac{r}{2} \, 
  [S^\alpha(\vec{x})]^2 + \frac12 \, [\nabla S^\alpha(\vec{x})]^2 \nonumber \\
  &&\qquad\qquad + \frac{u}{4 !} \, [S^\alpha(\vec{x})]^2 \sum_\beta 
  [S^\beta(\vec{x})]^2 - h^\alpha(\vec{x}) \, S^\alpha(\vec{x}) \biggr] \ , 
\label{I1glwh} 
\end{eqnarray} 
where $h^\alpha(\vec{x})$ is the external field thermodynamically conjugate to
$S^\alpha(\vec{x})$, $u > 0$ denotes the strength of the nonlinearity that 
drives the phase transformation, and $r$ is the control parameter for the 
transition, i.e., $r \propto T - T_c^0$, where $T_c^0$ is the (mean-field)
critical temperature.
Spatial variations of the order parameter are energetically suppressed by
the term $\sim [\nabla S^\alpha(\vec{x})]^2$, and the corresponding positive 
coefficient has been absorbed into the fields $S^\alpha$.

We shall, however, not pursue the static theory further here, but instead
proceed to a full {\em dynamical description} in terms of nonlinear Langevin 
equations \cite{HohHalp,Janssen}.
We will then formulate the RG within this dynamic framework, and therein 
demonstrate the emergence of scaling laws and the computation of critical 
exponents in a systematic perturbative expansion with respect to the deviation
$\epsilon = d - d_c$ from the upper critical dimension.

In order to construct the desired effective stochastic dynamics near a 
critical point, we recall that correlated region of size $\xi$ become quite 
large in the vicinity of the transition.
Since the associated relaxation times for such clusters should grow with their
extent, one would expect the characteristic time scale for the relaxation of
the order parameter fluctuations to increase as well as $T \to T_c$, namely
\begin{equation}
  t_c(\tau) \sim \xi(\tau)^z \sim |\tau|^{- z \nu} \ ,
\label{I1dyex}
\end{equation}
which introduces the {\em dynamic critical exponent} $z$ that encodes the
{\em critical slowing down} at the phase transition; usually $z \geq 1$.
Since the typical relaxation rates therefore scale as 
$\omega_c(\tau) = 1 / t_c(\tau) \sim |\tau|^{z \nu}$, we may utilise the 
static scaling variable $\vec{p} = \vec{q} \, \xi$ to generalise the crucial 
observation (\ref{I1dyex}) and formulate a {\em dynamic scaling hypothesis}
for the wavevector-dependent dispersion relation of the order parameter 
fluctuations \cite{Ferretal,HalpHoh},
\begin{equation} 
  \omega_c(\tau,\vec{q}) = |\vec{q}|^z \, {\hat \omega}_\pm(\vec{q} \, \xi) 
  \ .
\label{I1dysc}
\end{equation}

We can then proceed to write down dynamical scaling laws by simply postulating
the additional scaling variables $s = t / t_c(\tau)$ or 
$\omega / \omega_c(\tau,\vec{q})$.
For example, as an immediate consequence we find for the time-dependent mean 
order parameter
\begin{equation}  
  \langle S(\tau,t) \rangle = |\tau|^\beta \, {\hat S}(t / t_c) \ , 
\label{I1opsc}
\end{equation}
with ${\hat S}(s \to \infty) = {\rm const.}$, but 
${\hat S}(s) \sim s^{- \beta / z \nu}$ as $s \to 0$ in order for the 
$\tau$ dependence to disappear.
At the critical point ($\tau = 0$), this yields the power-law decay 
$\langle S(t) \rangle \sim t^{- \alpha}$, with
\begin{equation}
  \alpha = \frac{\beta}{z \, \nu} = \frac{1}{2 \, z} \, (d - 2 + \eta) \ .
\label{I1opcd}
\end{equation}
Similarly, the scaling law for the {\em dynamic order parameter susceptibility
(response function)} becomes
\begin{equation}
  \chi(\tau,\vec{q},\omega) = |\vec{q}|^{- 2 + \eta} \,  
  {\hat \chi}_\pm(\vec{q} \, \xi , \omega \, \xi^z) \ ,
\label{I1susc}
\end{equation}
which constitutes the dynamical generalisation of Eq.~(\ref{I1cumq}), for
$\chi(\tau,\vec{q},0) = (k_{\rm B} T)^{-1} C(\tau,\vec{q})$.
Upon applying the {\em fluctuation--dissipation theorem}, valid in thermal
equilibrium, we therefrom obtain the {\em dynamic correlation function}
\begin{equation} 
  C(\tau,\vec{q},\omega) = \frac{2 k_{\rm B}T}{\omega} \ {\rm Im} \,  
  \chi(\tau,\vec{q},\omega) = |\vec{q}|^{- z - 2 + \eta} \,   
  {\hat C}_\pm\left( \vec{q} \, \xi , \omega \, \xi^z \right) \ , 
\label{I1corw}
\end{equation}
and for its Fourier transform in real space and time, 
\begin{equation}  
  C(\tau,\vec{x},t) = \int \! \frac{\D^dq}{(2 \pi)^d} \! \int \! 
  \frac{\D\omega}{2 \pi} \, \E^{\I (\vec{q} \cdot \vec{x} - \omega t)} \, 
  C(\tau,\vec{q},\omega) = |\vec{x}|^{-(d - 2 + \eta)} \,  
  {\widetilde C}_\pm\left( \vec{x} / \xi , t / \xi^z \right) \ , 
\label{I1cort}
\end{equation} 
which reduces to the static limit (\ref{I1cumx}) if we set $t = 0$.

The critical slowing down of the order parameter fluctuations near the
critical point provides us with a natural {\em separation of time scales}. 
Assuming (for now) that there are no other conserved variables in the system, 
which would constitute additional slow modes, we may thus resort to a 
{\em coarse-grained} long-wavelength and long-time description, focusing 
merely on the order parameter kinetics, while subsuming all other `fast' 
degrees of freedom in {\em random `noise'} terms. 
This leads us to a {\em mesoscopic Langevin equation} for the {\em slow} 
variables $S^\alpha(\vec{x},t)$ of the form 
\begin{equation}
  \frac{\partial S^\alpha(\vec{x},t)}{\partial t} = F^\alpha[S](\vec{x},t)  
  + \zeta^\alpha(\vec{x},t) \ .
\label{I1land}
\end{equation}
In the simplest case, the {\em systematic} force terms here just represent 
purely {\em relaxational} dynamics towards the equilibrium configuration
\cite{HaHoMa}, 
\begin{equation}
  F^\alpha[S](\vec{x},t) 
  = - D \, \frac{\delta {\cal H}[S]}{\delta S^\alpha(x,t)} \ ,
\label{I1reld}
\end{equation}
where $D$ represents the relaxation coefficient, and ${\cal H}[S]$ is again 
the effective Hamiltonian that governs the phase transition, e.g. given by
Eq.~(\ref{I1glwh}).
For the {\em stochastic forces} we may assume the most convenient form, and 
take them to simply represent Gaussian white noise with zero mean, 
$\left\langle \zeta^\alpha(\vec{x},t) \right\rangle = 0$, but with their 
second moment in thermal equilibrium fixed by {\em Einstein's relation}
\begin{equation}   
  \left\langle \zeta^\alpha(\vec{x},t) \, \zeta^\beta(\vec{x}',t') 
  \right\rangle = 2 k_{\rm B} T \, D \, \delta(\vec{x}-\vec{x}') \,
  \delta(t-t') \, \delta^{\alpha \beta} \ .
\label{I1eins} 
\end{equation} 
As can be verified by means of the associated Fokker--Planck equation for the 
time-dependent probability distribution ${\cal P}[S,t]$, Eq.~(\ref{I1eins})
guarantees that eventually 
${\cal P}[S,t \to \infty] \to {\cal P}_{\rm eq}[S]$, the canonical 
distribution (\ref{I1eqpd}).
The stochastic differential equation (\ref{I1land}), with (\ref{I1reld}), the 
Hamiltonian (\ref{I1glwh}), and the noise correlator (\ref{I1eins}), define 
the {\em relaxational model A} (according to the classification in 
Ref.~\cite{HohHalp}) for a nonconserved $O(n)$-symmetric order parameter.

If, however, the order parameter is {\em conserved}, we have to consider the
associated continuity equation 
$\partial_t \, S^\alpha + \vec{\nabla} \cdot \vec{J}^\alpha  = 0$, where
typically the conserved current is given by a gradient of the field 
$S^\alpha$: $\vec{J}^\alpha = - D \, \vec{\nabla} S^\alpha + \ldots$; as a 
consequence, the order parameter fluctuations will relax {\em diffusively} 
with diffusion coefficient $D$.
The ensuing {\em model B} \cite{HaHoMa,HohHalp} for the relaxational critical 
dynamics of a conserved order parameter can be obtained by replacing 
$D \to - D \, \vec{\nabla}^2$ in Eqs.~(\ref{I1reld}) and (\ref{I1eins}).
In fact, we will henceforth treat both models A and B simultaneously by 
setting $D \to D \, (\I \vec{\nabla})^a$, where $a = 0$ and $a = 2$ 
respectively represent the nonconserved and conserved cases.
Explicitly, we thus obtain
\begin{eqnarray} 
  \frac{\partial S^\alpha(\vec{x},t)}{\partial t} &=& - D \, 
  (\I \vec{\nabla})^a \, \frac{\delta {\cal H}[S]}{\delta S^\alpha(\vec{x},t)}
  + \zeta^\alpha(\vec{x},t) \nonumber \\
  &=& - D \, (\I \vec{\nabla})^a \Bigl[ r - \vec{\nabla}^2 + \frac{u}{6} 
  \sum_\beta [S^\beta(\vec{x})]^2 \Bigr] S^\alpha(\vec{x},t) \nonumber \\  
  &&+ D \, (\I \vec{\nabla})^a \, h^\alpha(\vec{x},t) + 
  \zeta^\alpha(\vec{x},t) \ ,
\label{I1lane}
\end{eqnarray}
with
\begin{equation}
  \left\langle \zeta^\alpha(\vec{x},t) \, \zeta^\beta(\vec{x}',t') 
  \right\rangle = 2 k_{\rm B} T \, D \, (\I \vec{\nabla})^a \,
  \delta(\vec{x}-\vec{x}') \, \delta(t-t') \, \delta^{\alpha \beta} \ .
\label{I1ncor}
\end{equation}
Notice already that the presence or absence of a conservation law for the 
order parameter implies different dynamics for systems described by identical
static behaviour.
Before proceeding with the analysis of the relaxational models, we remark that
in general there may exist additional {\em reversible} contributions to the
systematic forces $F^\alpha[S]$, see Sec.~\ref{ssec:I6}, and / or dynamical
mode-couplings to additional conserved, slow fields, which effect further 
splitting into several distinct {\em dynamic universality classes}
\cite{Cardy,HohHalp,UCT}.  
 
Let us now evaluate the dynamic response and correlation functions in the
{\em Gaussian} (mean-field) approximation in the high-temperature phase.
To this end, we set $u = 0$ and thus discard the nonlinear terms in the 
Hamiltonian (\ref{I1glwh}) as well as in Eq.~(\ref{I1lane}).
The ensuing Langevin equation becomes linear in the fields $S^\alpha$, and is 
therefore readily solved by means of Fourier transforms.
Straightforward algebra and regrouping some terms yields 
\begin{equation} 
  \left[ - \I \omega + D \vec{q}^a \left( r + \vec{q}^2 \right) \right] 
  S^\alpha(\vec{q},\omega) = D \vec{q}^a \, h^\alpha(\vec{q},\omega) 
  + \zeta^\alpha(\vec{q},\omega) \ . 
\label{I1gaud}
\end{equation}
With $\left\langle \zeta^\alpha(\vec{q},\omega) \right\rangle = 0$, this gives
immediately
\begin{equation}
  \chi_0^{\alpha\beta}(\vec{q},\omega) = \frac{\partial \langle 
  S^\alpha(\vec{q},\omega) \rangle}{\partial h^\beta(\vec{q},\omega)} 
  \bigg\vert_{h = 0} = D \vec{q}^a \, G_0(\vec{q},\omega) \, 
  \delta^{\alpha\beta} \ ,
\label{I1gsus}
\end{equation}  
with the {\em response propagator}
\begin{equation}
  G_0(\vec{q},\omega) = 
  \left[ - \I \omega + D \vec{q}^a \, (r + \vec{q}^2) \right]^{-1} \ . 
\label{I1resp}
\end{equation}
As is readily established by means of the residue theorem, its Fourier 
backtransform in time obeys {\em causality}, 
\begin{equation}
  G_0(\vec{q},t) = \Theta(t) \, \E^{- D \vec{q}^a \, (r + \vec{q}^2) \, t} \ .
\label{I1rest}
\end{equation} 
Setting $h^\alpha = 0$, and with the noise correlator (\ref{I1ncor}) in 
Fourier space 
\begin{equation}
  \left\langle \zeta^\alpha(\vec{q},\omega) \, \zeta^\beta(\vec{q}',\omega')
  \right\rangle = 2 k_{\rm B} T \, D \vec{q}^a \, (2 \pi)^{d+1} 
  \delta(\vec{q}+\vec{q}') \, \delta(\omega+\omega') \, 
  \delta^{\alpha\beta} \ ,
\label{I1ncof} 
\end{equation}
we obtain the Gaussian dynamic correlation function 
$\left\langle S^\alpha(\vec{q},\omega) \, S^\beta(\vec{q}',\omega') 
\right\rangle_0 = C_0(\vec{q},\omega) \, (2 \pi)^{d+1} 
\delta(\vec{q}+\vec{q}') \, \delta(\omega+\omega')$, where 
\begin{equation}
  C_0(\vec{q},\omega) = \frac{2 k_{\rm B}T D \vec{q}^a}{\omega^2 +  
  [D \vec{q}^a (r + \vec{q}^2)]^2} = 2 k_{\rm B}T \, D \vec{q}^a \, 
  |G_0(\vec{q},\omega)|^2 \ .
\label{I1gcor}
\end{equation}
The fluctuation--dissipation theorem (\ref{I1corw}) is of course satisfied; 
moreover, as function of wavevector and time, 
\begin{equation}
  C_0(\vec{q},t) = \frac{k_{\rm B} T}{r + \vec{q}^2} \, 
  \E^{- D \vec{q}^a (r + \vec{q}^2) \, |t|} \ . 
\end{equation}
In the Gaussian approximation, away from criticality ($r > 0$, 
$\vec{q} \not= 0$) the temporal correlations for models A and B decay 
exponentially, with the relaxation rate 
$\omega_c(r,\vec{q}) = D \vec{q}^{2 + a} (1 + r / \vec{q}^2)$.
Upon comparison with the dynamic scaling hypothesis (\ref{I1dysc}), we infer 
the mean-field scaling exponents $\nu_0 = 1/2$ and $z_0 = 2 + a$.
At the critical point, a nonconserved order parameter relaxes diffusively 
($z_0 = 2$) in this approximation, whereas the conserved order parameter 
kinetics becomes even slower, namely subdiffusive with $z_0 = 4$.
Finally, invoking Eqs.~(\ref{I1susc}), (\ref{I1corw}), (\ref{I1cort}), or 
simply the static limit $C_0(\vec{q},0) = k_{\rm B} T / (r + \vec{q}^2)$, we 
find $\eta_0 = 0$ for the Gaussian model.  
 
The full nonlinear Langevin equation (\ref{I1lane}) cannot be solved exactly.
Yet a perturbation expansion with respect to the coupling $u$ may be set up in
a slightly cumbersome, but straightforward manner by direct iteration of the 
equations of motion \cite{HaHoMa,DDBrZJ}.
More elegantly, one may utilise a path-integral representation of the Langevin
stochastic process \cite{Jan,DeDom}, which allows the application of all the
standard tools from statistical and quantum field theory 
\cite{Ramond,Amit,ItzDro,Bellac,Zinn,Cardy}, and has the additional advantage 
of rendering symmetries in the problem more explicit 
\cite{Janssen,BaJaWa,UCT}.

\subsection{Field theory representation of Langevin equations} 
\label{ssec:I2} 

Our starting point is a set of coupled Langevin equations of the form 
(\ref{I1land}) for mesoscopic, coarse-grained stochastic variables 
$S^\alpha(\vec{x},t)$.
For the stochastic forces, we make the simplest possible assumption of 
{\em Gaussian white noise},
\begin{equation} 
  \left\langle \zeta^\alpha(\vec{x},t) \right\rangle = 0 \ , \quad
  \left\langle \zeta^\alpha(\vec{x},t) \, \zeta^\beta(\vec{x}',t') 
  \right\rangle = 2 L^\alpha \, \delta(\vec{x}-\vec{x}') \, \delta(t-t') \, 
  \delta^{\alpha \beta} \ ,
\label{I2noic}
\end{equation}
where $L^\alpha$ may represent a differential operator (such as the Laplacian
$\vec{\nabla}^2$ for conserved fields), and even a functional of $S^\alpha$. 
In the time interval $0 \leq t \leq t_f$, the moments (\ref{I2noic}) are
encoded in the probability distribution
\begin{equation} 
  {\cal W}[\zeta] \propto \exp \biggl[ - \frac14 \int \! \D^dx \! 
  \int_0^{t_f} \! \D t \sum_\alpha \zeta^\alpha(\vec{x},t) 
  \left[ (L^\alpha)^{-1} \zeta^\alpha(\vec{x},t) \right] \biggr] \ . 
\label{I2nprd}
\end{equation} 
If we now switch variables from the stochastic noise $\zeta^\alpha$ to the
fields $S^\alpha$ by means of the equations of motion (\ref{I1land}), we 
obtain
\begin{equation}
  {\cal W}[\zeta] \, {\cal D}[\zeta] = {\cal P}[S] \, {\cal D}[S] 
  \propto \E^{- {\cal G}[S]} \, {\cal D}[S] \ ,
\label{I2vart}
\end{equation} 
with the statistical weight determined by the {\em Onsager--Machlup 
functional}
\cite{BaJaWa} 
\begin{equation}
  {\cal G}[S] = \frac14 \int \! \D^dx \! \int \! \D t \sum_\alpha  
  \left( \frac{\partial S^\alpha}{\partial t} - F^\alpha[S] \right)  
  \left[ (L^\alpha)^{-1} \left( \frac{\partial S^\alpha}{\partial t}  
  - F^\alpha[S] \right) \right] \ .
\label{I2onma}
\end{equation}

Note that the Jacobian for the nonlinear variable transformation 
$\{ \zeta^\alpha \} \to \{ S^\alpha \}$ has been omitted here.
In fact, the above procedure is properly defined through appropriately 
discretising time. 
If a {\em forward} (It\^o) discretisation is applied, then indeed the 
associated functional determinant is a mere constant that can be absorbed in 
the functional measure.
The functional (\ref{I2onma}) already represents a desired field theory 
action.
Since the probability distribution for the stochastic forces should be 
normalised, $\int \! {\cal D}[\zeta] \, W[\zeta] = 1$, the associated 
`partition function' is unity, and carries no physical information (as opposed
to static statistical field theory, where it determines the free energy and 
hence the entire thermodynamics).
The Onsager--Machlup representation is however plagued by technical problems: 
Eq.~(\ref{I2onma}) contains $(L^\alpha)^{-1}$, which for conserved variables 
entails the inverse Laplacian operator, i.e., a Green function in real space 
or the singular factor $1/\vec{q}^2$ in Fourier space; moreover the 
nonlinearities in $F^\alpha[S]$ appear {\em quadratically}.
Hence it is desirable to linearise the action (\ref{I2onma}) by means of a 
Hubbard--Stratonovich transformation \cite{BaJaWa}. 
 
We shall follow an alternative, more general route that completely avoids the 
appearance of the inverse operators $(L^\alpha)^{-1}$ in intermediate steps.
Our goal is to average over noise `histories' for observables $A[S]$ that need
to be expressible in terms of the stochastic fields $S^\alpha$: 
$\langle A[S] \rangle_\zeta \propto \int \! {\cal D}[\zeta] \, A[S(\zeta)] \, 
W[\zeta]$.
For this purpose, we employ the identity 
\begin{eqnarray} 
  1 &=& \int \! {\cal D}[S] \prod_\alpha \prod_{(\vec{x},t)} \, 
  \delta\left( \frac{\partial S^\alpha(\vec{x},t)}{\partial t} - 
  F^\alpha[S](\vec{x},t) - \zeta^\alpha(\vec{x},t) \right) \nonumber \\ 
  &=& \int \! {\cal D}[\I {\widetilde S}] \! \int \! {\cal D}[S] \exp \biggl[ 
  - \int \! \D^dx \! \int \! \D t \sum_\alpha {\widetilde S}^\alpha \left( 
  \frac{\partial S^\alpha}{\partial t} - F^\alpha[S] - \zeta^\alpha \right) 
  \biggr] \ , \quad
\label{I2fcid}
\end{eqnarray}  
where the first line constitutes a rather involved representation of the unity
(in a somewhat symbolic notation; again proper discretisation should be 
invoked here), and the second line utilises the Fourier representation of the 
(functional) delta distribution by means of the purely imaginary auxiliary 
fields ${\widetilde S}$ (and factors $2 \pi$ have been absorbed in its 
functional measure).

Inserting (\ref{I2fcid}) and the probability distribution (\ref{I2nprd}) into 
the desired stochastic noise average, we arrive at  
\begin{eqnarray}
  \langle A[S] \rangle_\zeta &\propto& \int \! {\cal D}[\I {\widetilde S}] \! 
  \int \! {\cal D}[S] \exp \! \biggl[ - \! \int \! \D^dx \! \int \! \D t 
  \sum_\alpha {\widetilde S}^\alpha \left(\frac{\partial S^\alpha}{\partial t}
  - F^\alpha[S] \right) \biggr] \, A[S] \nonumber \\ 
  &\times& \int \! {\cal D}[\zeta] \exp \biggl( - \int \! \D^dx \!
  \int \! \D t \sum_\alpha \biggl[ \, \frac14 \, \zeta^\alpha (L^\alpha)^{-1}
  \zeta^\alpha - {\widetilde S}^\alpha \, \zeta^\alpha \biggr] \biggr) \ .
\label{I2noia}
\end{eqnarray} 
We may now evaluate the Gaussian integrals over the noise $\zeta^\alpha$, 
which yields 
\begin{equation}
  \langle A[S] \rangle_\zeta = \int \! {\cal D}[S] \, A[S] \, {\cal P}[S] \ , 
  \quad {\cal P}[S] \propto \int \! {\cal D}[\I {\widetilde S}] \,  
  \E^{- {\cal A}[{\widetilde S},S]} \ ,
\label{I2noip}
\end{equation} 
with the statistical weight now governed by the {\em Janssen--De~Dominicis 
`response' functional} \cite{Jan,DeDom,BaJaWa} 
\begin{equation}  
  {\cal A}[{\widetilde S},S] = \int \! \D^dx \! \int_0^{t_f} \! \D t 
  \sum_\alpha \left[ {\widetilde S}^\alpha \left( 
  \frac{\partial S^\alpha}{\partial t} - F^\alpha[S] \right) 
  - {\widetilde S}^\alpha \, L^\alpha \, {\widetilde S}^\alpha \right] \ . 
\label{I2jade} 
\end{equation} 
Once again, we have omitted the functional determinant from the variable 
change $\{ \zeta^\alpha \} \to \{ S^\alpha \}$, and normalisation implies 
$\int \! {\cal D}[\I {\widetilde S}] \int \! {\cal D}[S] \, 
\E^{- {\cal A}[{\widetilde S},S]} = 1$.
The first term in the action (\ref{I2jade}) encodes the temporal evolution
according to the systematic terms in the Langevin equations (\ref{I1land}),
whereas the second term specifies the noise correlations (\ref{I2noic}).
Since the auxiliary variables ${\widetilde S}^\alpha$, often termed 
Martin--Siggia--Rose response fields \cite{MaSiRo}, appear only quadratically
here, they may be eliminated via completing the squares and Gaussian 
integrations; thereby one recovers the Onsager--Machlup functional 
(\ref{I2onma}). 

The Janssen--De~Dominicis functional (\ref{I2jade}) takes the form of a
($d+1$)-dimensional statistical field theory with two independent sets of 
fields $S^\alpha$ and ${\widetilde S}^\alpha$. 
We may thus bring the established machinery of statistical and quantum field
theory \cite{Ramond,Amit,ItzDro,Bellac,Zinn,Cardy} to bear here; it should
however be noted that the response functional formalism for stochastic 
Langevin dynamics incorporates causality in a nontrivial manner, which leads 
to important distinctions \cite{Janssen}.

Let us specify the Janssen--De~Dominicis functional for the purely 
{\em relaxational models} A and B \cite{HaHoMa,DDBrZJ}, see 
Eqs.~(\ref{I1lane}) and (\ref{I1ncor}), splitting it into the Gaussian and 
anharmonic parts ${\cal A} = {\cal A}_0 + {\cal A}_{\rm int}$ \cite{BaJaWa}, 
which read
\begin{eqnarray} 
  {\cal A}_0[{\widetilde S},S] &=& \int \! \D^dx \! \int \! \D t \sum_\alpha  
  \biggl( {\widetilde S}^\alpha \left[ \frac{\partial}{\partial t} + D \, 
  (\I \vec{\nabla})^a \, (r - \vec{\nabla}^2) \right] S^\alpha \nonumber \\ 
  &&\qquad\qquad\qquad\quad - D \, {\widetilde S}^\alpha \, 
  (\I \vec{\nabla})^a \, {\widetilde S}^\alpha - D \, {\widetilde S}^\alpha \,
  (\I \vec{\nabla})^a \, h^\alpha \biggr) \ , 
\label{I2hajd} \\ 
  {\cal A}_{\rm int}[{\widetilde S},S] &=& D \, \frac{u}{6} \int \! \D^dx \! 
  \int \! \D t \sum_{\alpha,\beta} {\widetilde S}^\alpha \, 
  (\I \vec{\nabla})^a \, S^\alpha \, S^\beta \, S^\beta \ . 
\label{I2ndjd}
\end{eqnarray} 
Since we are interested in the vicinity of the critical point $T \approx T_c$,
we have absorbed the constant $k_{\rm B} T_c$ into the fields.
The prescription (\ref{I2noip}) tells us how to compute time-dependent 
correlation functions 
$\left \langle S^\alpha(\vec{x},t) \, S^\beta(\vec{x}',t') \right\rangle$.
Using Eq.~(\ref{I2hajd}), the dynamic order parameter {\em susceptibility} 
follows from
\begin{equation}
  \chi^{\alpha \beta}(\vec{x}-\vec{x}',t-t') = \frac{\delta \langle 
  S^\alpha(\vec{x},t) \rangle}{\delta h^\beta(\vec{x}',t')} \bigg\vert_{h=0} 
  = D \left\langle S^\alpha(\vec{x},t) \, (\I \vec{\nabla})^a \, 
  {\widetilde S}^\beta(\vec{x}',t') \right\rangle \ ; \quad
\label{I2abrf}
\end{equation}
for the simple relaxational models (only), the response function is just given
by a correlator that involves an auxiliary variable, which explains why the 
${\widetilde S}^\alpha$ are referred to as `response' fields.
In equilibrium, one may employ the Onsager--Machlup functional (\ref{I2onma}) 
to derive the {\em fluctuation--dissipation theorem} \cite{BaJaWa}
\begin{equation} 
  \chi^{\alpha \beta}(\vec{x}-\vec{x}',t-t') = \Theta(t-t') \, 
  \frac{\partial}{\partial t'} \left\langle S^\alpha(\vec{x},t) \, 
  S^\beta(\vec{x}',t') \right\rangle \ ,
\label{I2fldt}
\end{equation}
which is equivalent to Eq.~(\ref{I1corw}) in Fourier space.

In order to access arbitrary correlators, we define the {\em generating 
functional}
\begin{equation} 
  {\cal Z}[{\tilde j},j] = \biggl\langle \exp \int \! \D^dx \! \int \! \D t  
  \sum_\alpha \left( {\tilde j}^\alpha \, {\widetilde S}^\alpha + 
  j^\alpha \, S^\alpha \right) \biggr\rangle \ ,
\label{I2genf}
\end{equation}  
wherefrom the correlation functions follow via functional derivatives,
\begin{equation}
  \biggl\langle \prod_{ij} S^{\alpha_i} \, {\widetilde S}^{\alpha_j}  
  \biggr\rangle = \prod_{ij} \frac{\delta}{\delta j^{\alpha_i}} \,  
  \frac{\delta}{\delta {\tilde j}^{\alpha_j}} \, {\cal Z}[{\tilde j},j] 
  \bigg\vert_{{\tilde j}=0=j} \ ,
\label{I2corf}
\end{equation}
and the {\em cumulants} or {\em connected} correlation functions via
\begin{equation}
  \biggl\langle \prod_{ij} S^{\alpha_i} \, {\widetilde S}^{\alpha_j}  
  \biggr\rangle_c = \prod_{ij} \frac{\delta}{\delta j^{\alpha_i}} \,  
  \frac{\delta}{\delta {\tilde j}^{\alpha_j}} \, 
  \ln {\cal Z}[{\tilde j},j] \bigg\vert_{{\tilde j}=0=j} \ . 
\label{I2cumf}
\end{equation}
In the harmonic approximation, setting $u = 0$, ${\cal Z}[{\tilde j},j]$ can
be evaluated explicitly (most directly in Fourier space) by means of Gaussian 
integration \cite{BaJaWa,UCT}; one thereby recovers (with $k_{\rm B} T = 1$) 
the Gaussian response propagator (\ref{I1resp}) and two-point correlation 
function (\ref{I1gcor}).
Moreover, as a consequence of causality, 
$\left\langle {\widetilde S}^\alpha(\vec{q},\omega) \, 
{\widetilde S}^\beta(\vec{q}',\omega') \right\rangle_0 = 0$.

\subsection{Outline of dynamic perturbation theory} 
\label{ssec:I3} 

Since we cannot evaluate correlation functions with the nonlinear action
(\ref{I2ndjd}) exactly, we resort to a perturbational treatment, assuming, for
the time being, a small coupling strength $u$.
The {\em perturbation expansion} with respect to $u$ is constructed by 
rewriting the desired correlation functions in terms of averages with respect 
to the Gaussian action (\ref{I2hajd}), henceforth indicated with index '$0$',
and then expanding the exponential of $- {\cal A}_{\rm int}$,
\begin{eqnarray} 
  \biggl\langle \prod_{ij} S^{\alpha_i} \, {\widetilde S}^{\alpha_j} 
  \biggr\rangle &=& \frac{\left\langle \prod_{ij} S^{\alpha_i} \,   
  {\widetilde S}^{\alpha_j} \, \E^{-{\cal A}_{\rm int}[{\widetilde S},S]}  
  \right\rangle_0}{\left\langle e^{-{\cal A}_{\rm int}[{\widetilde S},S]}  
  \right\rangle_0} \nonumber \\ 
  &=& \biggl\langle \prod_{ij} S^{\alpha_i} \, {\widetilde S}^{\alpha_j} 
  \sum_{l=0}^\infty \frac{1}{l!} \left( -{\cal A}_{\rm int}[{\widetilde S},S] 
  \right)^l \biggr\rangle_0 \ .
\label{I3pexp}
\end{eqnarray}
The remaining Gaussian averages, a series of polynomials in the fields 
$S^\alpha$ and ${\widetilde S}^\alpha$, can be evaluated by means of 
{\em Wick's theorem}, here an immediate consequence of the Gaussian 
statistical weight, which states that all such averages can be written as a 
sum over all possible factorisations into Gaussian two-point functions 
$\bigl\langle S^\alpha \, {\widetilde S}^\beta \bigr\rangle_0$, i.e., 
essentially the response propagator $G_0$, Eq.~(\ref{I1resp}), and 
$\bigl\langle S^\alpha \, S^\beta \bigr\rangle_0$, the Gaussian correlation
function $C_0$, Eq.~(\ref{I1gcor}).
Recall that the denominator in Eq.~(\ref{I3pexp}) is exactly unity as a 
consequence of normalisation; alternatively, this result follows from 
causality in conjunction with our forward descretisation prescription, which 
implies that we should identify $\Theta(0) = 0$.
(We remark that had we chosen another temporal discretisation rule, any 
apparent contributions from the denominator would be precisely cancelled by 
the in this case nonvanishing functional Jacobian from the variable 
transformation $\{ \zeta^\alpha \} \to \{ S^\alpha \}$.)
At any rate, our stochastic field theory contains {\em no `vacuum'
contributions}. 

\begin{figure} 
\centering 
\includegraphics[width = 10 truecm]{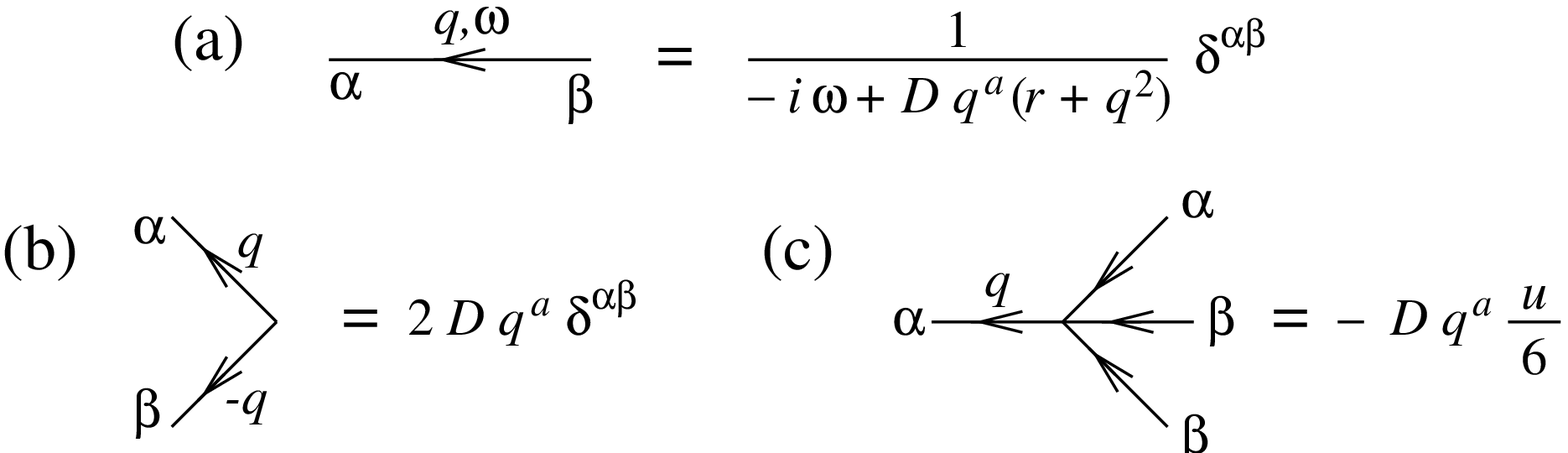} 
\caption{Elements of dynamic perturbation theory for the $O(n)$-symmetric 
  relaxational models: (a) response propagator; (b) noise vertex; 
  (c) anharmonic vertex.} 
\label{relgra} 
\end{figure} 
The many terms in the perturbation expansion (\ref{I3pexp}) are most lucidly
organised in a graphical representation, using {\em Feynman diagrams} with the
basic elements depicted in Fig.~\ref{relgra}.
We represent the response propagator (\ref{I1resp}) by a {\em directed line}
(here conventionally from right to left), which encodes its causal nature; the
noise by a two-point {\em `source' vertex}, and the anharmonic term in 
Eq.~(\ref{I2ndjd}) as a {\em four-point vertex}.
In the diagrams representing the different terms in the perturbation series, 
these vertices serve as links for the propagator lines, with the fields 
$S^\alpha$ being encoded as the `incoming', and the ${\widetilde S}^\alpha$ as
the `outgoing' components of the lines.
In Fourier space, translational invariance in space and time implies 
wavevector and frequency conservation at each vertex, see Fig.~\ref{temver} 
below.
An alternative, equivalent representation uses both the response and
correlation propagators as independent elements, the latter depicted as 
undirected line, thereby disposing of the noise vertex, and retaining the
nonlinearity in Fig.~\ref{relgra}(c) as sole vertex.

Following standard field theory procedures 
\cite{Ramond,Amit,ItzDro,Bellac,Zinn}, one establishes that the perturbation 
series for the {\em cumulants} (\ref{I2cumf}) is given in terms of 
{\em connected} Feynman graphs only (for a detailed exposition of this and the
following results, see Ref.~\cite{UCT}). 
An additional helpful reduction in the number of diagrams to be considered 
arises when one considers the {\em vertex functions}, which generalise the 
self-energy contributions $\Sigma(\vec{q},\omega)$ in the {\em Dyson equation}
for the response propagator, 
$G(\vec{q},\omega)^{-1} = D \vec{q}^a \, \chi(\vec{q},\omega)^{-1} 
= G_0(\vec{q},\omega)^{-1} - \Sigma(\vec{q},\omega)$.
To this end, we define the fields 
${\widetilde \Phi}^\alpha = \delta \ln {\cal Z} / \delta {\tilde j}^\alpha$ 
and  $\Phi^\alpha = \delta \ln {\cal Z} / \delta j^\alpha$, and introduce the 
new {\em generating functional}
\begin{equation} 
  \Gamma[{\widetilde \Phi},\Phi] = - \ln {\cal Z}[{\tilde j},j] + \int \!  
  \D^dx \! \int \! \D t \sum_\alpha \left( {\tilde j}^\alpha \,  
  {\widetilde \Phi}^\alpha + j^\alpha \, \Phi^\alpha \right) \ , 
\label{I3vgen}
\end{equation}
wherefrom the vertex functions are obtained via the functional derivatives
\begin{equation} 
  \Gamma^{({\widetilde N},N)}_{\{ \alpha_i \};\{ \alpha_j \}} = 
  \prod_i^{\widetilde N} \frac{\delta}{\delta {\widetilde \Phi}^{\alpha_i}} \,
  \prod_j^N \frac{\delta}{\delta \Phi^{\alpha_j}} \, 
  \Gamma[{\widetilde \Phi},\Phi] \bigg\vert_{{\tilde j}=0=j} \ .
\label{I3verf}
\end{equation} 
Diagrammatically, these quantities turn out to be represented by the possible
sets of {\em one-particle (1PI) irreducible Feynman graphs} with $N$ incoming 
and ${\widetilde N}$ outgoing `amputated' legs; i.e., these diagrams do not 
split into allowed subgraphs by simply cutting any single propagator line.
For example, for the two-point functions a direct calculation yields the 
relations
\begin{eqnarray} 
  \Gamma^{(1,1)}(\vec{q},\omega) 
  &=& D \vec{q}^a \, \chi(-\vec{q},-\omega)^{-1} 
  = G_0(-\vec{q},-\omega)^{-1} - \Sigma(-\vec{q},-\omega) \ , 
\label{I3gm11} \\
  \Gamma^{(2,0)}(\vec{q},\omega) &=& - \, \frac{C(\vec{q},\omega)}
  {|G(\vec{q},\omega)|^2} = - \frac{2 D \, \vec{q}^a}{\omega} \ 
  {\rm Im} \, \Gamma^{(1,1)}(\vec{q},\omega) \ ,
\label{I3gm20}   
\end{eqnarray}
where the second equation for $\Gamma^{(2,0)}$ follows from the 
fluctuation--dissipation theorem (\ref{I1corw}).
Note that $\Gamma^{(0,2)}(\vec{q},\omega) = 0$ vanishes because of causality.

\begin{figure}
\centering
\includegraphics[width = 7 truecm]{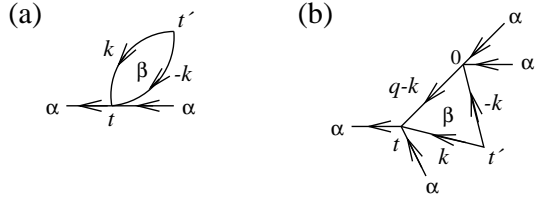} 
\caption{One-loop diagrams for (a) $\Gamma^{(1,1)}$ and (b) $\Gamma^{(1,3)}$ 
  in the time domain.} 
\label{temver} 
\end{figure} 
The perturbation series can then be organised graphically as an expansion in
successive orders with respect to the number of closed propagator {\em loops}.
As an example, Fig.~\ref{temver} depicts the one-loop contributions for the
vertex functions $\Gamma^{(1,1)}$ and $\Gamma^{(1,3)}$ in the time domain with
all required labels.
One may formulate general {\em Feynman rules} for the construction of the 
diagrams and their translation into mathematical expressions for the $l$th 
order contribution to the {\em vertex function} $\Gamma^{({\widetilde N},N)}$:

\begin{enumerate} 
\item Draw all topologically different, connected {\em one-particle 
  irreducible graphs} with ${\widetilde N}$ outgoing and $N$ incoming lines  
  connecting $l$ relaxation vertices $\propto u$.
  Do {\em not} allow closed response loops (since in the It\^o calculus
  $\Theta(0) = 0$). 
\item Attach wavevectors $\vec{q}_i$, frequencies $\omega_i$ or times $t_i$, 
  and component indices $\alpha_i$ to all directed lines, obeying `momentum 
  (and energy)' conservation at each vertex. 
\item Each directed line corresponds to a response propagator  
  $G_0(-\vec{q},-\omega)$ or $G_0(\vec{q},t_i-t_j)$ in the frequency and time
  domain, respectively, the two-point vertex to the noise strength 
  $2 D \, \vec{q}^a$, and the four-point relaxation vertex to 
  $- D \, \vec{q}^a \, u / 6$. 
  Closed loops imply integrals over the internal wavevectors and frequencies  
  or times, subject to causality constraints, as well as sums over the  
  internal vector indices. 
  Apply the residue theorem to evaluate frequency integrals. 
\item Multiply with $-1$ and the combinatorial factor counting all possible  
  ways of connecting the propagators, $l$ relaxation vertices, and $k$  
  two-point vertices leading to topologically identical graphs, including a  
  factor $1 / l! \, k!$ originating in the expansion of  
  $\exp(-{\cal A}_{\rm int}[{\widetilde S},S])$. 
\end{enumerate} 

\begin{figure}
\centering
\includegraphics[width = 11 truecm]{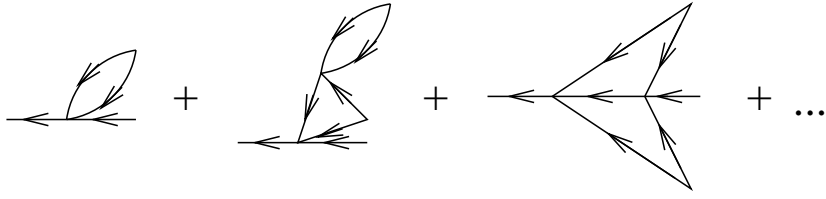} 
\caption{One-particle irreducible diagrams for 
  $\Gamma^{(1,1)}(\vec{q},\omega)$ to second order in $u$.} 
\label{renpro} 
\end{figure} 
For later use, we provide the explicit results for the two-point vertex 
functions to two-loop order.
After some algebra, the three diagrams in Fig.~\ref{renpro} give
\begin{eqnarray} 
  &&\Gamma^{(1,1)}(\vec{q},\omega) = \I \omega + D \vec{q}^a \biggl[ r + 
  \vec{q}^2 + \frac{n+2}{6} \, u \int_k \frac{1}{r+\vec{k}^2} \nonumber \\ 
  &&\qquad\qquad\qquad - \left( \frac{n+2}{6} \, u \right)^2 \int_k 
  \frac{1}{r+\vec{k}^2} \int_{k'} \frac{1}{(r+{\vec{k}'}^2)^2} \nonumber \\ 
  &&\qquad - \, \frac{n+2}{18} \, u^2 \int_k \frac{1}{r+\vec{k}^2} \int_{k'} 
  \frac{1}{r+{\vec{k}'}^2} \, \frac{1}{r+(\vec{q}-\vec{k}-\vec{k}')^2} 
  \nonumber \\ 
  &&\qquad\qquad\qquad \times \left( 1 - \frac{i \omega}{\I \omega  
  + \Delta(\vec{k}) + \Delta(\vec{k}') + \Delta(\vec{q}-\vec{k}-\vec{k}')} 
  \right) \biggr] \ ,  
\label{I3ga11}
\end{eqnarray} 
where we have separated out the dynamic part in the last line, and introduced 
the abbreviations $\Delta(\vec{q}) = D \vec{q}^a \, (r + \vec{q}^2)$ and
$\int_k = \int \! \D^dk / (2 \pi)^d$ \cite{UCT}.
\begin{figure}
\centering 
\includegraphics[width = 8 truecm]{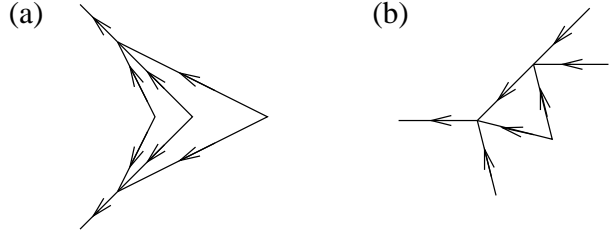} 
\caption{(a) Two-loop diagram for $\Gamma^{(2,0)}(\vec{q},\omega)$; 
  (b) one-loop graph for $\Gamma^{(1,3)}$.} 
\label{renver} 
\end{figure} 
For the noise vertex, Fig.~\ref{renver}(a) yields \cite{UCT}
\begin{eqnarray} 
  &&\Gamma^{(2,0)}(\vec{q},\omega) = - 2 D \vec{q}^a \biggl[ 1 + D \vec{q}^a 
  \, \frac{n+2}{18} \, u^2 \int_k \frac{1}{r+\vec{k}^2} \int_{k'} 
  \frac{1}{r+{\vec{k}'}^2} \nonumber \\ 
  &&\qquad \times \frac{1}{r+(\vec{q}-\vec{k}-\vec{k}')^2} \ {\rm Re} \, 
  \frac{1}{\I \omega + \Delta(\vec{k}) + \Delta(\vec{k}') + 
  \Delta(\vec{q}-\vec{k}-\vec{k}')} \biggr] \ ; \quad
\label{I3ga20}
\end{eqnarray}
notice that for model B, as a consequence of the conservation law for the 
order parameter and ensuing wavevector dependence of the nonlinear vertex, 
see Fig.~\ref{relgra}(c), to {\em all orders} in the perturbation expansion
\begin{equation}
  a = 2 \, : \ \Gamma^{(1,1)}(\vec{q}=0,\omega) = \I \omega \ , \quad 
  \frac{\partial}{\partial \vec{q}^2} \, \Gamma^{(2,0)}(\vec{q},\omega) 
  \bigg\vert_{\vec{q}=0} = - 2 D \ .  
\label{I3mbvf}
\end{equation}
At last, with the shorthand notation ${\underline k} = (\vec{q},\omega)$,  
the analytical expression corresponding to the graph in Fig.~\ref{renver}(b) 
for the four-point vertex function at symmetrically chosen external wavevector
labels reads
\begin{eqnarray}
  &&\Gamma^{(1,3)}(-3{\underline k}/2;\{ {\underline k}/2 \}) = D \left(  
  \frac{3}{2} \, \vec{q} \right)^a u \biggl[ 1 - \frac{n+8}{6} \, u \nonumber 
  \\ &&\qquad 
  \times \int_k \frac{1}{r+\vec{k}^2} \, \frac{1}{r+(\vec{q}-\vec{k})^2}
  \left( 1 - \frac{\I \omega}{\I \omega + \Delta(\vec{k}) 
  + \Delta(\vec{q}-\vec{k})} \right) \biggr] \ .
\label{I3ga13}
\end{eqnarray}

\subsection{Renormalisation}
\label{ssec:I4}

Consider a typical loop integral, say the correction in Eq.~(\ref{I3ga13}) to
the four-point vertex function $\Gamma^{(1,3)}$ at zero external frequency and
momentum, whose `bare' value, without any fluctuation contributions, is $u$.
In dimensions $d < 4$, one obtains, after introducing $d$-dimensional 
spherical coordinates and rendering the integrand dimensionless 
($x = |\vec{k}| / \sqrt{r}$):
\begin{equation}
  u \int \! \frac{\D^dk}{(2 \pi)^d} \, \frac{1}{(r + \vec{k}^2)^2} =   
  \frac{u \, r^{-2 + d/2}}{2^{d-1} \pi^{d/2} \Gamma(d/2)} \int_0^\infty \! 
  \frac{x^{d-1}}{(1 + x^2)^2} \, \D x \ , 
\label{I4flin}
\end{equation}
where we have inserted the surface area $S_d = 2 \pi^{d/2} / \Gamma(d/2)$ of 
the $d$-dimensional unit sphere, with Euler's Gamma function,
$\Gamma(1 + x) = x \, \Gamma(x)$.
Note that the integral on the right-hand side is finite.
Thus, we see that the {\em effective} expansion parameter in perturbation
theory is not just $u$, but the combinaton $u_{\rm eff} = u \, r^{(d-4)/2}$.
Far away from $T_c$, it is small, and the perturbation expansion well-defined.
However, $u_{\rm eff} \to \infty$ as $r \to 0$ for $d < 4$: we are facing  
{\em infrared (IR) divergences}, induced by the strong critical fluctuations 
that render the loop corrections singular.
A straightforward application of perturbation theory will therefore not 
provide meaningful results, and we must expect the fluctuation contributions 
to modify the critical power laws.

Conversely, for dimensions $d \geq 4$, the integral in (\ref{I4flin}) develops
{\em ultraviolet (UV) divergences} as the upper integral boundary is sent to 
infinity ($k = |\vec{k}|$),
\begin{equation}
  \int_0^\Lambda \! \frac{k^{d-1}}{(r + k^2)^2} \, \D k \sim \left\{  
  \begin{array}{cc} \ln (\Lambda^2 / r) & \quad d = 4 \\ \Lambda^{d-4} & \quad
  d > 4 \end{array} \right\} \to \infty \quad {\rm as} \ \Lambda \to \infty
  \ . 
\label{I4uvsi}
\end{equation} 
In lattice models, there is a finite wavevector cutoff, namely the Brillouin
zone boundary, $\Lambda \sim (2 \pi / a_0)^d$ for a hypercubic lattice with
lattice constant $a_0$, whence physically these UV problems do not emerge.
Yet we shall see that a formal treatment of these unphysical UV divergences 
will allow us to infer the correct power laws for the physical IR
singularities associated with the critical point. 
The borderline dimension that separates the IR and UV singular regimes is
referred to as {\em upper critical dimension} $d_c$; here $d_c = 4$.
Note that at $d_c$, UV and IR singularities are intimately connected and
appear in the form of {\em logarithmic divergences}, see Eq.~(\ref{I4uvsi}).
The situation is summarised in Table~\ref{iruvst}, where we have also stated
that models with continuous order parameter symmetry, such as the Hamiltonian
(\ref{I1glwh}) with $n \geq 2$, do not allow long-range order in dimensions
$d \leq d_{\rm lc} = 2$ ({\em Mermin--Wagner--Hohenberg} theorem
\cite{Wagner,MerWag,Hohenb}).
Here, $d_{\rm lc}$ is called the {\em lower critical dimension}; for the Ising
model represented by Eq.~(\ref{I1glwh}) with $n = 1$, of course 
$d_{\rm lc} = 1$. 
\begin{table}
\centering
\caption{Mathematical and physical distinctions of the regimes $d < d_c$,  
  $d = d_c$, and $d > d_c$, for the $O(n)$-symmetric models A and B   
  (or static $\Phi^4$ field theory).}
\label{iruvst} 
\begin{tabular}{llll}  
\hline \noalign{\smallskip}
dimension & perturbation & model A / B or & critical \\   
interval & series & $\Phi^4$ field theory & behaviour \\
\noalign{\smallskip}\hline\noalign{\smallskip}
$d \leq d_{lc} = 2$ & IR-singular & ill-defined & no long-range \\  
& UV-convergent & $u$ relevant & order ($n \geq 2$) \\
$2 < d < 4$ & IR-singular & super-renormalisable & nonclassical \\  
& UV-convergent & $u$ relevant & exponents \\
$d = d_c = 4$ & logarithmic IR-/ \ & renormalisable & logarithmic \\  
& UV-divergence & $u$ marginal & corrections \\  
$d > 4$ & IR-regular & nonrenormalisable & mean-field \\  
& UV-divergent & $u$ irrelevant & exponents \\ 
\noalign{\smallskip}\hline 
\end{tabular}  
\end{table} 

The upper critical dimension can be obtained in a more direct manner through
simple {\em power counting}.
To this end, we introduce an arbitrary momentum scale $\mu$, i.e., define the
{\em scaling dimensions} $[x] = \mu^{-1}$ and $[q] = \mu$.
If in addition we choose $[t] = \mu^{-2-a}$, or $[\omega] = \mu^{2+a}$, then
the relaxation constant becomes dimensionless, $[D] = \mu^0$.
For the deviation from the critical point, we obtain $[r] = \mu^2$, and the
{\em positive} exponent indicates that this control parameter constitutes a
{\em relevant} coupling in the theory; as we shall see below, its renormalised
counterpart grows under subsequent RG transformations.
For the nonlinear coupling, one finds $[u] = \mu^{4-d}$, so it is relevant
for $d < 4$: nonlinear thermal fluctuations will qualitatively affect the
physical properties at the phase transition; but $u$ becomes irrelevant for 
$d > 4$: one then expects mean-field (Gaussian) critical exponents.
At the upper critical dimension $d_c = 4$, the nonlinear coupling $u$ is
{\em marginally relevant}: this will induce logarithmic corrections to the
mean-field scaling laws, see Table~\ref{iruvst}.  

It is obviously not a simple task to treat the IR-singular perturbation
expansion in a meaningful, well-defined manner, and thus allow nonanalytic 
modifications of the critical power laws (note that mean-field scaling is
completely determined by dimensional analysis or power counting).
The key of the success of the RG approach is to focus on the very specific
symmetry that emerges near critical points, namely {\em scale invariance}.
There are several (largely equivalent) versions of the RG method; we shall 
here formulate and employ the field-theoretic variant 
\cite{Ramond,Amit,ItzDro,Bellac,Zinn,Cardy,Janssen,UCT}.   
In order to proceed, it is convenient to evaluate the loop integrals in 
momentum space by means of {\em dimensional regularisation}, whereby one 
assigns finite values even to UV-divergent expressions, namely the 
analytically continued values from the UV-finite range.
For example, even for noninteger dimensions $d$ and $\sigma$, we set
\begin{equation} 
  \int \! \frac{\D^dk}{(2 \pi)^d} \, 
  \frac{\vec{k}^{2 \sigma}}{\left( \tau + \vec{k}^2 \right)^s} = 
  \frac{\Gamma(\sigma + d/2) \, \Gamma(s - \sigma - d/2)}{2^d \, \pi^{d/2} \,
  \Gamma(d/2) \, \Gamma(s)} \ \tau^{\sigma - s + d/2} \ .  
\label{I4dimr} 
\end{equation} 
The {\em renormalisation} program then consists of the following steps: 
\begin{enumerate}
\item We aim to carefully keep track of formal, unphysical UV divergences.   
  In dimensionally regularised integrals (\ref{I4dimr}), these appear as 
  poles in $\epsilon = d_c - d$; their residues characterise the asymptotic 
  UV behaviour of the field theory under consideration.
\item Therefrom we may infer the (UV) scaling properties of the control 
  parameters of the model under a RG transformation, namely essentially a 
  change of the momentum scale $\mu$, while keeping the form of the action
  invariant.
  This will allow us to define suitable {\em running couplings}. 
\item We seek {\em fixed points} in parameter space where certain marginal 
  couplings ($u$ here) do not change anymore under RG transformations.
  This describes a {\em scale-invariant} regime for the model under
  consideration, where the UV and IR scaling properties become intimately 
  linked. 
  Studying the parameter flows near a stable RG fixed point then allows us
  to extract the asymptotic IR power laws.  
\end{enumerate}

As a preliminary step, we need to take into account that the fluctuations will
also shift the critical point downwards from the mean-field phase transition
temperature $T_c^0$; i.e., we expect the transition to occur at $T_c < T_c^0$.
This fluctuation-induced $T_c$ shift can be determined by demanding that the
inverse static susceptibility vanish at $T_c$: 
$\chi(\vec{q} = 0,\omega = 0)^{-1} = \tau = r - r_c$, where 
$\tau \sim T - T_c$ and thus $r_c = T_c - T_c^0$.
Using our previous results (\ref{I3gm11}) and (\ref{I3ga11}), we find to first
order in $u$ (and with finite cutoff $\Lambda$), 
\begin{equation}
  r_c = - \frac{n+2}{6}\, u \int_k^\Lambda \frac{1}{r_c + \vec{k}^2} + O(u^2) 
  = - \frac{n+2}{6} \, \frac{u \, S_d \, \Lambda^{d-2}}{(2 \pi)^d \, (d-2)} 
  + O(u^2) \ .
\label{I4cpsh}
\end{equation}
Notice that this quantity depends on microscopic details (the lattice 
structure enters the cutoff $\Lambda$) and is thus not universal; moreover
it diverges for $d \geq 2$ (quadratically near $d_c = 4$) as 
$\Lambda \to \infty$.
We next use $r = \tau + r_c$ to write physical quantities as functions of the
true distance $\tau$ from the critical point, which technically amounts to an
{\em additive renormalisation}; e.g., the dynamic response function becomes
to one-loop order
\begin{equation}   
  \chi(\vec{q},\omega)^{-1} = - \frac{i \omega}{D \vec{q}^a} + \vec{q}^2 + 
  \tau \left[ 1 - \frac{n+2}{6} \, u \int_k 
  \frac{1}{\vec{k}^2 (\tau + \vec{k}^2)} \right] + O(u^2) \ .  
\label{I4dsus}
\end{equation}
The remaining loop integral is UV-singular in dimensions $d \geq d_c = 4$.

We may now formally absorb the remaining UV divergences into 
{\em renormalised} fields and parameters, a procedure called 
{\em multiplicative renormalisation}.
For the renormalised fields, we use the convention
\begin{equation} 
  S_R^\alpha = Z_S^{1/2} \, S^\alpha \ , \quad
  {\widetilde S}_R^\alpha = Z_{\widetilde S}^{1/2} {\widetilde S}^\alpha \ ,
\label{I4fren}
\end{equation}
where we have exploited the $O(n)$ rotational symmetry in using identical
{\em renormalisation constants ($Z$ factors)} for each component. 
The renormalised cumulants with $N$ order parameter fields $S^\alpha$ and 
${\widetilde N}$ response fields ${\widetilde S}^\alpha$ naturally involve
the product $Z_S^{N/2} \, Z_{\widetilde S}^{{\widetilde N}/2}$, whence
\begin{equation}
  \Gamma_R^{({\widetilde N},N)} = Z_{\widetilde S}^{-{\widetilde N}/2} \,
  Z_S^{-N/2} \, \Gamma^{({\widetilde N},N)} \ .
\label{I4gren}
\end{equation}
In a similar manner, we relate the `bare' parameters of the theory via $Z$
factors to their renormalised counterparts, which we furthermore render
dimensionless through appropriate momentum scale factors,
\begin{equation}
  D_R = Z_D \, D \ , \quad \tau_R = Z_\tau \, \tau \, \mu^{-2} \ , \quad  
  u_R = Z_u \, u \, A_d \, \mu^{d-4} \ ,
\label{I4pren}
\end{equation}
where we have separated out the factor 
$A_d = \Gamma(3 - d/2) / 2^{d-1} \, \pi^{d/2}$ for convenience.
In the {\em minimal subtraction} scheme, the $Z$ factors contain {\em only}
the UV-singular terms, which in dimensional regularisation appear as poles at
$\epsilon = 0$, and their residues, evaluated at $d = d_c$.

These renormalisation constants are not all independent, however; since the
equilibrium fluctuation--dissipation theorem (\ref{I2fldt}) or (\ref{I1corw})
must hold in the renormalised theory as well, we infer that necessarily
\begin{equation}
  Z_D = \left( Z_S / Z_{\widetilde S} \right)^{1/2} \ ,
\label{I4zrel}
\end{equation}
and consequently from Eq.~(\ref{I3gm11})
\begin{equation}
  \chi_R = Z_S \, \chi \ .
\label{I4cren}
\end{equation}
Moreover, for model B with conserved order parameter Eq.~(\ref{I3mbvf})
implies that to all orders in the perturbation expansion
\begin{equation}
  a = 2 \, : \ Z_{\widetilde S} \, Z_S = 1 \ , \quad Z_D = Z_S \ . 
\label{I4brel}
\end{equation}
For the following, it is crucial that the theory is {\em renormalisable}, 
i.e., a {\em finite} number of reparametrisations suffice to formally rid it
of all UV divergences.
Indeed, for the relaxational models A and B, and the static 
Ginzburg--Landau--Wilson Hamiltonian (\ref{I1glwh}), all higher vertex 
function beyond the four-point function are UV-convergent near $d_c$, and
there are only the {\em three} independent static renormalisation factors 
$Z_S$, $Z_\tau$, and $Z_u$, and in addition $Z_D$ for nonconserved order 
parameter dynamics.
As we shall see, these directly translate into the {\em two} independent
static critical exponents and the unrelated dynamic scaling exponent $z$ for
model A; for model B with conserved order parameter, Eq.~(\ref{I4brel}) will 
yield a scaling relation between $z$ and $\eta$. 

In order to explicitly determine the renormalisation constants, we need to
ensure that we stay away from the IR-singular regime.
This is guaranteed by selecting as {\em normalisation point} either
$\tau_R = 1$ (i.e., $Z_\tau \, \tau = \mu^2$) or $q = \mu$.
Inevitably therefore, the renormalised theory depends on the corresponding
arbitrary momentum scale $\mu$.  
Since there are no fluctuation contributions to order $u$ to either
$\partial \Gamma^{(1,1)}(\vec{q},0) / \partial \vec{q}^2$ or
$\partial \Gamma^{(1,1)}(0,\omega) / \partial \omega$ (at $\tau_R = 1$), we
find $Z_S = 1$ and $Z_D = 1$ within the one-loop approximation.
Expressions (\ref{I4dsus}) and (\ref{I3ga13}) then yield with the formula
(\ref{I4dimr})
\begin{equation}
  Z_\tau = 1 - \frac{n + 2}{6} \, \frac{u_R}{\epsilon} \ , \quad  
  Z_u = 1 - \frac{n + 8}{6} \, \frac{u_R}{\epsilon} \ . 
\label{I41lpz}
\end{equation}
To two-loop order, we may infer the field renormalisation $Z_S$ from the 
static susceptibility as the singular contributions to 
$\partial \chi_R(\vec{q},0) / \partial \vec{q}^2 |_{\vec{q}=0}$,
and $Z_D$ for model A, through a somewhat lengthy calculation \cite{UCT},
from either $\Gamma^{(2,0)}_R(0,0)$ or $\Gamma^{(1,1)}_R(0,\omega)$, with the
results  
\begin{equation}
  Z_S = 1 + \frac{n + 2}{144} \, \frac{u_R^2}{\epsilon} \ , \quad  
  a = 0 \, : \ Z_D = 1 - \frac{n + 2}{144} 
  \Bigl( 6 \, \ln \, \frac43 - 1 \Bigr) \frac{u_R^2}{\epsilon} \ .
\label{I42lpz}
\end{equation}

\subsection{Scaling laws and critical exponents} 
\label{ssec:I5}

We now wish to related the renormalised vertex functions at different 
inverse length scales $\mu$.
This is accomplished by simply recalling that the {\em unrenormalised}
vertex functions obviously do {\em not} depend on $\mu$,
\begin{equation}
  0 = \mu \, \frac{\D}{\D \mu} \Gamma^{({\widetilde N},N)}(D,\tau,u) = 
  \mu \, \frac{\D}{\D \mu} \left[ Z_{\widetilde S}^{{\widetilde N}/2} \,
  Z_S^{N/2} \, \Gamma_R^{({\widetilde N},N)}\left( \mu,D_R,\tau_R,u_R \right) 
  \right] \ .
\label{I5unrm}
\end{equation}
In the second step, the bare quantities have been replaced with their
renormalised counterparts.
The innocuous statement (\ref{I5unrm}) then implies a very nontrivial 
partial differential equation for the renormalised vertex functions, the 
desired {\em renormalisation group equation},
\begin{eqnarray} 
  &&\biggl[ \mu \, \frac{\partial}{\partial \mu} +  
  \frac{{\widetilde N} \, \gamma_{\widetilde S} + N \, \gamma_S}{2}    
  + \gamma_D \, D_R \, \frac{\partial}{\partial D_R} 
  + \gamma_\tau \, \tau_R \, \frac{\partial}{\partial \tau_R} 
  + \beta_u \, \frac{\partial}{\partial u_R} \biggr] \nonumber \\ 
  &&\qquad \times \
  \Gamma_R^{({\widetilde N},N)}\left( \mu,D_R,\tau_R,u_R \right) = 0 \ . 
\label{I5rgeq}
\end{eqnarray}
Here we have defined {\em Wilson's flow functions} (the index `0' indicates 
that the derivatives with respect to $\mu$ are to be taken with fixed 
unrenormalised parameters)
\begin{eqnarray}
  \gamma_{\widetilde S} &=& \mu \, \frac{\partial}{\partial \mu} \bigg\vert_0
  \, \ln Z_{\widetilde S} \ , \quad
  \gamma_S = \mu \, \frac{\partial}{\partial \mu} \bigg\vert_0 \, \ln Z_S \ ,
\label{I5wfss} \\
  \gamma_\tau &=& \mu \, \frac{\partial}{\partial \mu} \bigg\vert_0 \, 
  \ln (\tau_R / \tau) = - 2 + \mu \, \frac{\partial}{\partial \mu} 
  \bigg\vert_0 \ln Z_\tau \ ,
\label{I5wfta} \\
  \gamma_D &=& \mu \, \frac{\partial}{\partial \mu} \bigg\vert_0 \, 
  \ln (D_R / D) = \frac12 \left( \gamma_S - \gamma_{\widetilde S} \right) \ , 
\label{I5wfrd}
\end{eqnarray}
where we have used the relation (\ref{I4zrel}); for model B, 
Eq.~(\ref{I4brel}) gives in addition
\begin{equation}
  \gamma_D = \gamma_S = - \gamma_{\widetilde S} \ . 
\label{I45wfr}
\end{equation} 
We have also introduced the {\em RG beta function} for the nonlinear
coupling $u$,
\begin{equation}
  \beta_u = \mu \, \, \frac{\partial}{\partial \mu} \bigg\vert_0 u_R 
  = u_R \left( d - 4 + \mu \, \, \frac{\partial}{\partial \mu} \bigg\vert_0
  \ln Z_u \right) \ .
\label{I5beta} 
\end{equation}
Explicitly, Eqs.~(\ref{I42lpz}) and (\ref{I41lpz}) yield to lowest nontrivial 
order, with $\epsilon = 4 - d$, 
\begin{eqnarray} 
  \gamma_S &=& - \frac{n + 2}{72} \, u_R^2 + O(u_R^3) \ ,  
\label{I5gams} \\
  a = 0 \, : \ \gamma_D &=& \frac{n + 2}{72}  
  \left( 6 \ln \frac{4}{3} - 1 \right) u_R^2 + O(u_R^3) \ ,  
\label{I5gamd} \\
  \gamma_\tau &=& - 2 + \frac{n + 2}{6} \, u_R + O(u_R^2) \ , 
\label{I5gamt} \\
 \beta_u &=& u_R \left[ - \epsilon + \frac{n + 8}{6} \, u_R + O(u_R^2) \right]
 \ .
\label{I5betu}
\end{eqnarray} 

In the RG equation for the renormalised dynamic susceptibility, 
Eq.~(\ref{I4cren}) tells us that the second term in Eq.~(\ref{I5rgeq}) is to 
be replaced with $- \gamma_S$.
Its explicit dependence on the scale $\mu$ can be factored out via
$\chi_R(\mu,D_R,\tau_R,u_R,\vec{q},\omega) = \mu^{-2} \, {\hat \chi}_R\left( 
\tau_R, u_R, \vec{q}/\mu, \omega/D_R \, \mu^{2+a} \right)$, see 
Eq.~(\ref{I4dsus}), whence
\begin{equation} 
  \biggl[ - 2 - \gamma_S + \gamma_D \, D_R \, \frac{\partial}{\partial D_R} 
  + \gamma_\tau \, \tau_R \, \frac{\partial}{\partial \tau_R} 
  + \beta_u \, \frac{\partial}{\partial u_R} \biggr] 
  {\hat \chi}_R(D_R,\tau_R,u_R) = 0 \ . 
\label{I5rgsu}
\end{equation}
This {\em linear partial differential equation} is readily solved by means of 
the {\em method of characteristics}, as is Eq.~(\ref{I5rgeq}) for the vertex
functions.
The idea is to find a curve parametrisation $\mu(\ell) = \mu \, \ell$ in the
space spanned by the parameters ${\widetilde D}$, ${\widetilde \tau}$, and 
${\widetilde u}$ such that
\begin{equation}  
  \ell \, \frac{\D {\widetilde D}(\ell)}{\D \ell} = {\widetilde D}(\ell) \,  
  \gamma_D(\ell) \ , \quad 
  \ell \, \frac{\D {\widetilde \tau}(\ell)}{\D \ell} =  
  {\widetilde \tau}(\ell) \, \gamma_\tau(\ell) \ , \quad 
  \ell \, \frac{\D {\widetilde u}(\ell)}{\D \ell} = \beta_u(\ell) \ ,
\label{I5runc}
\end{equation}
with initial values $D_R$, $\tau_R$, and $u_R$, respectively at $\ell = 1$.
The {\em first-order ordinary differential equations} (\ref{I5runc}), with
$\gamma_D(\ell) = \gamma_D\Bigl( {\widetilde u}(\ell) \Bigr)$ etc. define
{\em running couplings} that describe how the parameters of the theory change
under scale transformations $\mu \to \mu \, \ell$.
The formal solutions for the ${\widetilde D}(\ell)$ and 
${\widetilde D}(\ell)$ read
\begin{equation} 
  {\widetilde D}(\ell) = D_R \, \exp \left[ \int_1^\ell \! \gamma_D(\ell') \, 
  \frac{\D \ell'}{\ell'} \right] \ , \quad 
  {\widetilde \tau}(\ell) = \tau_R \, \exp \left[ \int_1^\ell \!  
  \gamma_\tau(\ell') \, \frac{\D \ell'}{\ell'} \right] \ .
\label{I5irun} 
\end{equation} 
For the function ${\hat \chi}(\ell) = {\hat \chi}_R\Bigl( 
{\widetilde D}(\ell),{\widetilde \tau}(\ell),{\widetilde u}(\ell) \Bigr)$, we 
then obtain another ordinary differential equation, namely
\begin{equation}
  \ell \, \frac{\D {\hat \chi}(\ell)}{\D \ell} = \left[ 2 + \gamma_S(\ell)
  \right] {\hat \chi}(\ell) \ ,
\label{I5odfs}
\end{equation}
which is solved by
\begin{equation}
  {\hat \chi}(\ell) = {\hat \chi}(1) \, \ell^2 \, \exp\left[ \! \int_1^\ell 
  \gamma_S(\ell') \, \frac{\D \ell'}{\ell'} \right] \ . 
\label{I5rgsl}
\end{equation}
Collecting everything, we finally arrive at
\begin{eqnarray}
  \chi_R(\mu,D_R,\tau_R,u_R,\vec{q},\omega) &=& (\mu \, \ell)^{-2} \, \exp
  \left[ - \int_1^\ell \! \gamma_S(\ell') \, \frac{\D \ell'}{\ell'} \right]
  \nonumber \\
  &&\times {\hat \chi}_R\left( {\widetilde \tau}(\ell), {\widetilde u}(\ell),
  \frac{|\vec{q}|}{\mu \, \ell}, \frac{\omega}{{\widetilde D}(\ell) \,
  (\mu \, \ell)^{2+a}} \right) \ .
\label{I5rchi}
\end{eqnarray}

The solution (\ref{I5rchi}) of the RG equation (\ref{I5rgsu}), along with the
flow equations (\ref{I5runc}), (\ref{I5irun}) for the running couplings tell 
us how the dynamic susceptibility depends on the (momentum) scale 
$\mu \, \ell$ at which we consider the theory.
Similar relations can be obtained for arbitrary vertex functions by solving 
the associated RG equations (\ref{I5rgeq}) \cite{UCT}.
The point here is that the right-hand side of Eq.~(\ref{I5rchi}) may be
evaluated outside the IR-singular regime, by fixing one of its arguments at a
finite value, say $|\vec{q}| / \mu \, \ell = 1$.
The function ${\hat \chi}_R$ is regular, and can be calculated by means of 
perturbation theory.
A {\em scale-invariant} regime is characterised by the renormalised nonlinear 
coupling $u_R$ becoming independent of the scale $\mu \, \ell$, or 
${\widetilde u}(\ell) \to u^* = {\rm const}$.
For an {\em RG fixed point} to be {\em infrared-stable}, we thus require 
\begin{equation}
  \beta_u(u^*) = 0 \ , \quad \beta_u'(u^*) > 0 \ ,
\label{I5rgfp}
\end{equation}
since Eq.~(\ref{I5runc}) then implies that 
${\widetilde u}(\ell \to 0) \to u^*$.
Taking the limit $\ell \to 0$ thus provides the desired mapping of physical
observables such as (\ref{I5rchi}) onto the critical region.
In the vicinity of an IR-stable RG fixed point, Eq.~(\ref{I5irun}) yields the
power laws ${\widetilde D}(\ell) \approx D_R \, \ell^{\gamma_D^*}$, where
$\gamma_D^* = \gamma_D(\ell \to 0) = \gamma_D(u^*)$, etc.
Consequently, Eq.~(\ref{I5rchi}) reduces to 
\begin{equation}
  \chi_R(\tau_R,\vec{q},\omega) \approx \mu^{-2} \, \ell^{- 2 - \gamma_S^*} \,
  {\hat \chi}_R\!\left( \tau_R \, \ell^{\gamma_\tau^*}, u^*,  
  \frac{|\vec{q}|}{\mu \, \ell}, \frac{\omega}{D_R \, \mu^{2+a} \,  
  \ell^{2 + a + \gamma_D^*}} \right) \ ,
\label{I5chsc}
\end{equation} 
and upon {\em matching} $\ell = |\vec{q}| / \mu$ we recover the {\em dynamic 
scaling law} (\ref{I1susc}) with the {\em critical exponents} 
\begin{equation} 
  \eta = - \gamma_S^* \ , \quad \nu = - 1 / \gamma_\tau^* \ , \quad   
  z = 2 + a + \gamma_D^* \ .  
\label{I5crex} 
\end{equation} 

To one-loop order, we obtain from the RG beta function (\ref{I5betu})
\begin{equation}
  u_H^* = \frac{6 \, \epsilon}{n + 8} + O(\epsilon^2) \ .
\label{I5ufix}
\end{equation}
Here we have indicated that our perturbative expansion for small $u$ has 
effectively turned into a {\em dimensional expansion} in $\epsilon = d_c - d$.
In dimensions $d < 4$, the {\em Heisenberg} fixed point $u_H^*$ is IR-stable, 
since $\beta_u'(u_H^*) = \epsilon > 0$.
With Eqs.~(\ref{I5gams}) and (\ref{I5gamt}), the identifications 
(\ref{I5crex}) then give us explicit results for the static scaling exponents,
as mere functions of dimension $d = 4 - \epsilon$ and the number of order 
parameter components $n$,
\begin{equation}  
  \eta = \frac{n + 2}{2 \, (n + 8)^2} \, \epsilon^2 + O(\epsilon^3) \ , \quad
  \frac{1}{\nu} = 2 - \frac{n + 2}{n + 8} \, \epsilon + O(\epsilon^2) \ .
\label{I5stex}
\end{equation} 
For model A with nonconserved order parameter, the two-loop result 
(\ref{I5gamd}) yields the independent dynamic critical exponent
\begin{equation}
  a = 0 \, : \ z = 2 + c \, \eta \ , \quad c = 6 \ln \frac{4}{3} - 1 
  + O(\epsilon) \ ;
\label{I5mdaz}
\end{equation}
for model B with conserved order parameter, instead 
$\gamma_D^* = \gamma_S^* = - \eta$, whence we arrive at the {\em exact} 
scaling relation
\begin{equation}   
  a = 2 \, : \ z = 4 - \eta \ .
\label{I5mdbz}
\end{equation}

In dimensions $d > d_c = 4$, the {\em Gaussian} fixed point $u_0^* = 0$ is
stable ($\beta_u'(0) = - \epsilon > 0$).
Therefore all anomalous dimensions disappear, i.e., 
$\gamma_S^* = 0 = \gamma_D^*$ and $\gamma_\tau^* = - 2$, and we are left with 
the mean-field critical exponents $\eta_0 = 0$, $\nu_0 = 1/2$, and 
$z_0 = 2 + a$.
Precisely at the upper critical dimension $d_c = 4$, the RG flow equation for
the nonlinear coupling becomes 
\begin{equation}
  \ell \, \frac{\D {\widetilde u}(\ell)}{\D \ell} = \frac{n + 8}{6} \, 
  {\widetilde u}(\ell)^2 + O\Bigl( {\widetilde u}(\ell)^3 \Bigr) \ ,
\label{I5logu}
\end{equation}
which is solved by
\begin{equation}
  {\widetilde u}(\ell) = \frac{u_R}{1 - \frac{n+8}{6} \, u_R \, \ln \ell} \ .
\label{I5lgus}
\end{equation}
In four dimensions, ${\widetilde u}(\ell) \to 0$, but only logarithmically
slowly, which causes {\em logarithmic corrections} to the mean-field critical
power laws.
For example, upon inserting Eq.~(\ref{I5lgus}) into the flow equation 
(\ref{I5runc}), one finds ${\widetilde \tau}(\ell) \sim \tau_R \, \ell^{-2}  
(\ln |\ell|)^{- (n + 2) / (n + 8)}$; with 
${\widetilde \tau}(\ell = \xi^{-1}) = O(1)$, iterative inversion yields
\begin{equation}  
  \xi(\tau_R) \sim \tau_R^{-1/2} \left( \ln \tau_R \right)^{(n+2)/2(n+8)} \ .
\label{I5lgxi}  
\end{equation}

This concludes our derivation of asymptotic scaling laws for the critical 
dynamics of the purely relaxational models A and B, and the explicit 
computation of the scaling exponents in powers of $\epsilon = d_c - d$.
In the following sections, I will briefly sketch how the response functional 
formalism and the dynamic renormalisation group can be employed to study the
critical dynamics of systems with reversible mode-coupling terms, the `ageing'
behaviour induced by quenching from random initial conditions to the critical 
point, the effects of violating the detailed balance constraints on universal
dynamic critical properties, and the generically scale-invariant features of
nonequilibrium systems such as driven diffusive Ising lattice gases.

\subsection{Critical dynamics with reversible mode-couplings} 
\label{ssec:I6} 

In the previous chapters, we have assumed purely relaxational dynamics for the
order parameter, see Eq.~(\ref{I1reld}). 
In general, however, there are also {\em reversible} contributions to the 
systematic force terms $F^\alpha$ that enter its Langevin equation 
\cite{HohHalp,ChaiLub}.
Consider the Hamiltonian dynamics of {\em microscopic} variables, say, local 
spin densities, at $T = 0$:
$\partial_t \, S_m^\alpha(x,t) = \Bigl\{ H[S_m],S_m^\alpha(x,t) \Bigr\}$.
Here, the {\em Poisson brackets} $\{ A,B \}$ constitute the classical analog 
of the quantum-mechanical commutator $\frac{i}{\hbar} [A,B]$ (correspondence 
principle).  
Upon {\em coarse-graining}, the microscopic variables $S_m^\alpha$ become the
{\em mesoscopic} hydrodynamic fields $S^\alpha$.
Since the set of {\em slow} modes should provide a complete description of
the critical dynamics, we may formally expand 
\begin{equation}  
  \Bigl\{ {\cal H}[S] , S^\alpha(\vec{x}) \Bigr\} = \int \! \D^dx' \, 
  \sum_\beta \frac{\delta {\cal H}[S]}{\delta S^\beta(\vec{x}')} \, 
  Q^{\beta \alpha}(\vec{x}',\vec{x}) \ , 
\label{I6come}
\end{equation}
with the mutual Poisson brackets of the hydrodynamic variables
\begin{equation}
  Q^{\alpha \beta}(\vec{x},\vec{x}') = \Bigl\{ S^\alpha(\vec{x}) , 
  S^\beta(\vec{x}') \Bigr\} = - Q^{\beta \alpha}(\vec{x}',\vec{x}) \ . 
\label{I6pobr}
\end{equation} 

By inspection of the associated Fokker--Planck equation, one may then 
establish an additional {\em equilibrium condition} in order for the
time-dependent probability distribution to reach the canonical limit
(\ref{I1eqpd}): ${\cal P}[S,t] \to {\cal P}_{\rm eq}[S]$ as $t \to \infty$ 
{\em provided} the probability current is {\em divergence-free} in the space 
spanned by the stochastic fields $S^\alpha(\vec{x})$: 
\begin{equation} 
  \int \! \D^dx \sum_\alpha \frac{\delta}{\delta S^\alpha(\vec{x})} \left(  
  F_{\rm rev}^\alpha[S] \ e^{- {\cal H}[S] / k_{\rm B} T} \right) = 0 \ .  
\label{I6divf} 
\end{equation} 
It turns out that this equilibrium condition is often more crucial than the 
Einstein relation (\ref{I1eins}). 
In order to satisfy Eq.~(\ref{I6divf}) at $T \not= 0$, we must supplement 
Eq.~(\ref{I6come}) by a finite-temperature correction, whereupon the 
{\em reversible mode-coupling} contributions to the systematic forces become
\begin{equation}
  F_{\rm rev}^\alpha[S](\vec{x}) = 
  - \int \! \D^dx' \sum_\beta \left[ Q^{\alpha \beta}(\vec{x},\vec{x}') 
  \frac{\delta {\cal H}[S]}{\delta S^\beta(\vec{x}')} - k_{\rm B} T \, 
  \frac{\delta Q^{\alpha \beta}(\vec{x},\vec{x}')}{\delta S^\beta(\vec{x}')}  
  \right] \ ,
\label{I6revm}
\end{equation} 
and the complete coupled set of stochastic differential equations reads 
\begin{equation} 
  \frac{\partial S^\alpha(\vec{x},t)}{\partial t} = 
  F_{\rm rev}^\alpha[S](\vec{x},t) - D^\alpha (\I \vec{\nabla})^{a_\alpha}   
  \frac{\delta {\cal H}[S]}{\delta S^\alpha(\vec{x},t)} 
  + \zeta^\alpha(\vec{x},t) \ , 
\label{I6lang} 
\end{equation} 
where as before the $D^\alpha$ denote the relaxation coefficients, and
$a^\alpha = 0$ or $2$ respectively for nonconserved and conserved modes. 

As an instructive example, let us consider the {\em Heisenberg model} for 
{\em isotropic ferromagnets}, 
$H[\{ {\vec S}_j \}] = - \frac{1}{2} \sum_{j,k=1}^N J_{jk} \, \vec{S}_j \cdot 
\vec{S}_k$, where the spin operators satisfy the usual commutation relations
$\left[ S_j^\alpha , S_k^\beta \right] = i \hbar \sum_\gamma 
\epsilon^{\alpha \beta \gamma} S_j^\gamma \, \delta_{jk}$.
The corresponding Poisson brackets for the magnetisation density read 
\begin{equation}
  Q^{\alpha \beta}(\vec{x},\vec{x}') = - g \sum_\gamma 
  \epsilon^{\alpha \beta \gamma} S^\gamma(\vec{x}) \, \delta(\vec{x}-\vec{x}')
  \ ,
\label{I6mdjp}
\end{equation} 
where the purely {\em dynamical coupling} $g$ incorporates various factors
that emerge upon coarse-graining and taking the continuum limit.
The second contribution in Eq.~(\ref{I6revm}) vanishes, since it reduces to a
contraction of the antisymmetric tensor $\epsilon^{\alpha \beta \gamma}$ with 
the Kronecker symbol $\delta^{\beta \gamma}$, whence we arrive at the Langevin
equations governing the critical dynamics of the three order parameter 
components for isotropic ferromagnets \cite{MaMaz}
\begin{equation}  
  \frac{\partial \vec{S}(\vec{x},t)}{\partial t} = - g \, \vec{S}(\vec{x},t) 
  \times \frac{\delta {\cal H}[\vec{S}]}{\delta \vec{S}(\vec{x},t)} 
  + D \, \vec{\nabla}^2 \, 
  \frac{\delta {\cal H}[{\vec S}]}{\delta {\vec S}(\vec{x},t)} 
  + \vec{\zeta}(\vec{x},t) \ ,
\label{I6modj}
\end{equation}
with $\langle \vec{\zeta}(\vec{x},t) \rangle = 0$.
Since $\Bigl[ H[\{ {\vec S}_j \}] , \sum_k S_k^\alpha \Bigr] = 0$, the total 
magnetisation is conserved, whence the noise correlators should be taken as
\begin{equation}
  \left\langle \zeta^\alpha(\vec{x},t) \, \zeta^\beta(\vec{x}',t') 
  \right\rangle = - 2 D \, k_{\rm B} T \, \vec{\nabla}^2 
  \delta(\vec{x}-\vec{x}') \, \delta(t-t') \, \delta^{\alpha \beta} \ .
\label{I6corj}
\end{equation}
The vector product term in Eq.~(\ref{I6modj}) describes the spin precession in
the local effective magnetic field $\delta {\cal H}[\vec{S}]/\delta \vec{S}$, 
which includes a contribution induced by the exchange interaction. 

\begin{figure} 
\centering 
\includegraphics[width = 11.8 truecm]{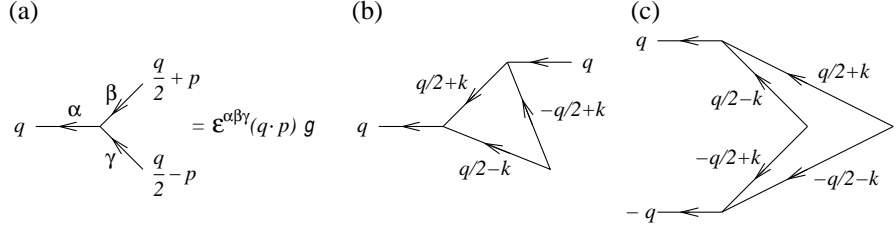} 
\caption{(a) Mode-coupling three-point vertex for model J. One-loop Feynman  
  diagrams for the propagator (b) and noise vertex (c) renormalisations in 
  model J. The same graphs (b), (c) apply for driven diffusive systems 
 (Sec.~\ref{ssec:I9}).} 
\label{modelj} 
\end{figure}

The Langevin equation (\ref{I6modj}) and (\ref{I6corj}) with the Hamiltonian 
(\ref{I1glwh}) for $n = 3$ define the so-called {\em model J} \cite{HohHalp}. 
In addition to the model B response functional (\ref{I2hajd}) and 
(\ref{I2ndjd}) with $a = 2$ (setting $k_{\rm B} T = 1$ again), the reversible 
force in Eq.~(\ref{I6modj}) leads to an additional contribution to the action
\begin{equation}
  {\cal A}_{\rm mc}[{\widetilde S},S] = - g \int \! \D^dx \! \int \! \D t  
  \sum_{\alpha,\beta,\gamma} \epsilon^{\alpha \beta \gamma}  
  {\widetilde S}^\alpha S^\beta \left( \nabla^2 S^\gamma + h^\gamma \right)\ ,
\label{I6mcpl}
\end{equation}
which gives rise to an additional {\em mode-coupling vertex}, as depicted in  
Fig.~\ref{modelj}(a). 
Power counting yields the scaling dimension $[g] = \mu^{3-d/2}$ for the 
associated coupling strength, whence we expect a {\em dynamical} upper 
critical dimension $d_c' = 6$. 
However, since we are investigating a system in thermal equilibrium, we can
treat its thermodynamics and static properties separately from its dynamics.
Obviously therefore, the static critical exponents must still be given (to
lowest nontrivial order and for $d < d_c = 4$) by Eqs.~(\ref{I5stex}) for the 
three-component Heisenberg model with $O(3)$ rotational symmetry.
Therefore our sole task is to find the dynamic critical exponent $z$.

Remarkably, $z$ is entirely fixed by the symmetries of the problem and can be
determined exactly.
To this end, we exploit the fact that the $S^\alpha$ are the generators of the
rotation group; indeed, it follows from Eq.~(\ref{I6mcpl}) that applying a
time-dependent external field $h^\gamma(t)$ induces a contribution
\begin{equation}
  \Bigl\langle S^\alpha(\vec{x},t) \Bigr\rangle_h = g \int_0^t \! \D t' 
  \sum_\beta \epsilon^{\alpha \beta \gamma} \Bigl\langle S^\beta(\vec{x},t') 
  \Bigr\rangle_h h^\gamma(t)
\label{I6hmag}
\end{equation} 
to the average magnetisation.
As a consequence, we obtain for the {\em nonlinear susceptibility} 
$R^{\alpha ; \beta \gamma} = \delta^2 \langle S^\alpha \rangle / 
\delta h^\beta \, \delta h^\gamma \vert_{h=0}$, 
\begin{equation}
  \int \! \D^dx' \, R^{\alpha ; \beta \gamma}(\vec{x},t;\vec{x}-\vec{x}',t-t')
  = g \, \epsilon^{\alpha \beta \gamma} \, \chi^{\beta \beta}(\vec{x},t) \, 
  \Theta(t) \, \Theta(t-t') \ .
\label{I6nlsu}
\end{equation}
An analogous expression must hold after renormalisation as well. 
If we define the dimensionless renormalised mode-coupling according to
\begin{equation}
  g_R^2 = Z_g \, g^2 \, B_d \, \mu^{d-6} \ , \quad f_R = g_R^2 / D_R^2 \ ,
\label{I6gren}
\end{equation} 
where $B_d = \Gamma(4 - d/2) / 2^d \, d \, \pi^{d/2}$, Eq.~(\ref{I6nlsu}) 
implies the identity \cite{BaJaWa}
\begin{equation} 
  Z_g = Z_S \ . 
\label{I6jzid}
\end{equation} 
For the RG beta function associated with the effective coupling entering the
loop corrections, we thus infer
\begin{equation}
  \beta_f = \mu \, \frac{\partial}{\partial \mu} \bigg\vert_0 f_R = f_R  
  \left( d - 6 + \gamma_S - 2 \, \gamma_D \right) \ . 
\label{I6betg}
\end{equation} 
Consequently, at any {\em nontrivial} IR-stable RG fixed point 
$0 < f^* < \infty$, we have the {\em exact} scaling relation, valid to 
{\em all} orders in perturbation theory, 
\begin{equation}
  d < 6 \, : \ z = 4 + \gamma_D^* = 4 + \frac{d - 6 + \gamma_S^*}{2} =  
  \frac{d + 2 - \eta}{2} \ .
\label{I6mdjz}
\end{equation}
Since the resulting value for the dynamic exponent, $z \approx 5/2$ in three 
dimensions, is markedly smaller than the model B mean-field $z_0 = 4$, we 
conclude that the reversible spin precession kinetics speeds up the order
parameter dynamics considerably \cite{MaMaz,HohHalp,BaJaWa}.

An explicit one-loop calculation, either for the propagator self-energy
$\Gamma^{(1,1)}(\vec{q},\omega)$, depicted in Fig.~\ref{modelj}(b), or the
noise vertex $\Gamma^{(2,0)}(\vec{q},\omega)$, shown in Fig.~\ref{modelj}(c), 
yields \cite{BaJaWa,UCT}
\begin{equation}
  \gamma_D = - f_R + O(u_R^2,f_R^2) \ , 
\label{I6gamd}
\end{equation}
which along with $\gamma_S = 0 + O(u_R^2,f_R^2)$ confirms that there exists a 
nontrivial mode-coupling RG fixed point  
\begin{equation}
  f_J^* = \frac{\varepsilon}{2} + O(\varepsilon^2) \ ,
\label{I6mcfp}
\end{equation}
where $\varepsilon  = 6 - d$, which is IR-stable for $d < 6$. 
As $\eta = 0$ for $d > 4$, we indeed recover the mean-field dynamic exponent
$z_0 = 4$ in $d \geq 6$ dimensions.
With the leading singularity thus isolated, the regular scaling functions can 
be computed numerically to high accuracy within a self-consistent one-loop 
approximation that also goes under the name {\em mode-coupling theory}.
Details of this procedure, an alternative derivation, and many results of 
mode-coupling theory as applied to the critical dynamics of magnets and 
comparisons with experimental data can be found in Ref.~\cite{FreySchw}.
 
Typically, reversible force terms of the form (\ref{I6revm}) involve dynamical
couplings of the order parameter to {\em other} conserved, slow variables.
In addition, there may also be static couplings to conserved fields in the
Hamiltonian.
These various possibilities give rise to a range of different {\em dynamic
universality classes} for near-equilibrium critical dynamics \cite{HohHalp}.
We shall not pursue these further here (for a partial account within the
field-theoretic RG approach, see Ref.~\cite{UCT}), but instead proceed and now
consider nonequilibrium effects.

\subsection{Critical relaxation, initial slip, and ageing} 
\label{ssec:I7}
 
We begin with a brief discussion of the {\em coarsening} dynamics of systems 
described by model A / B kinetics that are rapidly {\em quenched} from a 
disordered state at $T \gg T_c$ to the critical point $T \approx T_c$ 
\cite{JaSchSch,JanRel}.
The situation may be modeled as a relaxation from {\em Gaussian random initial
conditions}, i.e., the probability distribution for the order parameter at 
$t = 0$ can be taken as
\begin{equation}
  {\cal P}[S,t=0] \propto e^{-{\cal H}_0[S]} = \exp \left( - \frac{\Delta}{2} 
  \int \! \D^dx \sum_\alpha \left[ S^\alpha(\vec{x},0) - a^\alpha(\vec{x}) 
  \right]^2 \right) \ ,
\label{I7init}
\end{equation}
where the functions $a^\alpha(\vec{x})$ specify the most likely initial
configurations.
Power counting for the parameter $\Delta$ gives $[\Delta] = \mu^2$, whence it
is a relevant perturbation that will flow to $\Delta \to \infty$ under the RG.
Asymptotically, therefore, the system will be governed by sharp {\em Dirichlet
boundary conditions}.
Whereas the response propagators remains a causal function of the time 
difference between applied perturbation and effect, $G_0(\vec{q},t-t') 
= \Theta(t-t') \, \E^{- D \vec{q}^a \, (r + \vec{q}^2) \, (t-t')}$, see
Eq.~(\ref{I1rest}), time translation invariance is broken by the initial state
in the {\em Dirichlet correlator} of the Gaussian model,
\begin{equation}  
  C_D(\vec{q};t,t') = \frac{1}{r + \vec{q}^2} \left( 
  \E^{- D \vec{q}^a \, (r + \vec{q}^2) \, |t - t'|} 
  - \E^{- D \vec{q}^a \, (r + \vec{q}^2) \, (t + t')} \right) \ .
\label{I7dirc}
\end{equation} 
Away from criticality, i.e., for $r > 0$ and $\vec{q} \not= 0$, temporal 
correlations decay exponentially fast, and the system quickly approaches the
stationary equilibrium state.
However, as $T \to T_c$, the equilibration time diverges according to
$t_c \sim |\tau|^{- z \nu} \to \infty$, and the system never reaches thermal
equilibrium.
Two-time correlation functions will then depend on both times separately, in a
specific manner to be addressed below, a phenomenon termed critical 
{\em `ageing'} (for more details, see Refs.~\cite{CalGam,Andrea}). 
 
The field-theoretic treatment of the model A / B dynamical action 
(\ref{I2hajd}), (\ref{I2ndjd}) with the initial term (\ref{I7init}) follows 
the theory of {\em boundary critical phenomena} \cite{Diehl}. 
However, it turns out that additional singularities on the temporal `surface' 
at $t + t' = 0$ appear only for model A, and can be incorporated into a single
new renormalisation factor; to one-loop order, one finds 
\cite{JaSchSch,JanRel}
\begin{equation}  
  a = 0 \, : \ {\widetilde S}_R^\alpha(\vec{x},0) = 
  (Z_0 \, Z_{\widetilde S})^{1/2} \, {\widetilde S}^\alpha(\vec{x},0) \ , 
  \quad Z_0 = 1 - \frac{n + 2}{6} \, \frac{u_R}{\epsilon} \ .
\label{I7iren}
\end{equation}
This in turn leads to a {\em single} independent critical exponent associated
with the initial time relaxation, the {\em initial slip exponent}, which 
becomes for the purely relaxational models A and B with nonconserved and
conserved order parameter:
\begin{equation} 
  a = 0 \, : \ \theta = \frac{\gamma_0^*}{2 \, z} = 
  \frac{n + 2}{4 \, (n + 8)} \, \epsilon + O(\epsilon^2) \ , \quad 
  a = 2 \, : \ \theta = 0 \ . 
\label{I7isle}
\end{equation}
In order to obtain the {\em short-time} scaling laws for the dynamic response 
and correlation functions in the {\em ageing limit} $t' / t \to 0$, one 
requires additional information that can be garnered from the short-distance
{\em operator product expansion} for the fields, 
\begin{equation}  
  t \to 0 \, : \ {\widetilde S}(\vec{x},t) = 
  {\tilde \sigma}(t) \, {\widetilde S}_0(\vec{x}) \ , \quad 
  S(\vec{x},t) = \sigma(t) \, {\widetilde S}_0(\vec{x}) \ .
\label{I7opex}
\end{equation}
Subsequent analysis then yields eventually \cite{JaSchSch,JanRel}
\begin{eqnarray} 
  \chi(\vec{q};t, t' \to 0) &=& |\vec{q}|^{z - 2 + \eta} \left( \frac{t}{t'} 
  \right)^\theta {\hat \chi}_0(\vec{q} \, \xi, |\vec{q}|^z \, D t) \ , 
\label{I7suag} \\ 
  C(\vec{q};t,t' \to 0) &=& |\vec{q}|^{- 2 + \eta} \left( \frac{t}{t'} 
  \right)^{\theta - 1} {\hat C}_0(\vec{q} \, \xi, |\vec{q}|^z \, D t) \ , 
\label{I7coag}
\end{eqnarray}
and for the time dependence of the mean order parameter 
\begin{eqnarray}
&&\langle S(t) \rangle = S_0 \, t^{\theta'} {\hat S}\!\left( S_0 \,  
  t^{\theta' + \beta / z \nu} \right) \ , 
\label{I7opag} \\ 
  &&a = 0 \, : \ \theta' = \theta - \frac{z - 2 + \eta}{z} \ , \quad  
  a = 2 \, : \ \theta' = \theta  = 0 \ . 
\label{I7opse}
\end{eqnarray}

One may also compute the universal {\em fluctuation--dissipation ratios} in
this nonequilibrium ageing regime \cite{CalGam,Andrea}.
It emerges, though, that these depend on the quantity under investigation, 
which prohibits a unique definition of an effective nonequilibrium temperature
for critical ageing.
The method sketched above can be extended to models with reversible 
mode-couplings \cite{OerJan}. 
For model J capturing the critical dynamics of isotropic ferromagnets, one 
finds
\begin{equation}
  \theta = \frac{z - 4 + \eta}{z} = - \frac{6 - d - \eta}{d + 2 - \eta} \ ;
\label{I7mjis}
\end{equation} 
in systems where a {\em nonconserved} order parameter is dynamically coupled
to other conserved modes, the initial slip exponent $\theta$ is actually 
{\em not} a universal number, but depends on the width of the initial 
distribution \cite{OerJan}.

\subsection{Nonequilibrium relaxational critical dynamics} 
\label{ssec:I8}
 
Next we address the question \cite{TaAkSa}, What happens if the detailed 
balance conditions (\ref{I1eins}) and (\ref{I6divf}) are violated?
To start, we change the noise strength $D \to {\widetilde D}$ in the purely
relaxational models A and B, which (in our units) violates the Einstein 
relation (\ref{I1eins}).  
However, this modification can obviously be absorbed into a rescaled 
{\em effective temperature}, 
$k_{\rm B} T \to k_{\rm B} T' = {\widetilde D} / D$.
Formally this is established by means of the dynamical action (\ref{I2jade}),
which now reads 
\begin{eqnarray} 
  {\cal A}[{\widetilde S},S] &=& \int \! \D^dx \! \int \! \D t \sum_\alpha  
  {\widetilde S}^\alpha \biggl[ \partial_t \, S^\alpha + D \, 
  (i \vec{\nabla})^a \left( r - \vec{\nabla}^2 \right) S^\alpha \nonumber \\ 
  &&\qquad\qquad - {\widetilde D} \, (i \vec{\nabla})^a \, 
  {\widetilde S}^\alpha + D \, \frac{u}{6} \, (i \vec{\nabla})^a \, S^\alpha  
  \sum_\beta S^\beta S^\beta \biggr] \ . 
\label{I8neab}
\end{eqnarray} 
Upon simple rescaling ${\widetilde S}^\alpha \to {\widetilde S}^{' \alpha} =  
{\widetilde S}^\alpha \, \sqrt{{\widetilde D} / D}$, $S^\alpha \to {S'}^\alpha
= S^\alpha \, \sqrt{D / {\widetilde D}}$, the response functional 
(\ref{I8neab}) recovers its equilibrium form, albeit with modified nonlinear
coupling $u \to {\widetilde u} = u \, {\widetilde D} / D$.
However, the {\em universal asymptotic} properties of these models are 
governed by the Heisenberg fixed point (\ref{I5ufix}), and the specific value
of the (renormalised) coupling, which only serves as the initial condition for
the RG flow, does not matter.
In fact, the relaxational dynamics of the kinetic Ising model with Glauber
dynamics (model A with $n = 1$) is known to be quite stable against 
nonequilibrium perturbations \cite{HaLeWi,GrJaHe}, even if these break the 
Ising $Z_2$ symmetry \cite{BaSchm}.
For model J the above rescaling modifies in a similar manner merely the
mode-coupling strength in Eq.~(\ref{I6mcpl}), namely 
$g \to {\widetilde g} = g \, \sqrt{{\widetilde D} / D}$ \cite{TaRacz}. 
Again, since the dynamic critical behaviour is governed by the universal fixed
point (\ref{I6mcfp}), thermal equilibrium becomes effectively {\em restored} 
at criticality.
More generally, it has been established that {\em isotropic} detailed balance
violations do not affect the universal properties in other models for critical
dynamics that contain additional conserved variables either: the equilibrium 
RG fixed points tend to be asymptotically stable \cite{TaAkSa}.

In systems with {\em conserved} order parameter, however, we may in addition
introduce spatially {\em anisotropic} violations of Einstein's relation; for
example, in model B one can allow for anisotropic relaxation 
$- D \, \vec{\nabla}^2 \to - D_\perp \, \vec{\nabla}_\perp^2
- D_\parallel \, \vec{\nabla}_\parallel^2$, with different rates in two 
spatial subsectors and concomitantly anisotropic noise correlations 
$- {\widetilde D} \, \vec{\nabla}^2 \to - {\widetilde D}_\perp \, 
\vec{\nabla}_\perp^2- {\widetilde D}_\parallel \, \vec{\nabla}_\parallel^2$.
We have thus produced a truly nonequilibrium situation provided 
${\widetilde D}_\perp / D_\perp \not= {\widetilde D}_\parallel / D_\parallel$,
which we may interpret as having effectively coupled the longitudinal and 
transverse spatial sectors to heat baths with different temperatures
$T_\perp < T_\parallel$, say \cite{TaSaRa}. 
 
Evaluating the fluctuation-induced shift of the transition temperature, see
Eq.~(\ref{I4cpsh}) one finds not surprisingly that the {\em transverse} sector
softens first, while the longitudinal sector remains noncritical.
This suggests that we can neglect the nonlinear longitudinal fluctuations as
well as the $\vec{\nabla}_\parallel^4$ term in the propagator.
These features are indeed encoded in the corresponding {\em anisotropic} 
scaling: $[q_\perp] = \mu$, $[q_\parallel] = \mu^2$, $[\omega] = \mu^4$, 
whence $[{\widetilde D}_\perp] = [D_\perp] = \mu^0$, and 
$[{\widetilde D}_\parallel] = [D_\parallel] = \mu^{-2}$ become irrelevant.
Upon renaming $D = D_\perp$ and $c = r_\parallel D_\parallel / D_\perp$, this
ultimately leads to the {\em randomly driven} or {\em two-temperature model B}
\cite{SchmZia,Beate} as the effective theory describing the phase transition: 
\begin{eqnarray} 
  \frac{\partial S^\alpha(\vec{x},t)}{\partial t} &=& D \left[ 
  \vec{\nabla}_\perp^2 \left( r - \vec{\nabla}_\perp^2 \right) + c \,
  \vec{\nabla}_\parallel^2 \right] S^\alpha(\vec{x},t) \nonumber \\ 
  &&+ \frac{D \, \widetilde u}{6} \, \vec{\nabla}_\perp^2 \, 
  S^\alpha(\vec{x},t) \sum_\beta [S^\beta(\vec{x},t)]^2 
  + \zeta^\alpha(\vec{x},t) \ ,
\label{I8ttmb}
\end{eqnarray} 
with the noise correlations
\begin{equation}
  \left\langle \zeta^\alpha(\vec{x},t) \, \zeta^\beta(\vec{x}',t') 
  \right\rangle = - 2 D \, \vec{\nabla}_\perp^2 \, \delta(\vec{x}-\vec{x}') \,
  \delta(t-t') \, \delta^{\alpha \beta} \ .
\label{I8ttnc} 
\end{equation} 
Quite remarkably, the Langevin equation (\ref{I8ttmb}) can be derived as an
{\em equilibrium} diffusive relaxational kinetics  
\begin{equation}
  \frac{\partial S^\alpha(\vec{x},t)}{\partial t} = D \, \vec{\nabla}_\perp^2 
  \, \frac{\delta {\cal H}_{\rm eff}[S]}{\delta S^\alpha(\vec{x},t)}  
  + \zeta^\alpha(\vec{x},t)
\label{I8ttrd}
\end{equation} 
from an effective {\em long-range} Hamiltonian 
\begin{eqnarray} 
  {\cal H}_{\rm eff}[S] &=& \int \frac{\D^dq}{(2 \pi)^d}
  \frac{\vec{q}_\perp^2 (r + \vec{q}_\perp^2) + c \, \vec{q}_\parallel^2}{2 \,
  \vec{q}_\perp^2} \, \sum_\alpha \big| S^\alpha(\vec{q}) \big|^2 \nonumber \\
  &&+ \frac{\widetilde u}{4 !} \int \! \D^dx \sum_{\alpha,\beta} 
  [S^\alpha(\vec{x})]^2 \, [S^\beta(\vec{x})]^2 \ . 
\label{I8effh}
\end{eqnarray} 
Power counting gives $[{\widetilde u}] = \mu^{4 - d_\parallel - d}$: the
spatial anisotropy suppresses longitudinal fluctuations and lower the upper 
critical dimension to $d_c = 4 - d_\parallel$. 
The anisotropic correlations encoded in Eq.~(\ref{I8effh}) also reduce the 
lower critical dimension and affect the nature of the ordered phase
\cite{SchmZia,BasRacz}.
 
The scaling law for, e.g., the dynamic response function takes the form 
\begin{equation}
  \chi(\tau_\perp,\vec{q}_\perp,\vec{q}_\parallel,\omega) = 
  |\vec{q}_\perp|^{-2+\eta} \, 
  {\hat \chi}\!\left( \frac{\tau}{|\vec{q}_\perp|^{1/\nu}}, \frac{\sqrt{c} \, 
  |\vec{q}_\parallel|}{|\vec{q}_\perp|^{1 + \Delta}}, \frac{\omega}{D \,
  |\vec{q}_\perp|^z} \right) \ ,
\label{I8ansc}
\end{equation}
where we have introduced a new {\em anisotropy exponent} $\Delta$.
Since the nonlinear coupling ${\widetilde u}$ only affects the transverse
sector, we find to {\em all} orders in the perturbation expansion: 
\begin{equation}
  \Gamma^{(1,1)}(\vec{q}_\perp = 0, \vec{q}_\parallel, \omega) 
  = \I \omega + D \, c \, \vec{q}_\parallel^2 \ ,
\label{I8ga11}
\end{equation}
and consequently obtain the $Z$ factor identity
\begin{equation}
  Z_c = Z_D^{-1} = Z_S^{-1} \ ,
\label{I8zfid}
\end{equation}
which at any IR-stable RG fixed point implies the {\em exact} scaling 
relations
\begin{equation}  
  z = 4 - \eta \ , \quad \ \Delta = 1 - \frac{\gamma_c^*}{2} =  
  1 - \frac{\eta}{2} = \frac{z}{2} - 1 \ ,
\label{I8scrl}
\end{equation}
whereas the scaling exponents for the longitudinal sector read
\begin{equation}
  z_\parallel = \frac{z}{1 + \Delta} = 2 \ , \quad  
  \nu_\parallel = \nu \, (1 + \Delta) = \frac{\nu}{2} \, (4 - \eta) \ .
\label{I8loex}
\end{equation} 
As for the equilibrium model B, the only independent critical exponents to be
determined are $\eta$ and $\nu$.
To one-loop order, only the combinatorics of the Feynman diagrams (see
Fig.~\ref{temver}) enters their explicit values, whence one finds for 
$d < d_c = 4 - d_\parallel$ formally identical results as for the usual 
Ginzburg--Landau--Wilson Hamiltonian (\ref{I1glwh}),
\begin{equation}
  \eta = 0 + O(\epsilon^2) \ , \quad 
  \frac{1}{\nu} = 2 - \frac{n + 2}{n + 8} \, \epsilon + O(\epsilon^2) \ ,
\label{I81lex}
\end{equation}
albeit with {\em different} $\epsilon = 4 - d - d_\parallel$.
To two-loop order, however, the anisotropy manifestly affects the evaluation 
of the loop contributions, and the value for $\eta$ deviates from the 
expression in Eq.~(\ref{I5stex}) \cite{Beate}.

Interestingly, an analogously constructed nonequilibrium {\em two-temperature 
model J} with reversible mode-coupling vertex {\em cannot} be cast into a form
that is equivalent to an equilibrium system, for owing to the emerging 
anisotropy, the condition (\ref{I6divf}) cannot be satisfied.
A one-loop RG analysis yields a runaway flow, and no stable RG fixed point is
found \cite{TaSaRa}. 
Similar behaviour ensues in other anisotropic nonequilibrium variants of 
critical dynamics models with conserved order parameter; the precise 
interpretation of the apparent instability is as yet unclear \cite{TaAkSa}.

\subsection{Driven diffusive systems} 
\label{ssec:I9}
 
Finally, we wish to consider Langevin representations of genuinely 
nonequilibrium systems, namely driven diffusive lattice gases (for a 
comprehensive overview, see Ref.~\cite{SchZiaB}).
First we address the coarse-grained continuum version of the asymmetric
exclusion process, i.e., hard-core repulsive particles that hop preferentially
in one direction.
We describe this system in terms of a {\em conserved} particle density, whose
fluctuations we denote with $S(\vec{x},t)$, such that $\langle S \rangle = 0$,
obeying a {\em continuity equation}
$\partial_t \, S(\vec{x},t) + \vec{\nabla} \cdot \vec{J}(\vec{x},t) = 0$.
We assume the system to be driven along the `$\parallel$' direction; in the
transverse sector (of dimension $d_\perp = d - 1$) we thus just have a noisy
{\em diffusion current} 
$\vec{J}_\perp = - D \, \vec{\nabla}_\perp S + \vec{\eta}$, whereas there is
a nonlinear term, stemming from the hard-core interactions, in the current 
along the direction of the external drive, with 
$J_{0 \, \parallel} = {\rm const.}$: $J_\parallel = J_{0 \, \parallel} 
- D \, c \, \nabla_\parallel S - \frac12 \, D g \, S^2 + \eta_\parallel$.
For the stochastic currents, we assume Gaussian white noise
$\langle \eta_i \rangle = 0 = \langle \eta_\parallel \rangle$ and  
$\left\langle \eta_i(\vec{x},t) \, \eta_j(\vec{x}',t') \right\rangle = 2 D \, 
\delta(\vec{x}-\vec{x}') \, \delta(t-t') \, \delta_{ij}$, 
$\left\langle \eta_\parallel(\vec{x},t) \, \eta_\parallel(\vec{x}',t') 
\right\rangle = 2 D \, {\tilde c}\, \delta(\vec{x}-\vec{x}') \, \delta(t-t')$.
Notice that since we are not in thermal equilibrium, Einstein's relation need 
not be fulfilled.
We can however always rescale the field to satisfy it in the transverse
sector; the ratio $w = {\tilde c} / c$ then measures the deviation from 
equilibrium.
These considerations yield the generic Langevin equation for the density 
fluctuations in {\em driven diffusive systems (DDS)} \cite{JanSch,LeuCar}
\begin{equation}
  \frac{\partial S(\vec{x},t)}{\partial t} = D \left( \vec{\nabla}_\perp^2  
  + c \, \nabla_\parallel^2 \right) S(\vec{x},t) + \frac{D \, g}{2} \, 
  \nabla_\parallel S(\vec{x},t)^2 + \zeta(\vec{x},t) \ ,
\label{I9ddsl}
\end{equation}
with conserved noise $\zeta = - \vec{\nabla}_\perp \cdot \vec{\eta} - 
\nabla_\parallel \eta_\parallel$, where $\langle \zeta \rangle = 0$ and
\begin{equation}
  \left\langle \zeta(\vec{x},t) \, \zeta(\vec{x}',t') \right\rangle = 
  - 2 D \left( \vec{\nabla}_\perp^2 + {\tilde c} \, \nabla_\parallel^2 \right)
  \delta(\vec{x}-\vec{x}') \, \delta(t-t') \ .
\label{I9noic}
\end{equation}
Notice that the drive term $\propto g$ breaks both the system's spatial 
reflection symmetry and the Ising $Z_2$ symmetry $S \to - S$.

The corresponding Janssen--De~Dominicis response functional (\ref{I2jade})
reads  
\begin{eqnarray} 
  {\cal A}[{\widetilde S},S] &=& \int \! \D^dx \! \int \!  \D t \,  
  {\widetilde S} \Biggl[ \frac{\partial S}{\partial t} - D \left( 
  \vec{\nabla}_\perp^2 + c \, \nabla_\parallel^2 \right) S \nonumber \\ 
  &&\qquad\qquad\qquad + D \left( \vec{\nabla}_\perp^2 + {\tilde c} \, 
  \nabla_\parallel^2 \right) {\widetilde S} - \frac{D \, g}{2} \, 
  \nabla_\parallel \, S^2 \Biggr] \ .
\label{I9ddsf}
\end{eqnarray} 
It describes a {\em `massless'} theory, hence we expect the system to be 
{\em generically scale-invariant}, without the need to tune it to a special
point in parameter space.
The nonlinear drive term will induce anomalous scaling in the drive direction,
different from ordinary diffusive behaviour.  
In the transverse sector, however, we have to {\em all} orders in the
perturbation expansion simply
\begin{equation}
  \Gamma^{(1,1)}(\vec{q}_\perp,q_\parallel = 0, \omega) 
  = \I \omega + D \, \vec{q}_\perp^2 \ , \quad
  \Gamma^{(2,0)}(\vec{q}_\perp, q_\parallel = 0, \omega) 
  = - 2 D \, \vec{q}_\perp^2 \ ,
\label{I9ga11}
\end{equation}
since the nonlinear three-point vertex, which is of the form depicted in 
Fig.~\ref{modelj}(a), is proportional to $\I q_\parallel$.
Consequently,
\begin{equation}
  Z_{\widetilde S} = Z_S = Z_D = 1 \ ,
\label{I9zfid} 
\end{equation}
which immediately implies
\begin{equation}
  \eta = 0 \ ,  \quad z = 2 \ .
\label{I9trex} 
\end{equation}

Moreover, the nonlinear coupling $g$ itself does not renormalise either as a
consequence of {\em Galilean invariance}.
Namely, the Langevin equation (\ref{I9ddsl}) and the action (\ref{I9ddsf})
are left invariant under Galilean transformations
\begin{equation} 
  S'(\vec{x}_\perp', x_\parallel', t') 
  = S(\vec{x}_\perp, x_\parallel - D g v \, t, t) - v \ ;
\label{I9galt}
\end{equation}
thus, the boost velocity $v$ must scale as the field $S$ under 
renormalisation, and since the product $D \, g \, v$ must be invariant under 
the RG, this leaves us with
\begin{equation}
  Z_g = Z_D^{-1} \, Z_S^{-1} = 1 \ .
\label{I9ginv}
\end{equation} 
The effective nonlinear coupling governing the perturbation expansion in 
terms of loop diagrams turns out to be $g^2 / c^{3/2}$; if we define its
renormalised counterpart as
\begin{equation}
  v_R = Z_c^{3/2} \, v \, C_d \, \mu^{d-2} \ ,
\label{I9renc}
\end{equation}
with the convenient choice $C_d = \Gamma(2-d/2) / 2^{d-1} \pi^{d/2}$, we see 
that the associated RG beta function becomes
\begin{equation}
  \beta_v = v_R \left( d - 2 - \frac32 \, \gamma_c \right) \ . 
\label{I9betv}
\end{equation}
At {\em any} nontrivial RG fixed point $0 < v^* < \infty$, therefore
$\gamma_c^* = \frac23 (d-2)$.
We thus infer that below the upper critical dimension $d_c = 2$ for DDS, the 
longitudinal scaling exponents are fixed by the system's symmetry 
\cite{JanSch,LeuCar},
\begin{equation}
  \Delta = - \frac{\gamma_c^*}{2} = \frac{2 - d}{3} \ , 
  \quad z_\parallel = \frac{2}{1 + \Delta} = \frac{6}{5 - d} \ .
\label{I9loex}
\end{equation}

An explicit one-loop calculation for the two-point vertex functions, see 
Fig.~\ref{modelj}(b) and (c), yields  
\begin{eqnarray} 
  \gamma_c &=& - \frac{v_R}{16} \left( 3 + w_R \right) \ , \quad 
  \gamma_{\tilde c} = - \frac{v_R}{32} \left( 3 w_R^{-1} + 2 + 3 w_R \right) 
  \ , 
\label{I9gamc} \\ 
  \beta_w &=& w_R \left( \gamma_{\tilde c} - \gamma_c \right) 
  = - \frac{v_R}{32} \left( w_R - 1 \right) \left( w_R - 3 \right) \ . 
\label{I9betw}
\end{eqnarray} 
This establishes that in fact the fixed point $w^* = 1$ is IR-stable (provided
$0 < v^* < \infty$), which means that asymptotically the Einstein relation is
satisfied in the longitudinal sector as well \cite{JanSch}.

In this context, it is instructive to make an intriguing connection with the
{\em noisy Burgers equation} \cite{FoNeSt}, describing simplified fluid 
dynamics in terms of a velocity field $\vec{u}(\vec{x},t)$:
\begin{eqnarray} 
  &&\frac{\partial \vec{u}(\vec{x},t)}{\partial t} + \frac{D g}{2} \, 
  \vec{\nabla} \left[ \vec{u}(\vec{x},t)^2 \right] = D \, \vec{\nabla}^2 
  \vec{u}(\vec{x},t) + \vec{\zeta}(\vec{x},t) \ , 
\label{I9nobg} \\ 
  &&\left\langle \zeta_i \right\rangle = 0 \, , \ \left\langle 
  \zeta_i(\vec{x},t) \, \zeta_j(\vec{x}',t') \right\rangle = - 2 D \, 
  \vec{\nabla}_i \vec{\nabla}_j \, \delta(\vec{x}-\vec{x}') \, \delta(t-t')\ .
\label{I9bgno}
\end{eqnarray} 
For $D g = 1$, the nonlinearity is just the usual fluid advection term.
In one dimension, the Burgers equation (\ref{I9nobg}) becomes {\em identical} 
with the DDS Langevin equation (\ref{I9ddsl}), so we immediately infer its 
anomalous dynamic critical exponent $z_\parallel = 3/2$. 
At least in one dimension therefore, it should represent an {\em equilibrium}
system which asymptotically approaches the canonical distribution 
(\ref{I1eqpd}), where the Hamiltonian is simply the fluid's kinetic energy 
(and we have set $k_{\rm B} T = 1$).
So let us check the equilibrium condition (\ref{I6divf}) with 
${\cal P}_{\rm eq}[\vec{u}] \propto \exp \left[ - \frac12 \int \!
\vec{u}(\vec{x})^2 \, \D^dx \right]$:
\begin{eqnarray*} 
  &&\int \! \D^dx \, {\delta \over \delta \vec{u}(\vec{x},t)} \cdot \left[ 
  \vec{\nabla} \vec{u}(\vec{x},t)^2 \right] \E^{- \frac12 \int \! 
  \vec{u}(\vec{x}',t)^2 \, \D^dx'} \\ 
  &&\quad = \int \! \left[ 2 \vec{\nabla} \cdot \vec{u}(\vec{x},t) - 
  \vec{u}(\vec{x},t) \cdot \vec{\nabla} \vec{u}(\vec{x},t)^2 \right] \D^dx
  \, \E^{- \frac12 \int \! \vec{u}(\vec{x}',t)^2 \, \D^dx'} \ .  
\end{eqnarray*} 
With appropriate boundary conditions, the first term here vanishes, but the 
second one does so {\em only} in $d = 1$: 
$- \int \! u \, (\D u^2 / \D x) \, \D x = \int \! u^2 \, (\D u / \D x) \, \D x 
= \frac13 \int (\D u^3 / \D x) \, \D x = 0$. 
Driven diffusive systems in one dimension are therefore subject to a 
{\em `hidden' fluctuation--dissipation theorem}.
 
To conclude this part on Langevin dynamics, let us briefly consider the
{\em driven model B} or {\em critical DDS} \cite{SchZiaB}, which corresponds 
to a driven Ising lattice gas near its critical point.
Here, a conserved scalar field $S$ undergoes a second-order phase transition, 
but similar to the randomly driven case, again only the {\em transverse} 
sector is critical.
Upon adding the DDS drive term from Eq.~(\ref{I9ddsl}) to the Langevin 
equation (\ref{I8ttmb}), we obtain
\begin{eqnarray} 
  \frac{\partial S(\vec{x},t)}{\partial t} &=& D \left[ \vec{\nabla}_\perp^2 
  \left( r - \vec{\nabla}_\perp^2 \right) + c \, \nabla_\parallel^2 \right] 
  S(\vec{x},t) + \frac{D \, \widetilde u}{6} \, \vec{\nabla}_\perp^2 \, 
  S(\vec{x},t)^3 \nonumber \\ 
  &&+ \frac{D \, g}{2} \, \nabla_\parallel \, S(\vec{x},t)^2 
  + \zeta(\vec{x},t) \ ,
\label{I9drmb}
\end{eqnarray} 
with the (scalar) noise specified in Eq.~(\ref{I8ttnc}).
The response functional thus becomes
\begin{eqnarray} 
  {\cal A}[{\widetilde S},S] &=& \int \! \D^dx \! \int \! \D t \, 
  {\widetilde S} \Biggl[ \frac{\partial S}{\partial t} - D \left[ 
  \vec{\nabla}_\perp^2 \left( r - \vec{\nabla}_\perp^2 \right) 
  + c \, \nabla_\parallel^2\right] S \nonumber \\ 
  &&\qquad\qquad\qquad + D \left( \vec{\nabla}_\perp^2 \, {\widetilde S}
  - \frac{\widetilde u}{6} \, \vec{\nabla}_\perp^2 \, S^3 - \frac{g}{2} \, 
  \nabla_\parallel \, S^2 \right) \Biggr] \ . 
\label{I9dmbf}
\end{eqnarray} 
Power counting gives $[g^2] = \mu^{5-d}$, so the upper critical dimension here
is $d_c = 5$, and $[{\widetilde u}] = \mu^{3-d}$.
The nonlinearity $\propto {\widetilde u}$ is thus {\em irrelevant} and can be
omitted if we wish to determine the asymptotic universal scaling laws; but
recall that it is responsible for the phase transition in the system.
The remaining vertex is then proportional to $\I q_\parallel$, whence 
Eqs.~(\ref{I9ga11}) and (\ref{I9zfid}) hold for critical DDS as well, and
the transverse critical exponents are just those of the Gaussian model B,
\begin{equation} 
  \eta = 0 \ , \quad \nu = 1/2 \ , \quad z = 4 \ .
\label{I9mbtr}
\end{equation}
In addition, Galilean invariance with respect to Eq.~(\ref{I9galt}) and
therefore Eq.~(\ref{I9ginv}) hold as before. 
With the renormalised nonlinear drive strength defined similarly to 
Eq.~(\ref{I9renc}), but a different geometric constant and the scale factor 
$\mu^{d-5}$, the associated RG beta function reads
\begin{equation}
  \beta_v = v_R \left( d - 5 - \frac32 \, \gamma_c \right) \ ,
\label{I9mbbv}
\end{equation} 
which again allows us to determine the longitudinal scaling exponents to 
{\em all} orders in perturbation theory, for $d < d_c = 5$,
\begin{equation}  
  \Delta = 1 - \frac{\gamma_c^*}{2} = \frac{8-d}{3} \ , 
  \quad z_\parallel = \frac{4}{1 + \Delta} = \frac{12}{11 - d} \ .
\label{I9mblo}
\end{equation}
It is worthwhile mentioning a few marked differences to the two-temperature
model B discussed in Sec.~\ref{ssec:I8}: 
In DDS, there are obviously nonzero {\em three-point} correlations, and in
the driven critical model B the upper critical dimension is $d_c = 5$ as
opposed to $d_c = 4 - d_\parallel$ for the randomly driven version.
Notice also that the latter is characterised by nontrivial static critical 
exponents, but the kinetics is purely diffusive along the drive direction,
$z_\parallel = 2$.
Conversely for the driven model B, only the longitudinal scaling exponents are
non-Gaussian.

\section{Reaction--Diffusion Systems}
\label{sec:II}
 
We now turn our attention to stochastic interacting particle systems, whose
microscopic dynamics is defined through a (classical) master equation.
Below, we shall see how the latter can be mapped onto a stochastic 
quasi-Hamiltonian in a second-quantised bosonic operator representation
\cite{JCardy,MatGlas,TauHowLee}.
Taking the continuum on the basis of coherent-state path integrals then yields
a field theory action that may be analysed by the very same RG methods as 
described before in Secs.~\ref{ssec:I3}--\ref{ssec:I5} (for more details, see
the recent overview \cite{TauHowLee}).

\subsection{Chemical reactions and population dynamics} 
\label{ssec:II1}
 
Our goal is to study systems of `particles' $A,B,\ldots$ that propagate 
through hopping to nearest neighbors on a $d$-dimensional lattice, or via 
diffusion in the continuum. 
Upon encounter, or spontaneously, with given stochastic rates, these particles
may undergo species changes, annihilate, or produce offspring. 
At large densities, the characteristic time scales of the kinetics will be 
governed by the reaction rates, and the system is said to be 
{\em reaction-limited}. 
In contrast, at low densities, any reactions that require at least two 
particles to be in proximity will be {\em diffusion-limited}: 
the basic time scale will be set by the hopping rate or diffusion coefficient.

As a first approximation to the dynamics of such `chemical' reactions, let us
assume homogeneous mixing of each species.
We may then hope to be able to capture the kinetics in terms of {\em rate
equations} for each particle concentration or mean density.
Note that such a description neglects any spatial fluctuations and 
correlations in the system, and is therefore in character a mean-field 
approximation.
As a first illustration consider the {\em annihilation} of $k - l > 0$ 
particles of species $A$ in the {\em irreversible} $k$th-order reaction 
$k \, A \to l \, A$, with rate $\lambda$.
The corresponding rate equation employs a factorisation of the probability of
encountering $k$ particles at the same point to simply the $k$th power of the 
concentration $a(t)$,
\begin{equation}
  \dot{a}(t) = - (k - l) \, \lambda \, a(t)^k \ .
\label{II1mfan}
\end{equation}
This ordinary differential equation is readily solved, with the result
\begin{eqnarray} 
  &k = 1 \, : \ &a(t) = a(0) \, e^{- \lambda \, t} \ , 
\label{II1mfrd} \\  
  &k \geq 2 \, : \ &a(t) = \left[ a(0)^{1-k} + (k-l) (k-1) \, \lambda \, t  
  \right]^{-1 /(k-1)} \ .
\label{II1mfpd}
\end{eqnarray}
For simple `radioactive' decay ($k = 1$), we of course obtain an exponential 
time dependence, as appropriate for statistically independent events. 
For pair ($k = 2$) and higher-order ($k \geq 3$) processes, however, we find 
algebraic long-time behaviour, $a(t) \to (\lambda \, t)^{-1 / (k-1)}$, with an
amplitude that becomes independent of the initial density $a(0)$.
The absence of a characteristic time scale hints at cooperative effects, and
we have to ask if and under which circumstances correlations might 
qualitatively affect the asymptotic long-time power laws.
For according to Smoluchowski theory \cite{TauHowLee}, we would expect the 
annihilation reactions to produce {\em depletion zones} in sufficiently low 
dimensions $d \leq d_c$, which would in turn induce a considerable 
{\em slowing down} of the density decay, see Sec.~\ref{ssec:II3}. 
For {\em two-species} pair annihilation $A + B \to \emptyset$ (without 
mixing), another complication emerges, namely particle species 
{\em segregation} in dimensions for $d \leq d_s$; the regions dominated by 
either species become largely inert, and the annihilation reactions are 
confined to rather sharp {\em fronts} \cite{TauHowLee}. 
 
{\em Competition} between particle decay and production processes, e.g., in 
the reactions $A \to \emptyset$ (with rate $\kappa$), 
$A \rightleftharpoons A + A$ (with forward and back rates $\sigma$ and 
$\lambda$, respectively), leads to even richer scenarios, as can already be 
inferred from the associated rate equation
\begin{equation}
  \dot{a}(t) = (\sigma - \kappa) \, a(t) - \lambda \, a(t)^2 \ .
\label{II1mfpt}
\end{equation}
For $\sigma < \kappa$, clearly $a(t) \sim \E^{- (\kappa - \sigma)\, t} \to 0$ 
as $t \to \infty$.
The system eventually enters an {\em inactive} state, which even in the fully
stochastic model is {\em absorbing}, since once there is no particle left, no 
process whatsoever can drive the system out of the empty state again.
On the other hand, for $\sigma > \kappa$, we encounter an {\em active} state
with $a(t) \to a_\infty = (\sigma - \kappa) / \lambda$ exponentially, with
rate $\sim \sigma - \kappa$.
We have thus identified a nonequilibrium {\em continuous} phase transition at 
$\sigma_c = \kappa$. 
Indeed, as in equilibrium critical phenomena, the critical point is governed 
by characteristic power laws; for example, the asymptotic particle density 
$a_\infty \sim (\sigma - \sigma_c)^\beta$, and the critical density decay 
$a(t) \sim (\lambda \, t)^{- \alpha}$  with $\beta_0 = 1 = \alpha_0$ in the 
mean-field approximation.
The following natural questions then arise: 
What are the {\em critical exponents} once statistical fluctuations are 
properly included in the analysis? 
Can we, as in equilibrium systems, identify and characterise certain
{\em universality classes}, and which microscopic or overall, global features 
determine them and their critical dimension? 

Already the previous set of reactions may also be viewed as a (crude) model 
for the {\em population dynamics} of a single species.
In the same language, we may also formulate a stochastic version of the 
classic {\em Lotka--Volterra predator--prey competition model} \cite{Murray}:
if by themselves, the `predators' $A$ die out according to $A \to \emptyset$,
with rate $\kappa$, whereas the prey reproduce $B \to B + B$ with rate 
$\sigma$, and thus proliferate with a Malthusian population explosion.
The predators are kept alive and the prey under control through 
{\em predation}, here modeled as the reaction $A + B \to A + A$: with rate
$\lambda$, a prey is `eaten' by a predator, who simultaneously produces an
offspring. 
The coupled kinetic rate equations for this system read
\begin{equation} 
  \dot{a}(t) = \lambda \, a(t) \, b(t) - \kappa \, a(t) \ , \quad 
  \dot{b}(t) = \sigma \, b(t) - \lambda \, a(t) \, b(t) \ .
\label{II1lvre}
\end{equation}
It is straightforward to show that the quantity 
$K(t) = \lambda [a(t) + b(t)] - \sigma \ln{a(t)} - \kappa \ln{b(t)}$ is a
constant of motion for this coupled system of differential equations, i.e.,
$\dot{K}(t) = 0$.  
As a consequence, the system is governed by regular population 
{\em oscillations}, whose frequency and amplitude are fully determined by the
{\em initial} conditions.
Clearly, this is not a very realistic feature (albeit mathematically 
appealing), and moreover Eqs.~(\ref{II1lvre}) are known to be quite unstable
with respect to model modifications \cite{Murray}.
Indeed, if one includes spatial degrees of freedom and takes account of the
full stochasticity of the processes involved, the system's behaviour turns out
to be much richer \cite{MoGeTa}:
In the species coexistence phase, one encounters for sufficiently large values
of the predation rate an incessant sequence of {\em `pursuit and evasion'} 
waves that form quite complex dynamical patterns.
In finite systems, these induce {\em erratic} population oscillations whose
features are however independent of the initial configuration, but whose 
amplitude vanishes in the thermodynamic limit.
Moreover, if locally the prey `carrying capacity' is limited (corresponding to
restricting the maximum site occupation number per site on a lattice), there
appears an {\em extinction threshold} for the predator population that 
separates the absorbing state of a system filled with prey from the active
coexistence regime through a continuous phase transition \cite{MoGeTa}. 

These examples all call for a systematic approach to include stochastic
fluctuations in the mathematical description of interacting 
reaction--diffusion systems that would be conducive to the application of 
field-theoretic tools, and thus allow us to bring the powerful machinery of 
the dynamic renormalisation group to bear on these problems.
In the following, we shall describe such a general method 
\cite{Doi,GraSch,Peliti} which allows a representation of the classical master
equation in terms of a coherent-state path integral and its subsequent 
analysis by means of the RG (for overviews, see 
Refs.~\cite{JCardy,MatGlas,TauHowLee}).

\subsection{Field theory representation of master equations} 
\label{ssec:II2}
 
The above interacting particle systems, when defined on a $d$-dimensional
lattice with sites $i$, are fully characterised by the set of occupation
integer numbers $n_i = 0,1,2,\ldots$ for each particle species.
The {\em master equation} then describes the temporal evolution of the
configurational probability distribution $P(\{ n_i \};t)$ through a 
{\em balance} of gain and loss terms.
For example, for the {\em binary annihilation} and {\em coagulation reactions}
$A + A \to \emptyset$ with rate $\lambda$ and $A + A \to A$ with rate 
$\lambda'$, the master equation on a specific site $i$ reads
\begin{eqnarray} 
  \frac{\partial P(n_i;t)}{\partial t} &=& \lambda \, (n_i + 2) \, (n_i + 1) 
  \, P(\ldots,n_i+2,\ldots;t) \nonumber \\ 
  &&+ \lambda' \, (n_i + 1) \, n_i \, P(\ldots,n_i+1,\ldots;t) \nonumber \\
  &&- (\lambda + \lambda') \, n_i \, (n_i-1) \, P(\ldots,n_i,\ldots;t) \ , 
\label{II2mast}
\end{eqnarray} 
with initially $P(\{ n_i \},0) = \prod_i P(n_i)$, e.g., a {\em Poisson}
distribution $P(n_i) = {\bar n}_0^{n_i} \, e^{-\bar n_0} / n_i !$. 
Since the reactions all change the site occupation numbers by integer values,
a {\em second-quantised Fock space representation} is particularly useful 
\cite{Doi,GraSch,Peliti}.
To this end, we introduce the {\em bosonic operator algebra}
\begin{equation}  
  \Bigl[ a_i , a_j \Bigr] = 0 = \Bigl[ a_i^\dagger , a_j^\dagger \Bigr] \ ,
  \quad \Bigl[ a_i , a_j^\dagger \Bigr] = \delta_{ij} \ .
\label{II2boso}
\end{equation}
From these commutation relations one establishes in the standard manner that
$a_i$ and $a_i^\dagger$ constitute lowering and raising ladder operators, from
which we may construct the particle number eigenstates $| n_i \rangle$,
\begin{equation}
  a_i \, |n_i \rangle = n_i \, |n_i-1 \rangle \ , \quad 
  a_i^\dagger \, |n_i \rangle = |n_i + 1 \rangle \ , \quad 
  a_i^\dagger \, a_i \, |n_i \rangle = n_i \, |n_i \rangle \ .
\label{II2boss}
\end{equation}
(Notice that we have chosen a different normalisation than in ordinary quantum
mechanics.)
A state with $n_i$ particles on sites $i$ is then obtained from the empty 
vaccum state $| 0 \rangle$, defined through $a_i \, | 0 \rangle = 0$, as the 
product state
\begin{equation}
  | \{ n_i \} \rangle = \prod_i \left( a_i^\dagger \right)^{n_i} | 0 \rangle 
  \ . 
\label{II2prst}
\end{equation}

To make contact with the time-dependent configuration probability, we 
introduce the formal {\em state vector} 
\begin{equation}
  | \Phi(t) \rangle = \sum_{\{ n_i \}} P(\{ n_i \};t) \, | \{ n_i \} \rangle 
  \ ,
\label{II2stvc}
\end{equation}
whereupon the linear time evolution according to the master equation is 
translated into an {\em `imaginary-time' Schr\"odinger equation}
\begin{equation}   
  \frac{\partial | \Phi(t) \rangle}{\partial t} = - H \, | \Phi(t) \rangle \ ,
  \quad | \Phi(t) \rangle = \E^{- H \, t} \, | \Phi(0) \rangle \ .
\label{II2schr}
\end{equation}
The {\em stochastic quasi-Hamiltonian} (rather, the time evolution or 
Liouville operator) for the on-site reaction processes is a sum of local 
terms, $H_{\rm reac} = \sum_i H_i(a_i^\dagger,a_i)$; e.g., for the binary 
annihilation and coagulation reactions,
\begin{equation}  
  H_i(a_i^\dagger,a_i) = - \lambda \Bigl( 1 - {a_i^\dagger}^2 \Bigr) a_i^2 
  - \lambda' \Bigl( 1 - a_i^\dagger \Bigr) a_i^\dagger \, a_i^2 \ .
\label{II2hrea}
\end{equation} 
The two contributions for each process may be physically interpreted as 
follows: 
The first term corresponds to the actual {\em process} under consideration, 
and describes how many particles are annihilated and (re-)created in each 
reaction.
The second term gives the {\em `order'} of each reaction, i.e., the number 
operator $a_i^\dagger \, a_i$ appears to the $k$th power, but in 
normal-ordered form as ${a_i^\dagger}^k \, a_i^k$, for a $k$th-order process.
Note that the reaction Hamiltonians such as (\ref{II2hrea}) are 
{\em non-Hermitean}, reflecting the particle creations and destructions.  
In a similar manner, hopping between neighbouring sites $\langle i j \rangle$
is represented in this formalism through
\begin{equation}  
  H_{\rm diff} = D \sum_{<ij>} \Bigl( a_i^\dagger - a_j^\dagger \Bigr) 
  \Bigl( a_i - a_j \Bigr) \ .
\label{II2hdif}
\end{equation} 

Our goal is of course to compute averages with respect to the configurational
probability distribution $P(\{ n_i \};t)$; this is achieved by means of the 
{\em projection state} $\langle {\cal P} | = \langle 0 | \prod_i \E^{a_i}$, 
which satisfies $\langle {\cal P} | 0 \rangle = 1$ and 
$\langle {\cal P} | a_i^\dagger = \langle {\cal P} |$, since 
$\Bigl[ \E^{a_i} , a_j^\dagger \Bigr] = \E^{a_i} \, \delta_{ij}$.
For the desired {\em statistical averages} of observables that must be
expressible in terms of the occupation numbers $\{ n_i \}$, we then obtain 
\begin{equation}
  \langle F(t) \rangle = \sum_{\{ n_i \}} F(\{ n_i \}) \, P(\{ n_i \};t) = 
  \langle {\cal P} | \, F(\{ a_i^\dagger \, a_i \}) \, | \Phi(t) \rangle \ .
\label{II2mean}
\end{equation}
Let first us explore the consequences of {\em probability conservation}, i.e.,
$1 = \langle {\cal P} | \Phi(t) \rangle = \langle {\cal P} | \E^{- H \, t} | 
\Phi(0) \rangle$.
This requires $\langle {\cal P} | H = 0$; upon commuting $\E^{\sum_i a_i}$ 
with $H$, effectively the creation operators become shifted 
$a_i^\dagger \to 1 + a_i^\dagger$, whence this condition is fulfilled provided
$H_i(a_i^\dagger \to 1,a_i) = 0$, which is indeed satisfied by our explicit 
expressions (\ref{II2hrea}) and (\ref{II2hdif}).
By this prescription, we may also in averages replace 
$a_i^\dagger \, a_i \to a_i$, i.e., the {\em particle density} becomes 
$a(t) = \langle a_i \rangle$, and the two-point operator  
$a_i^\dagger a_i \, a_j^\dagger a_j \to a_i \, \delta_{ij} + a_i \, a_j$. 
 
In the bosonic operator representation above, we have assumed that there exist
no restrictions on the particle occupation numbers $n_i$ on each site.
If, however, there is a maximum $n_i \leq 2 s + 1$, one may instead employ a 
representation in terms of spin $s$ operators.
For example, particle exclusion systems with $n_i = 0$ or $1$ can thus be
mapped onto non-Hermitean spin $1/2$ `quantum' systems.
Specifically in one dimension, such representations in terms of integrable
spin chains have proved a fruitful tool; for overviews, see 
Refs.~\cite{AlDrHeRi,HeOrSa,Schuetz,Stinch}.
An alternative approach uses the bosonic theory, but encodes the site
occupation restrictions through appropriate exponentials in the number
operators $\E^{- a_i^\dagger a_i}$ \cite{Wijland}.

We may now follow an established route in quantum many-particle theory 
\cite{NegOrl} and proceed towards a field theory representation through 
constructing the {\em path integral} equivalent to the `Schr\"odinger' 
dynamics (\ref{II2schr}) based on {\em coherent states}, which are right 
eigenstates of the annihilation operator, 
$a_i \, |\phi_i \rangle = \phi_i \, |\phi_i \rangle$, with complex
eigenvalues $\phi_i$.
Explicitly, one finds
\begin{equation}
  |\phi_i \rangle = \exp\left( - \frac12 \, |\phi_i|^2 + \phi_i \, a_i^\dagger
  \right) | 0 \rangle \ ,
\label{II2cost}
\end{equation}
satisfying the {\em overlap} and (over-){\em completeness relations}
\begin{equation}
  \langle \phi_j |\phi_i \rangle = \exp\left( - \frac12 |\phi_i|^2 
  - \frac12 |\phi_j|^2 + \phi_j^* \, \phi_i \right) \ , \
  \int \prod_i \frac{\D^2 \phi_i}{\pi} \, |\{ \phi_i \} \rangle \, 
  \langle \{ \phi_i \}| = 1 \ .
\label{II2ovcp}
\end{equation}
Upon splitting the temporal evolution (\ref{II2schr}) into infinitesimal 
steps, and inserting Eqs.~(\ref{II2ovcp}) at each time step, standard 
procedures (elaborated in detail in Ref.~\cite{TauHowLee}) yield eventually
\begin{equation} 
  \langle F(t) \rangle \propto \int \prod_i {\cal D}[\phi_i] \, 
  {\cal D}[\phi_i^*] \, F(\{ \phi_i \}) \, \E^{- {\cal A}[\phi_i^*,\phi_i]} 
  \ ,
\label{II2copi}
\end{equation}
with the {\em action}
\begin{equation}
  {\cal A}[\phi_i^*,\phi_i] = \sum_i \biggl( - \phi_i(t_f) 
  + \int_0^{t_f} \! \D t \left[ \phi_i^* \, \frac{\partial \phi_i}{\partial t}
  + H_i(\phi_i^*,\phi_i) \right] - {\bar n}_0 \, \phi^*_i(0) \biggr) \ ,
\label{II2cact} 
\end{equation}
where the first term originates from the projection state, and the last one
from the initial Poisson distribution.
Notice that in the Hamiltonian, the creation and annihilation operators 
$a_i^\dagger$ and $a_i$ are simply replaced with the complex numbers
$\phi_i^*$ and $\phi_i$, respectively.

Taking the {\em continuum limit}, $\phi_i(t) \to \psi(\vec{x},t)$,  
$\phi_i^*(t) \to {\hat \psi}(\vec{x},t)$, the `bulk' part of the action
becomes
\begin{equation} 
  {\cal A}[{\hat \psi},\psi] = \int \! \D^dx \! \int \! \D t \left[  
  {\hat \psi} \left( \frac{\partial}{\partial t} - D \, \vec{\nabla}^2 \right)
  \psi + {\cal H}_{\rm reac}({\hat \psi},\psi) \right] \ ,  
\label{II2maft} 
\end{equation}
where the hopping term (\ref{II2hdif}) has naturally turned into a diffusion 
propagator.
We have thus arrived at a {\em microscopic} stochastic field theory for
reaction--diffusion processes, with {\em no} assumptions whatsoever on the
form of the (internal) noise.
This is a crucial ingredient for nonequilibrium dynamics, and we may now use
Eq.~(\ref{II2maft}) as a basis for systematic coarse-graining and the 
renormalisation group analysis. 
Returning to our example of pair annihilation and coagulation, the reaction
part of the action (\ref{II2maft}) reads
\begin{equation}
  {\cal H}_{\rm reac}({\hat \psi},\psi) = - \lambda \left( 1 - {\hat \psi}^2
  \right) \psi^2 - \lambda' \left( 1 - {\hat \psi} \right) {\hat \psi} \,  
  \psi^2 \ ,
\label{II2anco}
\end{equation}
see Eq.~(\ref{II2hrea}). 
Let us have a look at the {\em classical field equations}, namely  
$\delta {\cal A} / \delta \psi = 0$, which is always solved by 
${\hat \psi} = 1$, reflecting probability conservation, and 
$\delta {\cal A} / \delta {\hat \psi} = 0$, which, upon inserting 
${\hat \psi} = 1$ gives here
\begin{equation}  
  \frac{\partial \psi(\vec{x},t)}{\partial t} = D \, \vec{\nabla}^2 \, 
  \psi(\vec{x},t) - (2 \lambda + \lambda') \, \psi(\vec{x},t)^2 \ ,
\label{II2dreq}
\end{equation}
i.e., essentially the mean-field rate equation for the local particle density
$\psi(\vec{x},t)$, see Eq.~(\ref{II1mfan}), supplemented with diffusion.
The field theory action (\ref{II2maft}), derived from the master equation 
(\ref{II2mast}), then provides a means of including fluctuations in our
analysis.

Before we proceed with this program, it is instructive to perform a shift in
the field ${\hat \psi}$ about the mean-field solution,  
${\hat \psi}(x,t) = 1 + {\widetilde \psi}(x,t)$, whereupon the reaction
Hamiltonian density (\ref{II2anco}) becomes
\begin{equation}
  {\cal H}_{\rm reac}({\widetilde \psi},\psi) = (2 \lambda + \lambda') \, 
  {\widetilde \psi} \, \psi^2 + (\lambda + \lambda') \, {\widetilde \psi}^2 \,
  \psi^2 \ . 
\label{II2shfa} 
\end{equation}
In addition to the diffusion propagator, the annihilation and coagulation
processes thus give {\em identical} three- and four-point vertices; aside 
from non-universal amplitudes, one should therefore obtain {\em identical} 
scaling behaviour for both binary reactions in the asymptotic long-time limit
\cite{PelRen}.
Lastly, we remark that if we interpret the action 
${\cal A}[{\widetilde \psi},\psi]$ as a response functional (\ref{I2jade}),
despite the fields ${\widetilde \psi}$ not being purely imaginary, our field
theory becomes formally equivalent to a `Langevin' equation, wherein additive 
noise is added to Eq.~(\ref{II2dreq}), albeit with {\em negative} correlator 
$L[\psi] = - (\lambda + \lambda') \, \psi^2$, which represents 
{\em `imaginary' multiplicative noise}. 
This Langevin description is thus not well-defined; however, one may render
the noise correlator positive through a nonlinear {\em Cole--Hopf 
transformation} ${\widetilde \psi} = \E^{\tilde \rho}$, 
$\psi = \E^{- \tilde \rho} \rho$ such that ${\widetilde \psi} \, \psi = \rho$,
with Jacobian $1$, but at the expense of `diffusion noise' 
$\propto D \, \rho \, (\vec{\nabla} {\tilde \rho})^2$ in the action 
\cite{Hannes}.
In summary, binary (and higher-order) annihilation and coagulation processes
cannot be cast into a Langevin framework in any simple manner.

\subsection{Diffusion-limited single-species annihilation processes} 
\label{ssec:II3}
 
We begin by analysing diffusion-limited single-species annihilation 
$k \, A \to \emptyset$ \cite{PelRen,Lee}.
The corresponding field theory action (\ref{II2maft}) reads
\begin{equation}
  {\cal A}[{\hat \psi},\psi] = \int \! \D^dx \! \int \! \D t  
  \left[ {\hat \psi} \left( \frac{\partial}{\partial t} - D \vec{\nabla}^2 
  \right) \psi - \lambda \left( 1 - {\hat \psi}^k \right) \psi^k \right] \ ,
\label{II3anac}
\end{equation}
which for $k \geq 3$ allows no (obvious) equivalent Langevin description.  
Straightforward power counting gives the scaling dimension for the 
annihilation rate, $[\lambda] = \mu^{2 - (k-1) d}$, which suggests the upper
critical dimension $d_c(k) = 2/(k-1)$.
Thus we expect mean-field behaviour $\sim (\lambda \, t)^{-1 / (k-1)}$, see 
Eq.~(\ref{II1mfpd}), in any physical dimension for $k > 3$, logarithmic 
corrections at $d_c = 1$ for $k = 3$ and at $d_c = 2$ for $k = 2$, and 
nonclassical power laws for pair annihilation only in one dimension. 
The field theory defined by the action (\ref{II3anac}) has two vertices, the
`annihilation' sink with $k$ incoming lines only, and the `scattering' vertex
with $k$ incoming and $k$ outgoing lines.
Neither allows for propagator renormalisation, hence the model remains 
massless with {\em exact} scaling exponents $\eta = 0$ and $z = 2$, i.e.,
diffusive dynamics.

\begin{figure} 
\centering 
\includegraphics[width = 10 truecm]{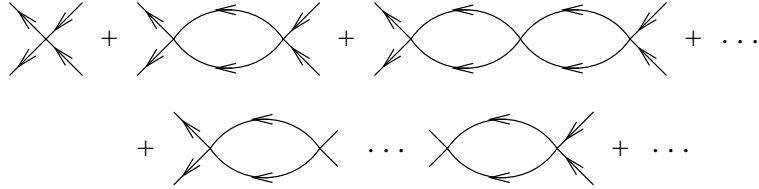} 
\caption{Vertex renormalisation for diffusion-limited binary annihilation 
  $A + A \to \emptyset$.} 
\label{versum} 
\end{figure} 
In addition, the entire perturbation expansion for the renormalisation for the
annihilation vertices is merely a geometric series of the one-loop diagram, 
see Fig.~\ref{versum} for the pair annihilation case ($k = 2$).
If we define the renormalised effective coupling according to  
\begin{equation}
  g_R = Z_g \, \frac{\lambda}{D} \, B_{kd} \, \mu^{- 2 (1 - d / d_c)} \ ,
\label{II3renc}
\end{equation}
where $B_{kd} = k! \, \Gamma(2 - d/d_c) \, d_c / k^{d/2} \, (4 \pi)^{d/d_c}$,
we obtain for the single nontrivial renormalisation constant
\begin{equation}
  Z_g^{-1} = 1 + \frac{\lambda \, B_{kd} \, \mu^{- 2 (1 - d / d_c)}} 
  {D \, (d_c - d)}
\label{II3zetg}
\end{equation}
to {\em all} orders.
Consequently, the associated RG beta function becomes
\begin{equation}
  \beta_g = \mu \, \frac{\partial}{\partial \mu} \bigg\vert_0 \, g_R =  
  - \frac{2 \, g_R}{d_c} \left( d - d_c + g_R \right) \ ,
\label{II3betg}
\end{equation}
with the Gaussian fixed point $g_0^* = 0$ stable for $d > d_c(k) = 2/(k-1)$,
leading to the mean-field power laws (\ref{II1mfpd}), whereas for $d < d_c(k)$
the flow approaches
\begin{equation}  
  g^* = d_c(k) - d \ .
\label{II3gfpt}  
\end{equation}

Since the particle density has scaling dimension $[a] = \mu^d$, we may write
$a_R(\mu,D_R,n_0,g_R) = \mu^d \, {\hat a}_R(D_R,n_0 \, \mu^{-d}, g_R)$, where 
we have retained the dependence on the initial density $n_0$.
Since the fields and the diffusion constant do not renormalise 
($\gamma_D = 0$ and $\gamma_{n_0} = - d$), the RG equation for the density 
takes the form
\begin{equation}
  \left[ d - d \, n_0 \, \frac{\partial}{\partial n_0} 
  + \beta_g \, \frac{\partial}{\partial g_R} \right] 
  {\hat a}_R\left( n_0 \, \mu^{-d} , g_R \right) = 0 \ , 
\label{II3rgeq}
\end{equation}
see Eq.~(\ref{I5rgsu}).
With the characteristics set equal to $\mu \, \ell = (D \, t)^{-1/2}$, the 
solution of the RG equation (\ref{II3rgeq}) near the IR-stable RG fixed point
$g^*$ becomes
\begin{equation}
  a_R(n_0,t) \sim (D \mu^2 \, t)^{- d/2} \, 
  {\hat a}_R\Bigl( n_0 \, (D \mu^2 \, t)^{d/2} , g^* \Bigr) \ .
\label{II3drgs}
\end{equation}
Under the RG, the first argument in Eq.~(\ref{II3drgs}) flows to infinity.
One therefore needs to establish that the result for the scaling function
${\hat a}$ is finite to {\em all} orders in the initial density 
\cite{Lee,TauHowLee}.
One then finds the following asymptotic long-time behaviour for pair
annihilation below the critical dimension \cite{PelRen,Lee},
\begin{equation}   
  k = 2 \, , \ d < 2 \, : \ a(t) \sim (D \, t)^{-d/2} \ . 
\label{II3pddc}
\end{equation}
At the critical dimension, ${\tilde g}(\ell) \to 0$ logarithmically slowly, 
and the process is still diffusion-limited; this gives
\begin{eqnarray}
  k = 2 \, , \ d = 2 \, : \ &&a(t) \sim (D \, t)^{-1} \ln (D \, t) \ , 
\label{II3pdlc} \\ 
  k = 3 \, , \ d = 1 \, : \ &&a(t) \sim \left[ (D \, t)^{-1} \ln (D \, t) 
  \right]^{1/2} \ .
\label{II3tdlc} 
\end{eqnarray}

\subsection{Segregation for multi-species pair annihilation} 
\label{ssec:II4}
 
In pair annihilation reactions of two {\em distinct} species 
$A + B \to \emptyset$, where {\em no} reactions between the same species are 
allowed, a novel phenomenon emerges in sufficiently low dimensions
$d \leq d_s$, namely particle segregation in separate spatial domains, with
the decay processes restricted to sharp reaction fronts on their boundaries
\cite{LeeCar}.
Note that the reaction $A + B \to \emptyset$ preserves the difference of
particle numbers (even locally), i.e., there is a {\em conservation law} for
$c(t) = a(t) - b(t) = c(0)$ \cite{TouWil}.
The rate equations for the concentrations
\begin{equation}
  \dot{a}(t) = - \lambda \, a(t) \, b(t) = \dot{b}(t) 
\label{II4mfre}
\end{equation}
are for {\em equal} initial densities $a(0) = b(0)$ solved by the 
single-species pair annihilation mean-field power law
\begin{equation}  
  a(t) = b(t) \sim (\lambda \, t)^{-1} \ ,
\label{II4edms}
\end{equation}
whereas for {\em unequal} initial densities $c(0) = a(0) - b(0) > 0$, say, the
majority species $a(t) \to a_\infty = c(0) > 0$ as $t \to \infty$, and the 
minority population disappears, $b(t) \to 0$. 
From Eqs.~(\ref{II4mfre}) we obtain for $d > d_c = 2$ the exponential approach
\begin{equation}
  a(t) - a_\infty \sim b(t) \sim \E^{- c(0) \, \lambda \, t} \ .
\label{II4udms}
\end{equation}

Mapping the associated master equation onto a continuum field theory 
(\ref{II2maft}), the reaction term now reads (with the fields $\psi$ and 
$\varphi$ representing the $A$ and $B$ particles, respectively) 
\cite{LeeJohn}
\begin{equation}
  {\cal H}_{\rm reac}({\hat \psi},\psi,{\hat \varphi},\varphi) = - \lambda
  \left( 1 - {\hat \psi} \, {\hat \varphi} \right) \psi \, \varphi  \ .
\label{II4abft}
\end{equation}
As in the single-species case, there is no propagator renormalisation, and
moreover the Feynman diagrams for the renormalised reaction vertex are of
precisely the same form as for $A + A \to \emptyset$, see Fig.~\ref{versum}.
Thus, for unequal initial densities, $c(0) > 0$, the mean-field power law
$\sim \lambda \, t$ in the exponent of Eq.~(\ref{II4udms}) becomes again 
replaced with $(D \, t)^{d/2}$ in dimensions $d \leq d_c = 2$, leading to
{\em stretched exponential} time dependence,
\begin{equation} 
  d < 2 \, : \ \ln b(t) \sim -t^{d/2} \, , \quad 
  d = 2 \, : \ \ln b(t) \sim - t / \ln (Dt) \ . 
\label{II4stre}
\end{equation}

However, species segregation for {\em equal} initial densities, $a(0) = b(0)$,
even supersedes the slowing down due to the reaction rate renormalisation.
As confirmed by a thorough RG analysis, this effect can be captured within 
the classical field equations \cite{LeeJohn}.
To this end, we add {\em diffusion} terms (with equal diffusivities) to the 
rate equations (\ref{II4mfre}) for the now {\em local} particle densities,
\begin{equation}
  \left( \frac{\partial}{\partial t} - D \, \vec{\nabla}^2 \right) 
  a(\vec{x},t) = - \lambda \, a(\vec{x},t) \, b(\vec{x},t) = \left( 
  \frac{\partial}{\partial t} - D \, \vec{\nabla}^2 \right) b(\vec{x},t) \ .
\label{II4rdab}
\end{equation}
The local concentration difference $c(\vec{x},t)$ thus becomes a purely
{\em diffusive mode}, 
$\partial_t \, c(\vec{x},t) = D \, \vec{\nabla}^2 c(\vec{x},t)$, and we employ
the {\em diffusion Green function}
\begin{equation}
  G_0(\vec{q},t) = \Theta(t) \, e^{- D \, \vec{q}^2 \, t} \ , \quad   
  G_0(\vec{x},t) = \frac{\Theta(t)}{(4 \pi \, D t)^{d/2}} \, 
  \E^{- \vec{x}^2 / 4 D t} \ , 
\label{II4gref}
\end{equation}
compare Eq.~(\ref{I1rest}), to solve the initial value problem,
\begin{equation}
  c(\vec{x},t) = \int \! \D^dx' \, G_0(\vec{x} - \vec{x}',t) \, c(\vec{x}',0)
  \ .  
\label{II4iniv}
\end{equation} 
Let us furthermore assume a {\em Poisson distribution} for the {\em initial}
density correlations (indicated by an overbar), 
$\overline{a(\vec{x},0) \, a(\vec{x}',0)} = a(0)^2 + a(0) \, 
\delta(\vec{x} - \vec{x}') = \overline{b(\vec{x},0) \, b(\vec{x}',0)}$ and 
$\overline{a(\vec{x},0) \, b(\vec{x}'(0)} = a(0)^2$, which implies 
$\overline{c(\vec{x},0) \, c(\vec{x}',0)} = 2 \, a(0) \, 
\delta(\vec{x} - \vec{x}')$.
Averaging over the initial conditions then yields with Eq.~(\ref{II4iniv}) 
\begin{equation}  
  \overline{c(\vec{x},t)^2} = 2 \, a(0) \int \! \D^dx' \, 
  G_0(\vec{x} - \vec{x}',t)^2 = 2 \, a(0) \, (8 \pi \, D t)^{-d/2} \ ;
\label{II4inav}
\end{equation} 
since the distribution for $c$ will be a Gaussian, we thus obtain for the
{\em local density excess} originating in a random initial fluctuation,
\begin{equation}
  \overline{|c(\vec{x},t)|} = \sqrt{\frac{2}{\pi}\, \overline{c(\vec{x},t)^2}}
  = 2 \, \sqrt{\frac{a(0)}{\pi}} \, (8 \pi \, D t)^{-d/4} \ . 
\label{II4segd}
\end{equation} 
In dimensions $d < d_s = 4$ these density fluctuations decay {\em slower} than
the overall particle number $\sim t^{-1}$ for $d > 2$ and $\sim t^{- d/2}$ for
$d < 2$ in a homogeneous system.
Species segregation into $A$- and $B$-rich domains renders the particle 
distribution {\em nonuniform}, and the density decay is governed by the slow
power law (\ref{II4segd}), $a(t) \sim b(t) \sim (D t)^{-d/4}$.

For very {\em special} initial states, however, the situation can be 
different.
For example, consider hard-core particles (or $\lambda \to \infty$) regularly
arranged in an alternating manner $\ldots ABABABABAB \ldots$ on a 
one-dimensional chain.
The reactions $A + B \to \emptyset$ preserve this arrangement, whence the
distinction between $A$ and $B$ particles becomes meaningless, and one indeed
recovers the $t^{-1/2}$ power law from the single-species pair annihilation 
reaction. 

Let us at last generalise to {\em $q$-species} annihilation 
$A_i + A_j \to \emptyset$, with $1 \leq i < j \leq q$, with equal initial
densities $a_i(0)$ as well as uniform diffusion and reaction rates.
For $q > 2$, there exists {\em no} conservation law in the stochastic system,
and one may argue, based on the study of fluctuations in the associated
Fokker--Planck equation, that segregation happens only for 
$d < d_s(q) = 4 / (q-1)$ \cite{DelHilTau}.
In any physical dimension $d \geq 2$, one should therefore see the same
behaviour as for the single-species reaction $A + A \to \emptyset$; this is
actually obvious for $q = \infty$, since in this case the probability for 
particles of the same species to ever meet is zero, whence the species 
labeling becomes irrelevant.
In one dimension, with its special topology, segregation does occur, and for
generic initial conditions one finds the the decay law \cite{DelHilTau}
\begin{equation}
  a_i(t) \sim t^{- \alpha(q)} + C \, t^{-1/2} \ , \quad  
  \alpha(q) = \frac{q - 1}{2 \, q} \ ,
\label{II4msan}
\end{equation}
which recovers $\alpha(2) = 1/4$ and $\alpha(\infty) = 1/2$.
Once again, in special situations, e.g., the alignment 
$\ldots ABCDABCDABCD \ldots$ for $q = 4$, the single-species scaling ensues.
There are also curious {\em cyclic} variants, for example if for four species
we only allow the reactions $A + B \to \emptyset$, $B + C \to \emptyset$,
$C + D \to \emptyset$, and $D + A \to \emptyset$.
We may then obviously identify $A = C$ and $B = D$, which leads back to the
case of two-species pair annihilation.  
Generally, within essentially mean-field theory one finds for cyclic
multi-species annihilation processes $a_i(t) \sim t^{- \alpha(q,d)}$, where 
for
\begin{equation}  
  2 < d_s(q) = \left\{ \begin{array}{cc} 4 & \quad q = 2,4,6,\ldots \\  
  4 \cos(\pi / q) & \quad q = 3,5,7,\ldots \end{array} \right. \, : \ 
  \alpha(q,d) = d / d_s(q) \ .
\label{II4cycd}
\end{equation} 
Remarkably, for five species this yields the borderline dimension 
$d_s(5) = 1 + \sqrt{5}$ for segregation to occur, hence nontrivial decay 
exponents $\alpha(5,2) = \frac12 (\sqrt{5} - 1)$ in $d = 2$ and 
$\alpha(5,3) = \frac34 (\sqrt{5} - 1)$ in $d = 3$ that involve the golden 
ratio \cite{HilWasTau}.

\subsection{Active to absorbing state transitions and directed percolation} 
\label{ssec:II5}
 
Let us now return to the {\em competing} single-species reactions 
$A \to \emptyset$ (rate $\kappa$), $A \to A + A$ (rate $\sigma$), and, in 
order to limit the particle density in the active phase, $A + A \to A$ 
(rate $\lambda$).
Adding diffusion to the rate equation (\ref{II1mfpt}), we arrive at the 
{\em Fisher--Kolmogorov equation} of biology and ecology \cite{Murray}, 
\begin{equation}
  \frac{\partial a(\vec{x},t)}{\partial t} = - D \left( r - \vec{\nabla}^2 
  \right) a(\vec{x},t) - \lambda \, a(\vec{x},t)^2 \ ,
\label{II5fiko}
\end{equation}
where $r = (\kappa - \sigma) / D$.
As discussed in Sec.~\ref{ssec:II1}, it predicts a {\em continuous} transition
from an active to an inactive, {\em absorbing} state to occur at $r = 0$. 
If we define the associated {\em critical exponents} in close analogy to 
equilibrium critical phenomena, see Sec.~(\ref{ssec:I1}), the partial 
differential equation (\ref{II5fiko}) yields the Gaussian exponent values 
$\eta_ 0 = 0$, $\nu_0 = 1/2$, $z_0 = 2$, and $\alpha_0 = 1 = \beta_0$. 
 
By the methods outlined in Sec.~\ref{ssec:II2}, we may construct the  
coherent-state path integral (\ref{II2maft}) for the associated master 
equation,
\begin{eqnarray} 
  {\cal A}[{\hat \psi},\psi] &=& \int \! \D^dx \! \int \! \D t \, \biggl[  
  {\hat \psi} \left( \frac{\partial}{\partial t} - D \, \vec{\nabla}^2 \right)
  \psi - \kappa \left( 1 - {\hat \psi} \right) \psi \nonumber \\ 
  &&\qquad\qquad\quad + \sigma \left( 1 - {\hat \psi} \right) {\hat \psi} \, 
  \psi - \lambda \left( 1 - {\hat \psi} \right) {\hat \psi} \, \psi^2 \biggr] 
  \ .
\label{II5mase} 
\end{eqnarray} 
Upon shifting the field ${\hat \psi}$ about its stationary value $1$ and 
rescaling according to ${\hat \psi}(\vec{x},t) = 1 + \sqrt{\sigma / \lambda} 
\, {\widetilde S}(\vec{x},t)$ and 
$\psi(\vec{x},t) = \sqrt{\lambda / \sigma} \, S(\vec{x},t)$, the action 
becomes 
\begin{equation}  
  {\cal A}[{\widetilde S},S] = \int \! \D^dx \! \int \! \D t \,
  \left[ {\widetilde S} \left( \frac{\partial}{\partial t} + D \left( 
  r - \vec{\nabla}^2 \right) \right) S - u \left( {\widetilde S} - S \right) 
  {\widetilde S} \, S + \lambda \, {\widetilde S}^2 \, S^2 \right] \ .
\label{II5regf} 
\end{equation} 
Thus, the three-point vertices have been scaled to identical coupling
strengths $u = \sqrt{\sigma \, \lambda}$, which represents the effective 
coupling of the perturbation expansion, see Fig.~\ref{dploop} below.
Its scaling dimension is $[u] = \mu^{2-d/2}$, whence we infer the upper 
critical dimension $d_c = 4$.
The four-point vertex $\propto \lambda$, with $[\lambda] = \mu^{2-d}$, is
thus {\em irrelevant} in the RG sense, and can be dropped for the computation
of universal, asymptotic scaling properties. 

\begin{figure} 
\centering
\includegraphics[width = 4.4 truecm]{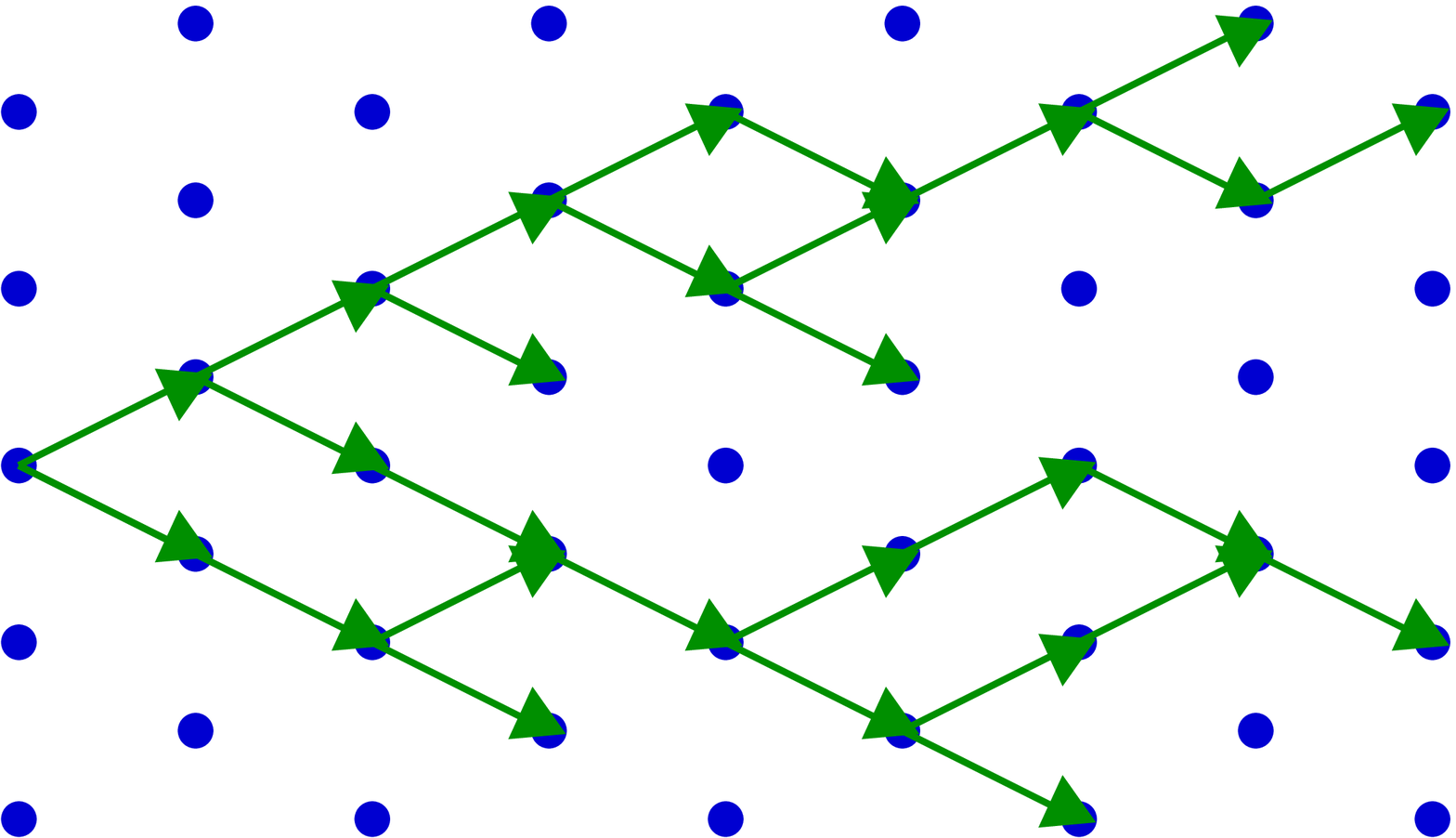} \quad $t \to$
\includegraphics[width = 6.3 truecm]{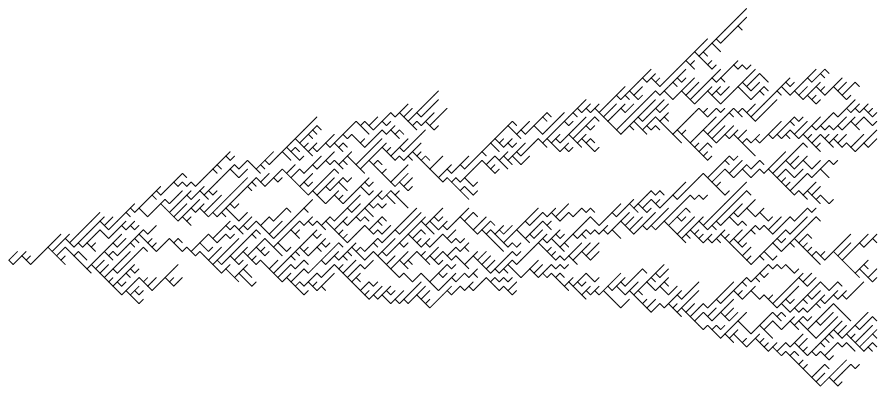} 
\caption{Directed percolation process (left) and critical DP cluster (right).}
\label{dirpcl} 
\end{figure}

The action (\ref{II5regf}) with $\lambda = 0$ is known as {\em Reggeon field 
theory} \cite{Moshe}, and its basic characteristic is its invariance under 
{\em rapidity inversion} 
$S(\vec{x},t) \leftrightarrow - {\widetilde S}(\vec{x},-t)$.
If we interpret Eq.~(\ref{II5regf}) as a response functional, we see that it
becomes formally equivalent to a stochastic process with multiplicative 
noise ($\langle \zeta(\vec{x},t) \rangle = 0$) captured by the Langevin 
equation \cite{GraSun,GraTor}
\begin{eqnarray} 
  &&\frac{\partial S(\vec{x},t)}{\partial t} = - D \left( r - \vec{\nabla}^2 
  \right) S(\vec{x},t) - u \, S(\vec{x},t)^2 + \zeta(\vec{x},t) \ , 
\label{II5dple}\\ 
  &&\langle \zeta(\vec{x},t) \, \zeta(\vec{x}',t') \rangle = 2 u \, 
  S(\vec{x},t) \, \delta(\vec{x} - \vec{x}') \, \delta(t - t')
\label{II5dpns}
\end{eqnarray} 
(for a more accurate mapping procedure, see Ref.~\cite{JanTau}), which is 
essentially a noisy Fisher--Kolmogorov equation (\ref{II5fiko}), with the 
noise correlator (\ref{II5dpns}) ensuring that the fluctuations indeed cease 
in the absorbing state where $\langle S \rangle = 0$.
It has moreover been established \cite{Obuk,CarSug,JanDP} that the action 
(\ref{II5regf}) describes the scaling properties of critical {\em directed 
percolation (DP)} clusters \cite{Kinzel}, illustrated in Fig.~\ref{dirpcl}.
Indeed, if the DP growth direction is labeled as `time' $t$, we see that the
structure of the DP clusters emerges from the basic decay, branching, and 
coagulation reactions encoded in Eq.~(\ref{II5mase}).
 
In fact, the field theory action should govern the scaling properties of
{\em generic} continuous nonequilibrium phase transitions from active to 
inactive, absorbing states, namely for an order parameter with Markovian
stochastic dynamics that is decoupled from any other slow variable, and in
the absence of quenched randomness \cite{JanDP,Grassb}.
This {\em DP conjecture} follows from the following {\em phenomenological} 
approach \cite{JanTau} to {\em simple epidemic processes (SEP)}, or epidemics 
with recovery \cite{Murray}: 
\begin{enumerate} 
\item A {\em `susceptible' medium} becomes locally {\em `infected'}, depending
  on the density $n$ of neighboring `sick' individuals. 
  The infected regions recover after a brief time interval. 
\item The state $n = 0$ is {\em absorbing}. 
  It describes the {\em extinction} of the `disease'. 
\item The disease spreads out {\em diffusively} via the short-range infection
  1. of neighboring susceptible regions. 
\item Microscopic fast degrees of freedom are incorporated as {\em local 
  noise} or stochastic forces that respect statement 2., i.e., the noise alone
  cannot regenerate the disease. 
\end{enumerate} 
These ingredients are captured by the coarse-grained mesoscopic Langevin 
equation $\partial_t \, n = D \left( \vec{\nabla}^2 - R[n] \right) n + \zeta$
with a reaction functional $R[n]$, and s stochastic noise correlator of the
form $L[n] = n \, N[n]$.
Near the extinction threshold, we may expand $R[n] = r + u \, n \, + \ldots$,
$N[n] = v + \ldots$, and higher-order terms turn out to be {\em irrelevant} in
the RG sense.
Upon rescaling, we recover the Reggeon field theory action (\ref{II5regf}) for
DP as the corresponding response functional (\ref{I2jade}). 

We now proceed to an explicit evaluation of the DP critical exponents to 
one-loop order, closely following the recipes given in 
Secs.~\ref{ssec:I3}--\ref{ssec:I5}.
The lowest-order fluctuation contribution to the two-point vertex function 
$\Gamma^{(1,1)}(\vec{q},\omega)$ (propagator self-energy) is depicted in 
Fig.~\ref{dploop}(a).
\begin{figure} 
\centering 
\includegraphics[width = 5 truecm]{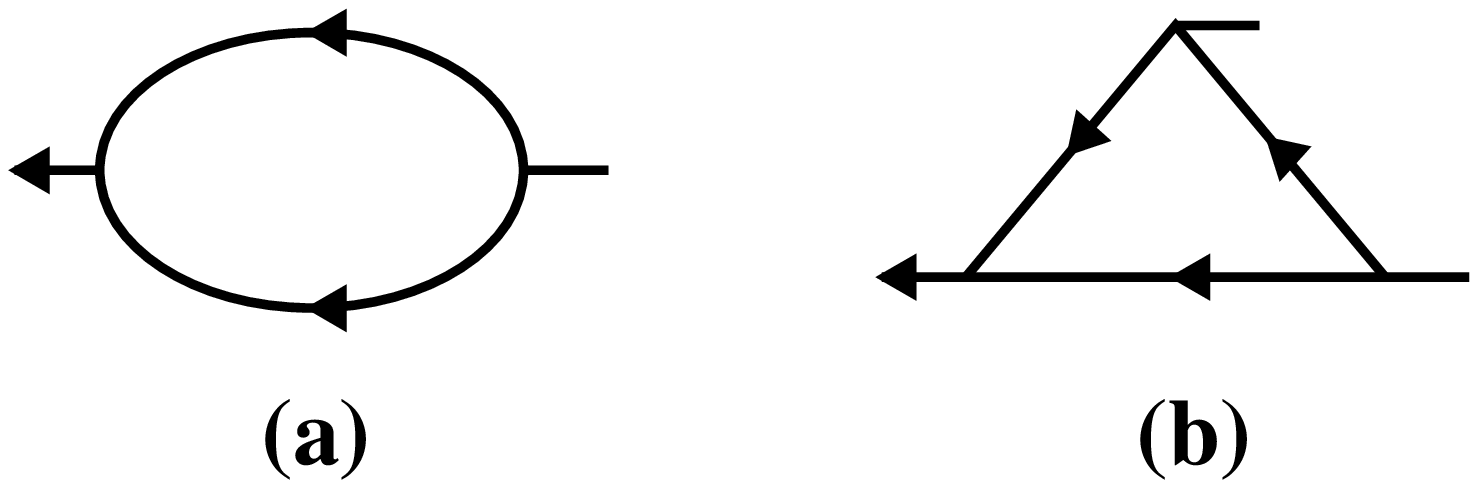} 
\caption{DP renormalisation: one-loop diagrams for the vertex functions  
  (a) $\Gamma^{(1,1)}$ (propagator self-energy), and 
  (b) $\Gamma^{(1,2)} = - \Gamma^{(2,1)}$ (nonlinear vertices).} 
\label{dploop} 
\end{figure} 
The Feynman rules of Sec.~\ref{ssec:I3} yield the corresponding analytic 
expression
\begin{equation}
  \Gamma^{(1,1)}(\vec{q},\omega) = \I \omega + D \, (r + \vec{q}^2) + 
  \frac{u^2}{D} \int_k \ \frac{1}{\I \omega/2 D + r + \vec{q}^2/4 + \vec{k}^2}
  \ .
\label{II5ga11}
\end{equation}
The {\em criticality condition} $\Gamma^{(1,1)}(0,0) = 0$ at $r = r_c$
provides us with the fluctuation-induced shift of the percolation threshold  
\begin{equation}
  r_c = - \frac{u^2}{D^2} \int_k^\Lambda \frac{1}{r_c + \vec{k}^2} + O(u^4)\ .
\label{II5pcsh}
\end{equation} 
Inserting $\tau = r - r_c$ into Eq.~(\ref{II5ga11}), we then find to this 
order
\begin{equation}
  \Gamma^{(1,1)}(\vec{q},\omega) = \I \omega + D \, (\tau + \vec{q}^2) 
  - \frac{u^2}{D} \int_k \frac{\I \omega / 2 D + \tau + \vec{q}^2/4}{\vec{k}^2
  \left( \I \omega / 2 D + \tau + \vec{q}^2/4 + \vec{k}^2 \right)} \ ,
\label{II5gm11}
\end{equation}
and the diagram in Fig.~\ref{dploop}(b) for the three-point vertex functions, 
evaluated at zero external wavevectors and frequencies, gives  
\begin{equation}
  \Gamma^{(1,2)}(\{ {\underline 0} \}) = -\Gamma^{(2,1)}(\{ {\underline 0} \})
  = - 2 u \left( 1 - \frac{2 u^2}{D^2} \int_k \frac{1}{\left( \tau + \vec{k}^2
  \right)^2} \right) \ .
\label{II5gm12} 
\end{equation}

For the renormalisation factors, we use again the conventions 
(\ref{I4fren}) and (\ref{I4pren}), but with
\begin{equation}
  u_R = Z_u \, u \, A_d^{1/2} \, \mu^{(d-4)/2} \ .
\label{II5uren}
\end{equation}
Because of rapidity inversion invariance, $Z_{\widetilde S} = Z_S$.
With Eq.~(\ref{I4dimr}) the derivatives of $\Gamma^{(1,1)}$ with respect to 
$\omega$, $\vec{q}^2$, and $\tau$, as well as the one-loop result for
$\Gamma^{(1,2)}$ in Eq.~(\ref{II5gm12}), all evaluated at the normalisation 
point $\tau_R = 1$, provide us with the $Z$ factors
\begin{eqnarray}  
  &&Z_S = 1 - \frac{u^2}{2 D^2} \, \frac{A_d \, \mu^{-\epsilon}}{\epsilon} \ ,
  \quad Z_D = 1 + \frac{u^2}{4 D^2} \, 
  \frac{A_d \, \mu^{-\epsilon}}{\epsilon} \ , \nonumber \\ 
  &&Z_\tau = 1 - \frac{3u^2}{4 D^2} \, \frac{A_d \, \mu^{-\epsilon}}{\epsilon}
  \ , \quad Z_u = 1 - \frac{5 u^2}{4 D^2} \,  
  \frac{A_d \, \mu^{-\epsilon}}{\epsilon} \ .
\label{II5dpzf}
\end{eqnarray}
From these we infer the RG flow functions
\begin{equation} 
  \gamma_S = v_R / 2 \ , \quad \gamma_D = - v_R / 4 \ , \quad 
  \gamma_\tau  = - 2 + 3 v_R / 4 \ ,
\label{II5dpwf}
\end{equation}
with the renormalised effective coupling
\begin{equation}
  v_R = \frac{Z_u^2}{Z_D^2} \, \frac{u^2}{D^2} \, A_d \, \mu^{d-4} \ ,
\label{II5dpcp}
\end{equation}
whose RG beta function is to this order
\begin{equation}
 \beta_v = v_R \left[ - \epsilon + 3 v_R + O(v_R^2) \right] \ . 
\label{II5dpbf}
\end{equation}
For $d > d_c = 4$, the Gaussian fixed point $v_0^* = 0$ is stable, and we 
recover the mean-field critical exponents.
For $\epsilon = 4 - d > 0$, we find the nontrivial IR-stable RG fixed point 
\begin{equation}
  v^* = \epsilon / 3 + O(\epsilon^2) \ .
\label{II5dpfp}
\end{equation} 

Setting up and solving the RG equation (\ref{I5rgeq}) for the vertex function 
proceeds just as in Sec.~\ref{ssec:I5}. 
With the identifications (\ref{I5crex}) (with $a = 0$) we thus obtain the DP 
{\em critical exponents} to first order in $\epsilon$, 
\begin{equation}  
  \eta = - \frac{\epsilon}{6} + O(\epsilon^2) \ , \quad 
  \frac{1}{\nu} = 2 - \frac{\epsilon}{4} + O(\epsilon^2) \ , \quad 
  z = 2 - \frac{\epsilon}{12} + O(\epsilon^2) \ .
\label{II5dpex}
\end{equation}
In the vicinity of $v^*$, the solution of the RG equation for the {\em order 
parameter} reads, recalling that $[S] = \mu^{d/2}$,
\begin{equation} 
  \left\langle S_R(\tau_R,t) \right\rangle \approx \mu^{d/2} \,  
  \ell^{(d - \gamma_S^*)/2} \, {\hat S}\left( \tau_R \,\ell^{\gamma_\tau^*},
  v_R^*, D_R \, \mu^2 \, \ell^{2 + \gamma_D^*} \, t \right) \ ,
\label{II5oprg}
\end{equation}
which leads to the following scaling relations and explicit exponent values,
\begin{equation} 
  \beta = \frac{\nu (d + \eta)}{2} = 1 - \frac{\epsilon}{6} + O(\epsilon^2) 
  \ , \quad 
  \alpha = \frac{\beta}{z \, \nu} = 1 - \frac{\epsilon}{4} + O(\epsilon^2) \ .
\label{II5dpba}
\end{equation}

\begin{table}  
\centering  
\caption{Comparison of the DP critical exponent values from Monte Carlo  
  simulations with the results from the $\epsilon$ expansion.} 
\label{dpexpt}
\begin{tabular}{llll}  
\hline\noalign{\smallskip}
Scaling exponent & $\ d = 1$ & $\ d = 2$ & $\ d = 4 - \epsilon$ \\
\noalign{\smallskip}\hline\noalign{\smallskip}
$\xi \sim |\tau|^{- \nu}$ & $\ \nu \approx 1.100$ & $\ \nu \approx 0.735$ &  
$\ \nu = 1/2 + \epsilon / 16 + O(\epsilon^2)$ \\
$t_c \sim \xi^z \sim |\tau|^{- z \nu}$ & $\ z \approx 1.576$ & 
$\ z \approx 1.73$ & $\ z = 2 - \epsilon / 12 + O(\epsilon^2)$ \\ 
$a_\infty \sim |\tau|^\beta$ & $\ \beta \approx 0.2765$ & 
$\ \beta \approx 0.584$ & $\ \beta = 1 - \epsilon / 6 + O(\epsilon^2)$ \\ 
$a_c(t) \sim t^{- \alpha}$ & $\ \alpha \approx 0.160$ & 
$\ \alpha \approx 0.46$ & $\ \alpha = 1 - \epsilon / 4 + O(\epsilon^2)$ \\ 
\noalign{\smallskip}\hline
\end{tabular}
\end{table}
The scaling exponents for critical directed percolation are known analytically
for a plethora of physical quantities (but the reader should beware that 
various different conventions are used in the literature); for the two-loop 
results to order $\epsilon^2$ in the perturbative dimensional expansion, see 
Ref.~\cite{JanTau}.
In Table~\ref{dpexpt}, we compare the $O(\epsilon)$ values with the results
from Monte Carlo computer simulations, which allow the DP critical exponents
to be measured to high precision (for recent overviews on simulation results
for DP and other absorbing state phase transitions, see 
Refs.~\cite{Haye,Odor}).
Yet unfortunately, there are to date hardly any real experiments that would
confirm the DP conjecture \cite{JanDP,Grassb} and actually measure the 
scaling exponents for this prominent nonequilibrium universality class.

\subsection{Dynamic isotropic percolation and multi-species variants} 
\label{ssec:II6}
 
An interesting variant of active to absorbing state phase transitions emerges
when we modify the SEP rules (1) and (2) in Sec.\ref{ssec:II5} to
\begin{enumerate} 
\item[1'.] The susceptible medium becomes infected, depending on the 
  densities $n$ {\em and} $m$ of sick individuals and the {\em `debris'}, 
  respectively.  
  After a brief time interval, the sick individuals decay into immune debris, 
  which ultimately stops the disease locally by exhausting the supply of 
  susceptible regions. 
\item[2'.] The states with $n = 0$ and any spatial distribution of $m$ are  
  {\em absorbing}, and describe the {\em extinction} of the disease. 
\end{enumerate}
Here, the debris is given by the accumulated decay products,
\begin{equation}
  m(\vec{x},t) = \kappa \int_{- \infty}^t n(\vec{x},t') \, \D t' \ .
\label{II6dipd}
\end{equation}
After rescaling, this {\em general epidemic process (GEP)} or epidemic with 
removal \cite{Murray} is described in terms of the mesoscopic Langevin 
equation \cite{GraPer}
\begin{equation} 
  \frac{\partial S(\vec{x},t)}{\partial t} = - D \left( r - \vec{\nabla}^2  
  \right) S(\vec{x},t) - D \, u \, S(\vec{x},t) \int_{-\infty}^t S(\vec{x},t')
  \, D t' + \zeta(\vec{x},t) \ , 
\label{II6dipe}
\end{equation}
with noise correlator (\ref{II5dpns}).
The associated response functional reads \cite{JanPer,CarGra}
\begin{equation}  
  {\cal A}[{\widetilde S},S] = \int \! \D^dx \! \int \! \D t \, \biggl[ 
  {\widetilde S} \left( \frac{\partial}{\partial t} + D \left( r - 
  \vec{\nabla}^2 \right) \right) S - u \, {\widetilde S}^2 \, S 
  + D \, u \, S \int^t S(t') \biggr] \ .
\label{II6dipf} 
\end{equation}

\begin{figure} 
\centering 
\includegraphics[width = 5 truecm]{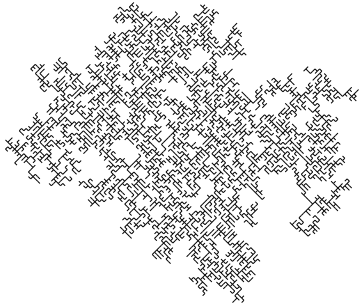} 
\caption{Isotropic percolation cluster.} 
\label{isopcl} 
\end{figure}

For the field theory thus defined, one may take the {\em quasistatic limit} by
introducing the fields
\begin{equation}
  {\tilde \varphi}(\vec{x}) = {\widetilde S}(\vec{x},t \to \infty) \ , \quad
  \varphi(\vec{x}) = D \int_{-\infty}^\infty S(\vec{x},t') \, \D t' \ .
\label{II6stal}
\end{equation}
For $t \to \infty$, the action (\ref{II6dipf}) thus becomes 
\begin{equation}
  {\cal A}_{\rm qst}[{\tilde \varphi},\varphi] = \int \! \D^dx \, 
  {\tilde \varphi} \Bigl[ r - \vec{\nabla}^2 - u \left( {\tilde \varphi} 
  - \varphi \right) \Bigr] \varphi \ ,
\label{II6qsta}
\end{equation}
which is known to describe the critical exponents of {\em isotropic
percolation} \cite{BenCar}.
An isotropic percolation cluster is shown in Fig.~\ref{isopcl}, to be 
contrasted with the anisotropic scaling evident in Fig.~\ref{dirpcl}(b). 
The upper critical dimension of isotropic percolation is $d_c = 6$, and an
explicit calculation, with the diagrams of Fig.~\ref{dploop}, but involving 
the static propagators $G_0(\vec{q}) = 1 / (r + \vec{q}^2)$, yields the 
following critical exponents for isotropic percolation, to first order in 
$\epsilon = 6 - d$,
\begin{equation}
  \eta = - \frac{\epsilon}{21} + O(\epsilon^2) \ , \quad  
  \frac{1}{\nu} = 2 - \frac{5 \, \epsilon}{21} + O(\epsilon^2) \ , \quad 
  \beta = 1 - \frac{\epsilon}{7} + O(\epsilon^2) \ . 
\label{II6ises}
\end{equation}
In order to calculate the dynamic critical exponent for this {\em dynamic 
isotropic percolation (dIP)} universality class, we must return to the full
action (\ref{II6dipf}). 
Once again, with the diagrams of Fig.~\ref{dploop}, but now involving a 
temporally nonlocal three-point vertex, one then arrives at
\begin{equation}
  z = 2 - \frac{\epsilon}{6} + O(\epsilon^2) \ .
\label{II6ised}
\end{equation} 
For a variety of two-loop results, the reader is referred to 
Ref.~\cite{JanTau}.
It is also possible to describe the {\em crossover} from isotropic to
directed percolation within this field-theoretic framework 
\cite{FreTauSch,JanSte}.

Let us next consider {\em multi-species} variants of directed percolation 
processes, which can be obtained in the particle language by coupling the DP 
reactions $A_i \to \emptyset$, $A_i \rightleftharpoons A_i + A_i$ via 
processes of the form $A_i \rightleftharpoons A_j + A_j$ (with $j \not= i$); 
or directly by the corresponding generalisation within the Langevin 
representation with $\langle \zeta_i(\vec{x},t) \rangle = 0$,
\begin{eqnarray} 
  &&\frac{\partial S_i}{\partial t} = D_i \left( \vec{\nabla}^2 - R_i[S_i] 
  \right) S_i + \zeta_i \ , \quad R_i[S_i] = r_i + \sum_j g_{ij} \, S_j 
  + \ldots \ , 
\label{II6msdp} \\ 
  &&\left\langle \zeta_i(\vec{x},t) \zeta_j(\vec{x}',t') \right\rangle = 
  2 S_i N_i[S_i] \, \delta(\vec{x} - \vec{x}') \, \delta(t - t') \, 
  \delta_{ij} \, , \ N_i[S_i] = u_i + \ldots \qquad\
\label{II6msdn}
\end{eqnarray} 
The ensuing renormalisation factors turn out to be precisely as for 
single-species DP, and consequently the {\em generical} critical behaviour 
even in such multi-species systems is governed by the DP universality class
\cite{Hannes}.
For example, the predator extinction threshold for the stochastic 
Lotka--Volterra system mentioned in Sec.~\ref{ssec:II1} is characterised by
the DP exponents as well \cite{MoGeTa}. 
But these reactions also {\em generate} $A_i \to A_j$, causing additional 
terms $\sum_{j \not= i} g_j \, S_j$ in Eqs.~(\ref{II6msdp}).
Asymptotically, the inter-species couplings become {\em unidirectional}, which
allows for the appearance of special {\em multicritical points} when several
$r_i = 0$ simultaneously \cite{TauHowHin}.
This leads to a {\em hierarchy} of order parameter exponents $\beta_k$ on the 
$k$th level of a unidirectional cascade, with 
\begin{equation}
  \beta_1 = 1 - \frac{\epsilon}{6} + O(\epsilon^2) \ , \quad  
  \beta_2 = \frac{1}{2} - \frac{13 \, \epsilon}{96} + O(\epsilon^2) \ , \ldots
  , \ \beta_k = \frac{1}{2^k} - O(\epsilon) \ ;
\label{II6mcdp}
\end{equation}
for the associated {\em crossover exponent}, one can show $\Phi = 1$ to 
{\em all} orders \cite{Hannes}. 
Quite analogous features emerge for {\em multi-species dIP processes} 
\cite{Hannes,JanTau}.

\subsection{Concluding remarks} 
\label{ssec:II7}
 
In these lecture notes, I have described how stochastic processes can be 
mapped onto field theory representations, starting either from a mesoscopic
Langevin equation for the coarse-grained densities of the relevant order
parameter fields and conserved quantities, or from a more microsopic master 
equation for interacting particle systems.
The dynamic renormalisation group method can then be employed to study and
characterise the universal scaling behaviour near continuous phase transitions
both in and far from thermal equilibrium, and for systems that generically
display scale-invariant features.
While the critical dynamics near equilibrium phase transitions has been
thoroughly investigated experimentally in the past three decades, regrettably 
such direct experimental verification of the by now considerable amount of 
theoretical work on {\em nonequilibrium} systems is largely amiss.
In this respect, applications of the expertise gained in the nonequilibrium
statistical mechanics of complex cooperative behaviour to biological systems 
might prove fruitful and constitutes a promising venture.
One must bear in mind, however, that nonuniversal features are often crucial
for the relevant questions in biology. 

In part, the lack of clearcut experimental evidence may be due to the fact
that asymptotic universal properties are perhaps less prominent in accessible 
nonequilibrium systems, owing to long crossover times.
Yet fluctuations do tend to play a more important role in systems that are 
driven away from thermal equilibrium, and the concept of universality classes,
despite the undoubtedly much increased richness in dynamical systems, should
still be useful.
For example, we have seen that the directed percolation universality class 
quite generically describes the critical properties of phase transitions from 
active to inactive, absorbing states, which abound in nature.
The few exceptions to this rule either require the coupling to another 
conserved mode \cite{KrSS,vWOeHi}; the presence, on a mesoscopic level, of
additional symmetries that preclude the spontaneous decay $A \to \emptyset$ as
in the so-called {\em parity-conserving (PC)} universality class, represented 
by branching and annihilating random walks $A \to (n + 1) \, A$ with $n$ 
{\em even}, and $A + A \to \emptyset$ \cite{CarTau} (for recent developments 
based on nonperturbative RG approaches, see Ref.~\cite{CanChDel}); or the 
absence of any first-order reactions, as in the (by now rather notorious) 
{\em pair contact process with diffusion (PCPD)} \cite{HenHin}, which has so 
far eluded a successful field-theoretic treatment \cite{JaWiDeTa}.
A possible explanation for the fact that DP exponents have not been measured
ubiquitously (yet) could be the instability towards {\em quenched disorder} in
the reaction rates \cite{JansDD}.

In reaction--diffusion systems, a complete classification of the scaling 
properties in {\em multi-species} systems remains incomplete, aside from pair
annihilation and DP-like processes, and still constitutes a quite formidable 
program (for a recent overview over the present situation from a 
field-theoretic viewpoint, see Ref.~\cite{TauHowLee}).
This is even more evident for nonequilibrium systems in general, even when 
maintained in driven {\em steady states}.
Field-theoretic methods and the dynamic renormalisation group represent
powerful tools that I believe will continue to crucially complement exact 
solutions (usually of one-dimensional models), other approximative approaches,
and computer simulations, in our quest to further elucidate the intriguing 
cooperative behaviour of strongly interacting and fluctuating many-particle 
systems.

\subsection*{Acknowledgements}

This work has been supported in part by the U.S. National Science Foundation
through grant NSF DMR-0308548.
I am indebted to many colleagues, students, and friends, from and with whom I
had the pleasure to learn and research the material presented here; 
specifically I would like to mention Vamsi Akkineni, Michael Bulenda, 
John Cardy, Olivier Deloubri\`ere, Daniel Fisher, Reinhard Folk, Erwin Frey, 
Ivan Georgiev, Yadin Goldschmidt, Peter Grassberger, Henk Hilhorst, 
Haye Hinrichsen, Martin Howard, Terry Hwa, Hannes Janssen, Bernhard Kaufmann, 
Mauro Mobilia, David Nelson, Beth Reid, Zolt\'an R\'acz, Jaime Santos, 
Beate Schmittmann, Franz Schwabl, Steffen Trimper, Ben Vollmayr-Lee, 
Mark Washenberger, Fred van Wijland, and Royce Zia.
Lastly, I would like to thank the organisers of the very enjoyable Luxembourg
Summer School on {\em Ageing and the Glass Transition} for their kind
invitation, and my colleagues at the Laboratoire de Physique Th\'eorique, 
Universit\'e de Paris-Sud Orsay, France and at the Rudolf Peierls Centre for 
Theoretical Physics, University of Oxford, U.K., where these lecture notes 
were conceived and written, for their warm hospitality.


\begin{thebibliography}{99.}

\bibitem{Ramond} 
  P.~Ramond:
  \textit{Field theory --- a modern primer},
  (Benjamin/Cummings, Reading 1981) 

\bibitem{Amit} 
  D.J.~Amit:
  \textit{Field theory, the renormalization group, and critical phenomena}
  (World Scientific, Singapore 1984)

\bibitem{ItzDro} 
  C.~Itzykson and J.M.~Drouffe:
  \textit{Statistical field theory}
  (Cambridge University Press, Cambridge 1989)

\bibitem{Bellac} 
  M.~Le~Bellac:
  \textit{Quantum and statistical field theory},
  (Oxford University Press, Oxford 1991)

\bibitem{Zinn} 
  J.~Zinn-Justin: 
  \textit{Quantum field theory and critical phenomena} 
  (Clarendon Press, Oxford 1993) 
 
\bibitem{Cardy} 
  J.~Cardy:
  \textit{Scaling and renormalization in statistical physics} 
  (Cambridge University Press, Cambridge 1996)

\bibitem{HohHalp} 
  P.C.~Hohenberg and B.I.~Halperin:  
  Rev. Mod. Phys. \textbf{49}, 435 (1977)
 
\bibitem{Janssen} 
  H.K.~Janssen:
  Field-theoretic methods applied to critical dynamics. 
  In: \textit{Dynamical critical phenomena and related topics,
  Lecture Notes in Physics}, vol. 104, ed by C.P.~Enz
  (Springer, Heidelberg 1979), pp. 26-47  
  
\bibitem{BaJaWa} 
  R.~Bausch, H.K.~Janssen, and H.~Wagner: 
  Z. Phys. B \textbf{24}, 113 (1976)

\bibitem{JCardy}
  J.L.~Cardy: 
  Renormalisation group approach to reaction-diffusion problems. 
  In: \textit{Proceedings of Mathematical Beauty of Physics}, 
  ed by J.-B.~Zuber, Adv. Ser. in Math. Phys. \textbf{24}, 113 (1997)
 
\bibitem{MatGlas}
  D.C.~Mattis and M.L.~Glasser: 
  Rev. Mod. Phys. \textbf{70}, 979 (1998)
 
\bibitem{TauHowLee} 
  U.C.~T\"auber, M.J.~Howard, and B.P.~Vollmayr-Lee:
  J. Phys. A: Math. Gen. \textbf{38}, R79 (2005) 

\bibitem{UCT}
  U.C.~T\"auber:
  \textit{Critical dynamics: a field theory approach to equilibrium and  
  non-equilibrium scaling behavior}, in preparation (to be published at 
  Cambridge University Press, Cambridge); for completed chapters, see: \\
  \texttt{http://www.phys.vt.edu/$\,\widetilde{}\,$tauber/utaeuber.html} 

\bibitem{Ferretal} 
  R.A.~Ferrell, N.~Menyh\`ard, H.~Schmidt, F.~Schwabl, and P.~Sz\'epfalusy: 
  Phys. Rev. Lett. \textbf{18}, 891 (1967); 
  Ann. Phys. (NY) \textbf{47}, 565 (1968)
 
\bibitem{HalpHoh} 
  B.I.~Halperin and P.C.~Hohenberg:
  Phys. Rev. \textbf{177}, 952 (1969)

\bibitem{HaHoMa} 
  B.I.~Halperin, P.C.~Hohenberg, and S.-k.~Ma:
  Phys. Rev. Lett. \textbf{29}, 1548 (1972)

\bibitem{DDBrZJ} 
  C.~De~Dominicis, E.~Br\'ezin, and J.~Zinn-Justin:
  Phys. Rev. B \textbf{12}, 4945 (1975)

\bibitem{Jan} 
  H.K.~Janssen:
  Z. Phys. B \textbf{23}, 377 (1976)

\bibitem{DeDom}
  C.~De~Dominicis:
  J. Phys. (France) Colloq. \textbf{37}, C2247 (1976)

\bibitem{MaSiRo} 
  P.C.~Martin, E.D.~Siggia, and H.A.~Rose:
  Phys. Rev. A \textbf{8}, 423 (1973)

\bibitem{Wagner}
   H.~Wagner: 
  Z. Phys. \textbf{195}, 273 (1966)
 
\bibitem{MerWag} 
  N.D.~Mermin and H.~Wagner: 
  Phys. Rev. Lett. \textbf{17}, 1133 (1966)
 
\bibitem{Hohenb} 
  P.C.~Hohenberg:
  Phys. Rev. \textbf{158}, 383 (1967)

\bibitem{ChaiLub}
  P.M.~Chaikin and T.C.~Lubensky: 
  \textit{Principles of condensed matter physics}, 
  (Cambridge University Press, Cambridge 1995)

\bibitem{MaMaz} 
  S.-k.~Ma and G.F.~Mazenko: 
  Phys. Rev. Lett. \textbf{33}, 1383 (1974); 
  Phys. Rev. B \textbf{11}, 4077 (1975)
  
\bibitem{FreySchw} 
  E.~Frey and F.~Schwabl:
  Adv. Phys. \textbf{43}, 577 (1994)  

\bibitem{JaSchSch} 
  H.K.~Janssen, B.~Schaub, and B.~Schmittmann: 
  Z. Phys. B \textbf{73}, 539 (1989)
 
\bibitem{JanRel}
  H.K.~Janssen: 
  On the renormalized field theory of nonlinear critical relaxation. 
  In: \textit{From phase transitions to chaos}, ed by G.~Gy\"orgyi, I.~Kondor,
  L.~Sasv\'ari, and T.~T\'el (World Scientific, Singapore 1992), pp.~68-91. 

\bibitem{CalGam}
  P.~Calabrese and A.~Gambassi:
  Phys. Rev. E \textbf{66}, 066101 (2002);
  J. Phys. A: Math. Gen. \textbf{38}, R133 (2005) 

\bibitem{Andrea}
  A.~Gambassi:
  In: \textit{Proceedings of the International Summer School
  ``Ageing and the Glass Transition''}, 
  to appear in J. Phys. A: Math. Gen. (2006)

\bibitem{Diehl}
  H.W.~Diehl:
  In: \textit{Phase Transitions and Critical Phenomena}, vol. 10,  
  ed by C.~Domb and J.L.~Lebowitz (Academic Press, London 1986)

\bibitem{OerJan} 
  K.~Oerding and H.K.~Janssen:
  J. Phys. A: Math. Gen. \textbf{26}, 5295 (1993)

\bibitem{TaAkSa} 
  U.C.~T\"auber, V.K.~Akkineni, and J.E.~Santos: 
  Phys. Rev. Lett. \textbf{88}, 045702 (2002)
 
\bibitem{HaLeWi}
  F.~Haake, M.~Lewenstein, and M.~Wilkens:
  Z. Phys. B \textbf{55}, 211 (1984)

\bibitem{GrJaHe} 
  G.~Grinstein, C.~Jayaprakash, and Y.~He:
  Phys. Rev. Lett. \textbf{55}, 2527 (1985)

\bibitem{BaSchm}
  K.E.~Bassler and B.~Schmittmann:
  Phys. Rev. Lett. \textbf{73}, 3343 (1994) 

\bibitem{TaRacz} 
  U.C.~T\"auber and Z.~R\'acz:
  Phys. Rev. E \textbf{55}, 4120 (1997)

\bibitem{TaSaRa} 
  U.C.~T\"auber, J.E.~Santos, and Z.~R\'acz:  
  Eur. Phys. J. B \textbf{7}, 309 (1999) 
 
\bibitem{SchmZia} 
  B.~Schmittmann and R.K.P.~Zia:
  Phys. Rev. Lett. \textbf{66}, 357 (1991)
 
\bibitem{Beate}
  B.~Schmittmann:
  Europhys. Lett. \textbf{24}, 109 (1993)
 
\bibitem{BasRacz}
  K.E.~Bassler and Z.~R\'acz:
  Phys. Rev. Lett. \textbf{73}, 1320 (1994);
  Phys. Rev. E \textbf{52}, R9 (1995)

\bibitem{SchZiaB} 
  B.~Schmittmann and R.K.P.~Zia:
  Statistical mechanics of driven diffusive systems.
  In: \textit{Phase Transitions and Critical Phenomena}, vol. 17,  
  ed by C.~Domb and J.L.~Lebowitz (Academic Press, London 1995)

\bibitem{JanSch} 
  H.K.~Janssen and B.~Schmittmann:
  Z. Phys. B {\bf 63}, 517 (1986)
 
\bibitem{LeuCar}
  K.-t.~Leung and J.L.~Cardy:
  J. Stat. Phys. \textbf{44}, 567 (1986)

\bibitem{FoNeSt} 
  D.~Forster, D.R.~Nelson, and M.J.~Stephen:
  Phys. Rev. A \textbf{16}, 732 (1977)

\bibitem{Murray}
  J.D.~Murray:
  \textit{Mathematical Biology}, vols.~I/II 
  (Springer, New York, 3rd ed 2002)

\bibitem{MoGeTa} 
  M.~Mobilia, I.T.~Georgiev, and U.C.~T\"auber: 
  e-print {\tt q-bio.PE/0508043} (2005) 
 
\bibitem{Doi}
  M.~Doi:
  J. Phys. A: Math. Gen. \textbf{9}, 1465 \& 1479 (1976) 
 
\bibitem{GraSch} 
  P.~Grassberger and M.~Scheunert:
  Fortschr. Phys. \textbf{28}, 547 (1980)
 
\bibitem{Peliti}
  L.~Peliti:
  J. Phys. (Paris) \textbf{46}, 1469 (1985)

\bibitem{AlDrHeRi}
  F.C.~Alcaraz, M.~Droz, M.~Henkel, and V.~Rittenberg:
  Ann. Phys. (NY) \textbf{230}, 250 (1994)

\bibitem{HeOrSa}
  M.~Henkel, E.~Orlandini, and J.~Santos:
  Ann. Phys. (NY) \textbf{259}, 163 (1997)

\bibitem{Schuetz} 
  G.M.~Sch\"utz:
  In: \textit{Phase Transitions and Critical Phenomena}, vol. 19,  
  ed by C.~Domb and J.L.~Lebowitz (Academic Press, London 2001)

\bibitem{Stinch}
  R.~Stinchcombe:
  Adv. Phys. \textbf{50}, 431 (2001)

\bibitem{Wijland} 
  F.~van~Wijland:
  Phys. Rev. E \textbf{63}, 022101 (2001) 
 
\bibitem{NegOrl}
  J.W.~Negele and H.~Orland:
  \textit{Quantum many-particle systems} 
  (Addison-Wesley, Redwood City 1988)

\bibitem{PelRen} 
  L.~Peliti: 
  J. Phys. A: Math. Gen. \textbf{19}, L365 (1986) 

\bibitem{Hannes} 
  H.K.~Janssen:
  J. Stat. Phys. \textbf{103}, 801 (2001) 
  
\bibitem{Lee} 
  B.P.~Lee:
  J. Phys. A: Math. Gen. {\bf 27}, 2633 (1994) 

\bibitem{LeeCar} 
  B.P.~Lee and J.~Cardy:
  Phys. Rev. E \textbf{50}, R3287 (1994) 

\bibitem{TouWil}
  D.~Toussaint and F.~Wilczek:
  J. Chem. Phys. \textbf{78}, 2642 (1983)

\bibitem{LeeJohn} 
  B.P.~Lee and J.~Cardy: 
  J. Stat. Phys. \textbf{80}, 971 (1995)

\bibitem{DelHilTau} 
  O.~Deloubri\`ere, H.J.~Hilhorst, and U.C.~T\"auber: 
  Phys. Rev. Lett. \textbf{89}, 250601 (2002); 
  H.J.~Hilhorst, O.~Deloubri\`ere, M.J.~Washenberger, and U.C.~T\"auber:  
  J. Phys. A: Math. Gen. \textbf{37}, 7063 (2004)
 
\bibitem{HilWasTau}
  H.J.~Hilhorst, M.J. Washenberger, and U.C.~T\"auber: 
  J. Stat. Mech. P10002 (2004)

\bibitem{Moshe}
  M.~Moshe:
  Phys. Rep. C \textbf{37}, 255 (1978)

\bibitem{GraSun}
  P.~Grassberger and K.~Sundermeyer: 
  Phys. Lett. B \textbf{77}, 220 (1978)

\bibitem{GraTor}
  P.~Grassberger and A.~De~La~Torre:
  Ann. Phys. (NY) \textbf{122}, 373 (1979)

\bibitem{JanTau} 
  H.K.~Janssen and U.C.~T\"auber:  
  Ann. Phys. (NY) \textbf{315}, 147 (2005) 

\bibitem{Obuk}
  S.P.~Obukhov:
  Physica A \textbf{101}, 145 (1980)

\bibitem{CarSug}
  J.L.~Cardy and R.L.Sugar:
  J. Phys. A: Math. Gen. \textbf{13}, L423 (1980)

\bibitem{JanDP} 
  H.K.~Janssen: 
  Z. Phys. B \textbf{42}, 151 (1981) 
 
\bibitem{Kinzel}
  W.~Kinzel:
  In: \textit{Percolation structures and processes}, 
  ed by G.~Deutsch, R.~Zallen, and J.~Adler (Hilger, Bristol 1983)
 
\bibitem{Grassb} 
  P.~Grassberger, 
  Z. Phys. B \textbf{47}, 365 (1982) 
 
\bibitem{Haye} 
  H.~Hinrichsen:
  Adv. Phys. \textbf{49}, 815 (2001) 
 
\bibitem{Odor} 
  G.~\'Odor: 
  Rev. Mod. Phys. \textbf{76}, 663 (2004)

\bibitem{GraPer}
  P.~Grassberger:
  Math. Biosc. \textbf{63}, 157 (1983)

\bibitem{JanPer}
  H.K.~Janssen:
  Z. Phys. B \textbf{58}, 311 (1985)

\bibitem{CarGra}
  J.L.~Cardy and P.~Grassberger:
  J. Phys. A: Math. Gen. \textbf{18}, L267 (1985)

\bibitem{BenCar}
  J.~Benzoni and J.L.~Cardy:
  J. Phys. A: Math. Gen. \textbf{17}, 179 (1984)

\bibitem{FreTauSch}
  E.~Frey, U.C.~T\"auber, and F.~Schwabl:
  Europhys. Lett. \textbf{26}, 413 (1994);
  Phys. Rev. E \textbf{49}, 5058 (1994)

\bibitem{JanSte}
  H.K.~Janssen and O.~Stenull:
  Phys. Rev. E \textbf{62}, 3173 (2000)

\bibitem{TauHowHin}
  U.C.~T\"auber, M.J.~Howard, and H.~Hinrichsen:  
  Phys. Rev. Lett. \textbf{80}, 2165 (1998);  
  Y.Y.~Goldschmidt, H.~Hinrichsen, M.J.~Howard, and U.C.~T\"auber:  
  Phys. Rev. E \textbf{59}, 6381 (1999)

\bibitem{KrSS}
  R.~Kree, B.~Schaub, and B.~Schmittmann:
  Phys. Rev. A \textbf{39}, 2214 (1989)

\bibitem{vWOeHi}
  F.~van~Wijland, K.~Oerding, and H.~Hilhorst:
  Physica A \textbf{251}, 179 (1998);
  K.~Oerding, F.~van~Wijland, J.P.~Leroy, and H.~Hilhorst:
  J. Stat. Phys. \textbf{99}, 1365 (2000)

\bibitem{CarTau} 
  J.~Cardy and U.C.~T\"auber:  
  Phys. Rev. Lett. \textbf{77}, 4780 (1996); 
  J. Stat. Phys. \textbf{90}, 1 (1998)

\bibitem{CanChDel} 
  L.~Canet, H.~Chat\'e, and B.~Delamotte: 
  Phys. Rev. Lett. \textbf{92}, 255703 (2004);
  L.~Canet, H.~Chat\'e, B.~Delamotte, I.~Dornic, and M.A.~Mu\~noz: 
  e-print {\tt cond-mat/0505170} (2005)

\bibitem{HenHin}
  M.~Henkel and H.~Hinrichsen:
  J. Phys. A: Math. Gen. \textbf{37}, R117 (2004)

\bibitem{JaWiDeTa} 
  H.K.~Janssen, F.~van~Wijland, O.~Deloubri\`ere, and U.C.~T\"auber: 
  Phys. Rev. E \textbf{70}, 056114 (2004)

\bibitem{JansDD}
  H.K.~Janssen:
  Phys. Rev. E \textbf{55}, 6253 (1997)

\end{thebibliography}
\end{document}